\pdfoutput=1 
\documentclass[a4paper,11pt]{article}

\usepackage{jheppub} 

\usepackage[T1]{fontenc} 

\usepackage{booktabs}
\usepackage{authblk}
\usepackage{lineno}
\usepackage{multirow}
\usepackage{bigstrut}
\usepackage{subfigure}
\usepackage{rotating}
\usepackage{placeins}
\usepackage{lscape}
\usepackage{grffile}
\usepackage{verbatim}
\usepackage{mathrsfs}
\usepackage{mathtools}
\usepackage{morefloats}
\usepackage{epstopdf}
\usepackage{afterpage}

\usepackage{atlasphysics} 

\newcommand\bb{\ensuremath{b\overline{b}}}
\newcommand\ttb{\ensuremath{t\overline{t}}}
\newcommand\qqb{\ensuremath{q\overline{q}}}
\newcommand\wzh{$(W/Z)H$}
\newcommand\wln{$W\to\ell\nu$}
\newcommand\zll{$Z\to\ell\ell$}
\newcommand\znn{$Z\to\nu\nu$}
\newcommand{\hbb}{\ensuremath{H\to b\overline{b}}}
\newcommand\zbb{$Z\to b\overline{b}$}

\def\vec#1{{\mbox{$\boldsymbol{#1}$}}}

\def\mh{\ensuremath{m_H}}
\def\met{\ensuremath{E_{\mathrm{T}}^{\mathrm{miss}}}}
\def\metvec{\ensuremath{{\vec{E_{\mathrm{T}}^{\mathrm{miss}}}}}}
\def\mht{\ensuremath{H_{\mathrm{T}}^{\mathrm{miss}}}}
\def\mpt{\ensuremath{p_{\mathrm{T}}^{\mathrm{miss}}}}
\def\mptvec{\ensuremath{{\vec{p_{\mathrm{T}}^{\mathrm{miss}}}}}}
\def\pt{\ensuremath{p_{\mathrm{T}}}}
\def\et{\ensuremath{E_{\mathrm{T}}}}
\def\mtw{\ensuremath{m_{\mathrm{T}}^W}}
\def\ptv{\ensuremath{p_{\mathrm{T}}^V}}
\def\ptw{\ensuremath{p_{\mathrm{T}}^W}}
\def\ptz{\ensuremath{p_{\mathrm{T}}^Z}}
\def\mbb{\ensuremath{m_{bb}}}
\def\dphi{\ensuremath{\Delta\phi(\mathrm{jet}_1,\mathrm{jet}_2)}}
\def\vec#1{{\mbox{$\boldsymbol{#1}$}}}
\def\ifb{\ensuremath{\mathrm{fb}^{-1}}}

\title{
Search for the \bb\ decay of the Standard Model Higgs boson in 
associated \wzh\ production with the ATLAS detector

}

\author{The ATLAS collaboration}
\affiliation{ATLAS, CERN}

\emailAdd{atlas.publications@cern.ch}

\abstract{
A search for the \bb\ decay of the Standard Model Higgs boson is performed with 
the ATLAS experiment using the full dataset recorded at the LHC in Run 1. 
The integrated luminosities used are 4.7 and 20.3~\ifb\ from $pp$ collisions at $\sqrt{s}=7$ 
and 8~TeV, respectively. The processes considered are associated \wzh\ 
production, where $W\to e\nu/\mu\nu$, $Z\to ee/\mu\mu$ 
and \znn. The observed (expected) deviation 
from the background-only hypothesis corresponds
to a significance of 1.4 (2.6) standard deviations and the ratio of the measured signal yield
to the Standard Model expectation is found to be 
$\mu = 0.52 \pm 0.32 \mathrm{(stat.)} \pm 0.24 \mathrm{(syst.)}$
for a Higgs boson mass of 125.36~GeV. 
The analysis procedure is validated by a measurement of the yield 
of $(W/Z)Z$ production with $Z\to\bb$ in the same final states as for the Higgs boson search,
from which the ratio of the observed signal yield to the Standard Model 
expectation is found to be $0.74 \pm 0.09 \mathrm{(stat.)} \pm 0.14 \mathrm{(syst.)}$.

}

\begin{document} 
\maketitle
\flushbottom

\section{Introduction}

For decades, the Higgs 
boson~\cite{Englert:1964aa,Higgs:1964aa,Higgs:1964ab,Guralnik:1964aa} 
of the Standard Model (SM) 
remained an unconfirmed prediction.
In July 2012, the ATLAS and CMS experiments at the LHC reported the observation 
of a new particle with a mass of about 125~GeV and with properties consistent 
with those expected for the SM Higgs boson~\cite{:2012gk,:2012gu}. 
Since then, more precise measurements have strengthened the hypothesis that 
the new particle is indeed a Higgs
boson~\cite{Aad:2013wqa,Aad:2013xqa,Chatrchyan:2013lba}. 
These measurements, however, have been mainly performed in the bosonic decay modes 
of the new particle 
($H\to\gamma\gamma$, $H\to ZZ$, and $H\to WW$).
It is also essential to verify whether it decays into fermions as
predicted by the Standard Model. 
Recently, the CMS Collaboration reported evidence for the $\tau\tau$ decay 
mode of the Higgs boson at a level of significance of 3.4 standard 
deviations ($\sigma$) for $\mh=125$~GeV~\cite{Chatrchyan:2014nva}.

The $H\to\bb$ decay mode is predicted in the SM to have a branching ratio 
of 58\% for $\mh=125$~GeV~\cite{Djouadi:1997yw}. 
Accessing $H\to\bb$ decays is therefore crucial for constraining,
under fairly general assumptions~\cite{Lafaye:2009vr}, the overall Higgs boson decay width 
and, in a global fit to all accessible combinations of Higgs boson production and decay modes, 
to allow for measurements of absolute Higgs boson couplings.
An inclusive search for $H\to\bb$ is not feasible at hadron colliders because
of the overwhelming background from multijet production. In spite of a cross 
section more than an order of magnitude lower than the dominant gluon-fusion 
process, associated production of a Higgs boson with a vector boson, 
$W$ or $Z$~\cite{PhysRevD.18.1724}, offers a viable alternative because leptonic decays 
of the vector boson, \wln, \zll\ ($\ell=e,\mu$), and \znn, can be efficiently used for 
triggering and background reduction purposes~\cite{Stange:1993ya,Stange:1994bb}. 
The CDF and D0 experiments at the Tevatron reported 
an excess of events in their search for associated \wzh\ production in the 
\hbb\ decay mode at a significance level of 2.8$\sigma$ 
for $\mh=125$~GeV~\cite{TevatronBB}. 
Recently, the CMS experiment reported an excess of events in the \hbb\ decay 
mode with a significance of 2.1$\sigma$ for 
$\mh=125$~GeV~\cite{Chatrchyan:2013zna}. 

In this paper, a search for associated \wzh\ production of 
the SM Higgs boson in the \bb\ decay mode is presented, using the 
full integrated luminosity accumulated by ATLAS during Run~1 of the LHC: 
4.7 and 20.3~\ifb\ from proton--proton ($pp$) collisions at
centre-of-mass energies of 7 and 8~TeV in 2011 and 2012, respectively. 
An analysis of the 7~TeV dataset has already been published by
ATLAS~\cite{atlas:2012zf}.
In addition to the increase in the amount of data analysed,
the update presented in this paper benefits from numerous analysis
improvements. Some of the improvements to the object 
reconstruction, however, are available only for the 8~TeV dataset, 
which leads to separate analysis strategies for the two datasets.

The analysis is performed for events containing zero, one, or two charged leptons 
(electrons or muons), targeting the \znn, \wln, or \zll\ decay modes of the 
vector boson, respectively. 
In addition to \znn\ decays, the 0-lepton channel has a 
smaller but not insignificant contribution from leptonic $W$ decays
when the lepton is produced outside of 
the detector acceptance or not identified. 
A $b$-tagging algorithm is used 
to identify the jets consistent with originating from an \hbb\ decay. 
To improve the sensitivity, the three channels
are each split according to the vector-boson transverse momentum, the
number of jets (two or three), and the number of $b$-tagged jets.   
Topological and kinematic selection criteria are applied
within each of the resulting categories. 

A binned maximum likelihood fit is used to extract the signal yield and the background 
normalisations. Systematic uncertainties on the signal and background 
modelling are implemented as deviations in their respective models in the form of 
"nuisance" parameters that are varied in the fit. Each nuisance parameter is 
constrained by a penalty term in the likelihood, associated with its uncertainty.
Two versions of the analysis are presented in this paper: in the first, 
referred to as the dijet-mass analysis, the mass of the dijet system of $b$-tagged jets is the final 
discriminating variable used in the statistical analysis;
in the other, a multivariate analysis (MVA) incorporating various kinematic 
variables in addition to the dijet mass, as well as $b$-tagging information, provides 
the final discriminating variable. Because the latter information is
not available in similar detail for the 7~TeV dataset, the MVA is used only for the 8~TeV dataset.
In both analyses, dedicated control samples, 
typically with loosened $b$-tagging requirements,  
constrain the contributions of the dominant background processes. 
The most significant background sources are $(W/Z)$+heavy-flavour-jet production 
and $\ttb$ production. The normalisations of these backgrounds are fully determined 
by the likelihood fit. Other significant background sources are single-top-quark and
diboson ($WZ$ and $ZZ$) production, with normalisations taken from theory, 
as well as multijet events, normalised using multijet-enriched control samples.
Since the MVA has higher expected sensitivity, it is chosen as the
nominal analysis for the 8~TeV dataset to extract the final results.
To validate the analysis procedures, both for the dijet-mass and MVA approaches, 
a measurement of the yield of $(W/Z)Z$ 
production is performed in the same final states and with the same event 
selection, with \hbb\ replaced by \zbb. 
 
This paper is organised as follows. A brief description of the ATLAS detector is 
given in section~\ref{sec:det}. 
Details of the data and simulated samples used in this 
analysis are provided in section~\ref{sec:samples}. 
This is followed by sections describing 
the dijet-mass and multivariate analyses applied to the 8~TeV data. 
The reconstruction of 
physics objects such as leptons and jets is addressed in section~\ref{sec:reco}. 
Section~\ref{sec:selec} details the event
selections applied to the dijet-mass and multivariate analyses, while 
section~\ref{sec:mva} explains the construction of the final discriminating 
variable of the MVA.
Section~\ref{sec:bkg} discusses the background composition in the various analysis regions,
while the systematic uncertainties are addressed in section~\ref{sec:sys}. 
The statistical procedure used to extract the results is described in 
section~\ref{sec:fit}. For
the 7~TeV data, only a dijet-mass analysis is used, and differences with respect
to the 8~TeV data analysis are specified in section~\ref{sec:seven}. 
The results are presented
and discussed in section~\ref{sec:results}, and a summary of the paper is given in 
section~\ref{sec:summary}.

\section{The ATLAS detector}\label{sec:det}
 
 The ATLAS detector~\cite{Aad:2008zzm} is cylindrically symmetric around 
 the beam axis and is structured in a barrel and two endcaps. It consists 
 of three main subsystems.  The inner tracking detector is immersed in 
 the 2~T axial magnetic field produced by a superconducting solenoid. 
 Charged-particle position and momentum
 measurements are made by pixel detectors followed by silicon-strip 
 detectors in the pseudorapidity\footnote{
ATLAS uses a right-handed coordinate system
with its origin at the nominal interaction point (IP) in the centre of
the detector and the $z$-axis coinciding with the axis of the beam
pipe.  The $x$-axis points from the IP towards the centre of the LHC ring,
and the $y$-axis points upward. Cylindrical coordinates ($r$,$\phi$)
are used in the transverse plane, $\phi$ being the azimuthal angle
around the $z$-axis. The pseudorapidity is defined in terms of the
polar angle $\theta$ as $\eta = - \ln \tan(\theta/2)$. 
The distance in ($\eta$,$\phi$) coordinates, 
$\Delta R = \sqrt{(\Delta\phi)^2+(\Delta\eta)^2}$, 
is also used to define cone sizes.
Transverse momentum and energy are defined as $\pt=p\sin\theta$ and 
$\et=E\sin\theta$, respectively.
For the purpose of object selections, $\eta$ is calculated relative 
to the geometric centre of the detector; otherwise, it is relative to the
reconstructed primary vertex of each event.
} 
 range $|\eta|<2.5$ and by a straw-tube transition-radiation tracker (TRT) in the range 
 $|\eta|<2.0$. The pixel detectors are crucial for $b$-tagging, 
 and the TRT also contributes to electron identification. 
 The calorimeters, located beyond the solenoid, cover the range $|\eta|<4.9$ with 
 a variety of detector technologies. 
 The liquid-argon electromagnetic calorimeters are divided into barrel 
 ($|\eta|<1.475$), endcap ($1.375<|\eta|<3.2$), and forward ($3.1<|\eta|<4.9$) 
 sections. The hadronic calorimeters
 (using scintillator tiles or liquid argon as active materials)
 surround the electromagnetic calorimeters with a coverage of $|\eta|< 4.9$.  The
 muon spectrometer measures the deflection of muon tracks in the field
 of three large air-core toroidal magnets, each containing eight
 superconducting coils. It is instrumented with separate trigger
 and high-precision tracking chambers covering the $|\eta|<2.4$ 
 and $|\eta|<2.7$ ranges, respectively.

The trigger system is organised in three levels. The first level is based
on custom-made hardware and uses coarse-granularity calorimeter and muon information.
The second and third levels are implemented as software algorithms and
use the full detector granularity. At the second level, only regions deemed 
interesting at the first level are analysed, while the third level, called
the event filter, makes use of the full detector read-out to reconstruct and
select events, which are then logged for offline analysis at a rate of up to 400~Hz averaged over an accelerator fill.

\section{Data and simulated samples}\label{sec:samples}

The datasets used in this analysis include only $pp$ collision data recorded in stable beam 
conditions and with all relevant sub-detectors providing 
high-quality data. The corresponding integrated luminosities are 4.7 and 
20.3~\ifb~\cite{Aad:2013ucp} for the 7 and 8 TeV data, respectively.

Events in the 0-lepton channel are selected by triggers based on the magnitude \met\ 
of the missing transverse momentum vector. 
The \met\ trigger configuration evolved during data taking
to cope with the increasing
luminosity, and the trigger efficiency was improved for the 8~TeV data.
The dependence of the \met\ trigger efficiency on the \met\ reconstructed offline is measured 
in $W\to\mu\nu$+jets and $Z\to\mu\mu$+jets events collected with single-muon triggers,
with the offline \met\ calculated without the muon contribution. 
As there was a brief period of data-taking in which the \met\ triggers were not 
available for the first bunch crossings of two bunch trains, the integrated luminosity 
for the 0-lepton channel in the 7~TeV dataset is reduced to 4.6~\ifb.
Events in the 1-lepton channel are primarily selected by single-lepton 
triggers. The \ET\ threshold of the single-electron trigger was raised from 20 
to 22~GeV during the 7~TeV data-taking period, and to 24~GeV for the 8~TeV data.
The \pt\ threshold of the single-muon trigger was similarly increased from 18 GeV
for the 7~TeV data to 24 GeV at 8~TeV. 
As the single-lepton triggers for the 8~TeV data include isolation criteria,
triggers with higher thresholds (60~GeV for electrons and 36~GeV for muons) 
but no isolation requirements are used in addition.
Single-lepton trigger efficiencies are measured using
a tag-and-probe method applied to $Z\to ee$ and $Z\to\mu\mu$ events.
In the 1-muon sub-channel, \met\ triggers are also used to compensate for the limited 
muon trigger-chamber coverage in some regions of the detector. 
Events in the 2-lepton channel are selected by a combination of single-lepton, dielectron
and dimuon triggers. The thresholds of the dilepton triggers are
12~GeV for electrons and 13~GeV for muons. 

Monte Carlo (MC) simulated samples are produced for signal and background
processes using the {\sc atlfast-II} simulation~\cite{atlas_simulation},
which includes a full simulation of the ATLAS detector based on 
the {\sc geant4} program~\cite{Agostinelli:2002hh}, 
except for the response of the calorimeters for which a parameterised 
simulation is used. A list of the generators used for signal and background simulations
is given in table~\ref{tab:samples}. 

\newcommand{\hsp}{\hspace*{0.5cm}}
\begin{table}[tb!]
\begin{center}{\small
\begin{tabular}{ll} \hline\hline
\hspace{0.8cm} Process \hspace{2.0cm} & Generator \hspace{2.5cm} \\ \hline
\hline
\hsp Signal$^{(\star)}$ \\
\hline
\hsp $\qqb\to ZH \to\nu\nu bb/\ell\ell bb$ & {\sc pythia8} \\
\hsp $gg  \to ZH \to\nu\nu bb/\ell\ell bb$ & {\sc powheg+pythia8} \\
\hsp $\qqb\to WH \to\ell\nu bb$ & {\sc pythia8} \\
\hline\hline
\hsp Vector boson + jets \\
\hline
\hsp $W\to\ell\nu$ & {\sc Sherpa 1.4.1} \\
\hsp $Z/\gamma*\to\ell\ell$ & {\sc Sherpa 1.4.1}  \\
\hsp $Z\to\nu\nu$ & {\sc Sherpa 1.4.1}  \\
\hline\hline
\hsp Top-quark  \\
\hline
\hsp $t\bar{t}$ &  {\sc powheg+pythia}\\
\hsp $t$-channel & {\sc AcerMC+pythia}   \\
\hsp $s$-channel & {\sc powheg+pythia}   \\
\hsp $Wt$ & {\sc powheg+pythia} \\
\hline\hline
\hsp Diboson$^{(\star)}$ & {\sc powheg+pythia8} \\
\hline
\hsp $WW$ & {\sc powheg+pythia8}  \\
\hsp $WZ$ & {\sc powheg+pythia8} \\
\hsp $ZZ$ &{\sc powheg+pythia8}  \\
\hline\hline
\end{tabular}
\caption
{The generators used for the simulation of the signal and background
processes.
$(\star)$ For the analysis of the 7~TeV data, {\sc pythia8} is used
for the simulation of the $gg\to ZH$ process, and {\sc herwig} for
the simulation of diboson processes.
\protect\label{tab:samples}}}
\end{center}
\end{table}

The MC generator used for $\qqb$-initiated $WH$ and $ZH$ production is 
{\sc pythia8}~\cite{Sjostrand:2007gs} with the CTEQ6L1~\cite{Pumplin:2002vw} 
parton distribution functions (PDFs). The AU2 tune~\cite{Aad:2011mctunes,ATL-PHYS-PUB-2011-008} 
is used for the parton shower, hadronisation, and multiple parton interactions.
The {\sc photos} program~\cite{Golonka:2005pn} is used for QED final-state radiation. 
The {\sc powheg} generator~\cite{Nason:2004rx,Frixione:2007vw,Alioli:2010xd} is used
within the MiNLO approach~\cite{Luisoni:2013cuh} 
with the CT10 PDFs~\cite{ct10}, interfaced to {\sc pythia8} with the AU2 tune, 
as a cross-check and to evaluate systematic uncertainties on the signal acceptance 
and kinematic properties. It is also used for the generation of
gluon-gluon-initiated $ZH$ production at leading order (LO) in QCD,
with results cross-checked by an independent computation~\cite{Englert:2013vua}.
(For the analysis of the 7~TeV data, the {\sc pythia8} generator is
used for $gg\to ZH$.) 
The transverse momentum 
distributions of the Higgs boson show substantial differences between the two $ZH$ 
production processes.
For $\qqb$-initiated $WH$ and $ZH$ production,
the total production cross sections and associated uncertainties are computed 
at next-to-next-to-leading order (NNLO) in QCD~\cite{Ohnemus:1992bd,Baer:1992vx,Brein:2003wg},
and with electroweak corrections at next-to-leading order (NLO)~\cite{Ciccolini:2003jy}. 
Additional normalisation-preserving differential
electroweak NLO corrections are applied as a function 
of the transverse momentum of the vector boson~\cite{Denner:2011rn}. 
For gluon-gluon-initiated $ZH$ production, NLO corrections~\cite{Altenkamp:2012sx},
which increase the total $ZH$ production cross section by about 5\%, are taken into account.
The Higgs boson decay branching ratios are calculated with {\sc hdecay}~\cite{Djouadi:1997yw}. 
Signal samples are simulated for Higgs boson masses from 100 to 150~GeV in steps of 5~GeV. 
All charged-lepton flavours are simulated in the $W$ and $Z$ decays, as leptonic decays of the
$\tau$ leptons can also be selected in the analysis. 
For the Higgs boson, only the \bb\ decay mode is considered in the analysis. 

The main background processes are $(W/Z)$+jets and \ttb\ production. 
Version 1.4.1 of the {\sc sherpa} generator~\cite{Gleisberg:2008ta} is used 
with the CT10 PDFs to simulate $W$+jets and $Z$+jets at 
leading-order in QCD, with massive $c$- and $b$-quarks. 
For \ttb\ production, the simulation is performed with the
{\sc powheg} generator with the CT10 PDFs, 
interfaced with {\sc pythia6}~\cite{pythia}, for which the CTEQ6L1 PDFs and the
Perugia2011C tune~\cite{Aad:2011mctunes,ATL-PHYS-PUB-2011-008} are used.
In this analysis, the final normalisations of these dominant backgrounds are 
constrained by the data, but theoretical cross sections are used to optimise the selection. 
The cross sections are calculated at NNLO for $(W/Z)$+jets~\cite{Melnikov:2006kv} and 
at NNLO, including resummations of next-to-next-to-leading logarithmic
(NNLL) soft gluon terms, for \ttb~\cite{Czakon:2013goa}.

Additional backgrounds arise from single-top-quark and diboson ($WW$, $WZ$, and 
$ZZ$) production. For single-top-quark production, the $s$-channel exchange process and $Wt$
production are simulated with {\sc powheg}, as for \ttb, 
while the $t$-channel exchange 
process is simulated with the {\sc AcerMC} generator~\cite{Kersevan:2004yg} 
interfaced with {\sc pythia6}, using the CTEQ6L1
PDFs and the Perugia2011C tune. The cross sections are taken from 
refs.~\cite{Kidonakis:2011wy,Kidonakis:2010tc,Kidonakis:2010ux}.
The {\sc powheg} generator with the CT10 PDFs, interfaced to {\sc pythia8} with the AU2 tune,
is used for diboson processes~\cite{Nason:2013ydw}.
(For the analysis of the 7~TeV data, the {\sc herwig} generator~\cite{herwig}
is used instead with the CTEQ6L1 PDFs 
and the AUET2 tune~\cite{Aad:2011mctunes,ATL-PHYS-PUB-2011-008}, and the cross sections
are obtained at NLO from {\sc mcfm}~\cite{Campbell:2010ff} 
with the MSTW2008NLO PDFs~\cite{mstw}.)

Events from minimum-bias interactions are simulated with the {\sc pythia8}
generator with the MSTW\-2008LO PDFs~\cite{Sherstnev:2007nd} and 
the A2 tune~\cite{Aad:2011mctunes,ATL-PHYS-PUB-2011-008}. They are
overlaid on the  simulated signal and background events according
to the luminosity profile of the recorded data. 
The contributions from these ``pile-up'' interactions are simulated both
within the same bunch crossing as the hard-scattering process
and in neighbouring bunch crossings. 
The resulting events are then processed through the
same reconstruction programs as the data. 

Additional generators are used for the assessment of systematic uncertainties 
as explained in section~\ref{sec:sys}.

Simulated jets are labelled according to which generated hadrons with $\pt>5$~GeV
are found within a cone of size $\Delta R = 0.4$ around the reconstructed jet axis. 
If a $b$-hadron is found, the jet is labelled as a $b$-jet. 
If not and a $c$-hadron is found, the jet is labelled as a $c$-jet. 
If neither a $b$- nor a $c$-hadron is found, the jet
is labelled as a light (i.e., $u$-, $d$-, or $s$-quark, or gluon) jet.
Simulated $V$+jet events, where $V$ stands for $W$ or $Z$, are then categorised 
according to the labels of the two jets that are used to reconstruct the 
Higgs boson candidate. 
If one of those jets is labelled as a $b$-jet, the event belongs to 
the $Vb$ category. If not and one of the jets is labelled as a $c$-jet, the event
belongs to the $Vc$ category. Otherwise, the event belongs to the $Vl$ category.
Further subdivisions are defined according to the flavour of the other jet from
the pair, using the same precedence order: $Vbb$, $Vbc$, $Vbl$, $Vcc$, $Vcl$.
The combination of $Vbb$, $Vbc$, $Vbl$ and $Vcc$ is denoted $V$+hf.
The $Vcl$ final state is not included in $V$+hf because the main 
production process is $gs\to Wc$ rather than gluon splitting.

\section{Object reconstruction}\label{sec:reco}

In this section, the reconstruction of physics objects used in the analysis of the 
8~TeV data is presented. Differences relevant for the analysis of the 7~TeV data are
reported in section~\ref{sec:seven}.

Charged-particle tracks are reconstructed with a \pt\ threshold of 400~MeV.
The primary vertex is selected from amongst all reconstructed vertices as the one 
with the largest sum of associated-track squared transverse momenta $\Sigma\pt^{2}$ 
and is required to have at least three associated tracks. 

Three categories of electrons~\cite{Aad:2014fxa,ATLAS-CONF-2014-032} and 
muons~\cite{Aad:2014zya} 
are used in the analysis, referred to as loose, medium and tight leptons
in order of increasing purity.
Loose leptons are selected with transverse energy $E_\mathrm{T} > 7$~GeV.
Loose electrons are required to have $|\eta| < 2.47$
and to fulfil the ``very loose likelihood'' identification 
criteria defined in ref.~\cite{ATLAS-CONF-2014-032}. 
The likelihood-based electron identification
combines shower-shape information, track-quality criteria, 
the matching quality between the track and its associated energy cluster in the calorimeter 
(direction and momentum/energy), TRT information 
and a criterion to help identify electrons originating from photon conversions.
The electron energies are calibrated by making use of reference
processes such as $Z\to ee$~\cite{Aad:2014nim}.
Three types of muons are included in the loose definition to maximise the 
acceptance: 
(1) muons reconstructed in both the muon spectrometer and the inner detector (ID); 
(2) muons with $\pt > 20$~GeV identified in the calorimeter and 
associated with an ID track with $|\eta| < 0.1$, 
where there is limited muon-chamber coverage; 
and (3) muons with $|\eta| > 2.5$ 
identified in the muon spectrometer, and which do not match full ID tracks due to the 
limited inner-detector coverage. 
For muons of the first and second type, the muon-track impact
parameters with respect to the primary vertex must be smaller
than 0.1~mm and 10~mm in the transverse plane and along the $z$-axis, respectively.
Finally, the scalar sum of the 
transverse momenta of tracks within a cone of size $\Delta R = 0.2$ centred 
on the lepton-candidate track, excluding the lepton track, is required to 
be less than 10\% of the transverse momentum of the lepton.

Medium leptons must meet the loose identification criteria and have $E_\mathrm{T} > 25$~GeV. 
Medium muons must be reconstructed in both the muon spectrometer and the inner detector 
and have $|\eta|<2.5$.
Tight electrons are required to additionally fulfil the ``very tight likelihood'' identification 
criteria~\cite{ATLAS-CONF-2014-032}.  
For both the tight electrons and the tight muons, more stringent isolation criteria must be satisfied:
the sum of the calorimeter energy deposits in a cone of size $\Delta R = 0.3$ around 
the lepton, excluding energy associated with the lepton candidate, must be less than 
4\% of the lepton energy, and the track-based isolation requirement is tightened
from 10\% to 4\%. 

Jets are reconstructed from noise-suppressed topological clusters of energy in the calori\-meters~\cite{Lampl:2008zz} using 
the anti-$k_t$ algorithm~\cite{AntiKt} with a radius parameter of 0.4. Jet 
energies are corrected for the contribution of pile-up interactions using a 
jet-area-based technique~\cite{Cacciari:2007fd} and calibrated using \pt- and $\eta$-dependent
correction factors determined from simulation, 
with residual corrections from in situ measurements applied to data~\cite{JES,Aad:2014bia}.
Further adjustments are made based on jet internal properties, which improve the energy 
resolution without changing the average calibration 
(global sequential calibration~\cite{JES}).
To reduce the contamination by jets from pile-up interactions, the scalar sum of the 
\pt\ of tracks matched to the jet and originating from the primary vertex must 
be at least 50\% of the scalar sum of the \pt\ of all tracks matched to the jet. 
This requirement is only applied to jets with 
$\pt < 50$~GeV and $|\eta| < 2.4$. Jets without any matched track are retained.
The jets kept for the analysis must have $\pt>20$~GeV and $|\eta| < 4.5$. 

To avoid double-counting, the following procedure is applied to loose leptons and jets.
First, if a jet and an electron are separated by $\Delta R < 0.4$, the jet is discarded.
Next, if a jet and a muon are separated by $\Delta R < 0.4$, the jet is discarded if it
has three or fewer matched tracks
since in this case it is likely
to originate from a muon having showered in the calorimeter; otherwise the muon is 
discarded. (Such muons are nevertheless included in the computation of the \met\ 
and in the jet energy corrections described in section~\ref{sec:selec}.)
Finally, if an electron and a muon are separated by $\Delta R < 0.2$, the muon is kept
unless it is identified only in the calorimeter, in which case the electron is kept.

The MV1c $b$-tagging algorithm is used to identify jets originating
from $b$-quark fragmentation. 
This algorithm combines in a neural network the information from various algorithms based on 
track impact-parameter significance or explicit reconstruction of $b$- and $c$-hadron 
decay vertices. It is an improved version of 
the MV1 algorithm~\cite{ATLAS-CONF-2011-102,btagnote2014b,Burmeister:1559721}
with  higher $c$-jet rejection.
Four $b$-tagging selection criteria (or operating points) are calibrated and used in the analysis, corresponding
to average efficiencies of 80\%, 70\%, 60\% and 50\% for 
$b$-jets with $\pt > 20$~GeV, as measured in simulated \ttb\ events.
In this analysis, the 80\%, 70\% and 50\% operating points are denoted 
loose, medium and tight, respectively. 
For the tight (loose) operating point, the rejection factors are 26 (3) and 1400 (30) 
against $c$-jets and light jets, respectively. For the tight operating point, the $c$-jet 
rejection factor is 1.9 times larger than obtained with the MV1 algorithm. 

The $b$-tagging efficiencies for $b$-jets, $c$-jets and light jets are measured in both data 
and simulation using dedicated event samples such as \ttb\ events for $b$-jets,
events with identified $D^*$ 
mesons for $c$-jets, or multijet events for light jets. 
The small differences observed
are used to correct the simulation by so-called ``scale factors''
(SFs) within intervals between two operating points. 
These SFs
are parameterised as a function of the jet \pt\ and, for light jets, also $|\eta|$. 
The SFs are, however, strictly valid only for the generator
used to derive them.
The differences observed when the efficiencies are measured
with different generators 
are taken into account by additional ``MC-to-MC'' 
SFs. 
Such differences can be caused by, e.g., different production
fractions of heavy-flavour hadrons or modelling of their decays.

Because of the large cross sections of $Vl$ and $Vc$ production, these backgrounds remain 
significant despite the powerful rejection of non-$b$-jets by the $b$-tagging algorithm. 
It is impractical to simulate a sufficiently large number of $Vl$ and $Vc$ events to provide 
a reliable description of these backgrounds in the analysis samples for which two $b$-tagged 
jets are required. An alternative procedure, parameterised tagging, is therefore used. 
Here, instead of directly tagging the $c$- and $l$-labelled jets with the MV1c algorithm, 
parameterisations as functions of \pt\ and $|\eta|$ of their probabilities to be $b$-tagged 
are used for the $Vl$, $Vc$ and $WW$ processes in all analysis samples in which two $b$-tagged 
jets are required.
These parameterisations are, however, integrated over other variables that can affect
the $c$- and light-jet tagging efficiencies. In particular, a strong dependence of these 
efficiencies is observed on $\Delta R$, the angular separation from the closest 
other jet, and a significant difference is seen between direct and parameterised tagging for $Vcc$ 
events with $\Delta R<1$. No such difference is seen for $Vcl$, $Vl$ and $WW$ events.
A dedicated correction, depending on $\Delta R$, is therefore applied to the $Vcc$ events. 

The missing transverse momentum vector
\metvec~\cite{Aad:2012re,ATLAS-CONF-2013-082} 
is measured as the negative vector sum of the transverse momenta associated 
with energy clusters in the calorimeters with $|\eta| < 4.9$. 
Corrections are applied to the energies of clusters associated 
with reconstructed objects (jets, electrons, $\tau$ leptons, and photons),
using the calibrations of these objects. The transverse momenta of 
reconstructed muons are included, with the energy deposited by these 
muons in the calorimeters properly removed to avoid double-counting.
In addition, a track-based 
missing transverse momentum vector, \mptvec, is calculated as the negative vector sum of 
the transverse momenta of tracks with 
$|\eta|<2.4$ associated with the primary vertex.

Additional corrections are applied to the simulation 
to account for small differences from data for trigger efficiencies, for lepton reconstruction 
and identification efficiencies, as well as for lepton energy and momentum resolutions.  
 
\clearpage

\section{Event selection}\label{sec:selec}

In this section, the event selection applied  in the analysis of the 
8~TeV data is presented. Differences in the analysis of the 
7~TeV data are reported in section~\ref{sec:seven}.

The analysis is optimised for a Higgs boson mass of 125~GeV.
Events are first categorised according to the numbers of leptons, jets, 
and $b$-tagged jets.

Events containing no loose leptons are assigned to the 0-lepton channel.  
Events containing one tight lepton and no additional loose leptons are 
assigned to the 1-lepton channel. 
Events containing one medium lepton and one additional loose lepton of the same flavour,
and no other loose leptons,
are assigned to the 2-lepton channel.
In the 1- and 2-lepton channels, for at least one of the lepton triggers by
which the event was selected, the objects that satisfied the trigger
are required to be associated with the selected leptons.

The jets used in this analysis, called ``selected jets'', must have $\pt > 20$~GeV 
and $|\eta| < 2.5$, the \eta\ range within which $b$-tagging can be applied. 
There must be exactly two or three such selected jets. 
Events 
containing an additional jet with $\pt > 30$~GeV and $|\eta| > 2.5$ are discarded to 
reduce the \ttb\ background. 
Only selected jets are
considered further, e.g., to define the jet multiplicity, or to calculate kinematic 
variables.

The $b$-tagging algorithm is applied to all selected jets. There must be no more 
than two such jets loosely $b$-tagged, and
3-jet events in which the lowest-\pt\ jet is loosely $b$-tagged are 
discarded.
At least one of the two $b$-tagged jets must have $\pt > 45$~GeV.
The following $b$-tagging categories are then defined as shown in figure~\ref{fig:tagging}.
Events with two jets satisfying the tight $b$-tagging criterion form the TT (or Tight) category; 
those not classified as TT, but with two jets satisfying the medium $b$-tagging criterion, 
form the MM (or Medium) category; 
those not classified as TT or MM, but with two jets satisfying the 
loose $b$-tagging criterion, form the LL (or Loose) category. This categorisation
improves the sensitivity with respect to what would be 
obtained using a single category, such as TT+MM, with the LL category providing constraints 
on the backgrounds not containing two real $b$-jets.  
Events with exactly one jet loosely $b$-tagged form the 1-tag category, and those with no 
loosely $b$-tagged jet form the 0-tag category.  
In the 3-jet categories, the dijet system is formed by the two $b$-tagged jets 
in any of the 2-tag categories, by the $b$-tagged jet and the leading (highest-\pt)
non-$b$-tagged jet for events in the 1-tag category, and by the two leading jets in the 
0-tag category.

\begin{figure}[tb!]
\begin{center}
\includegraphics[width=0.68\textwidth]{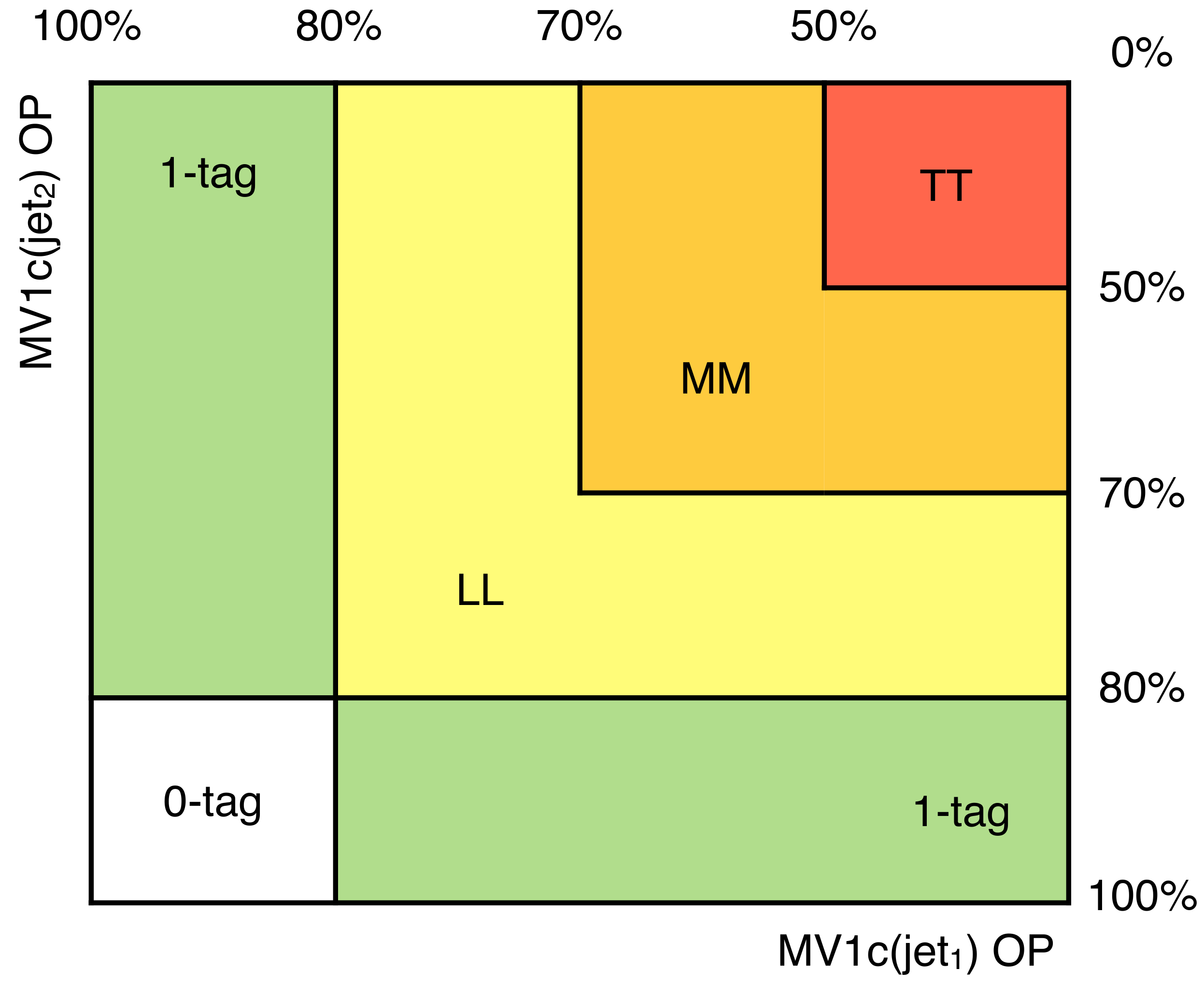}
\end{center}
\caption{Event classification as a function of the output of the MV1c $b$-tagging algorithm for the 
two highest \pt\ jets. The bin boundaries denote the operating points ({\it{MV1c(jet) OP}}) as defined 
in section~\ref{sec:reco}, corresponding to $b$-tagging efficiencies of 100\%, 80\%, 70\%, 50\%, 
i.e., the $b$-jet purity increases from left (bottom) to right (top). The event categories are 0-tag, 
1-tag, and TT, MM and LL for 2-tag, as explained in the text.}
\label{fig:tagging}
\end{figure}

Additional topological and kinematic criteria are applied to reject background events and enhance 
the sensitivity of the search. They are outlined in table~\ref{tab:selKinEvt} 
and detailed below. 
In general, the selection criteria are looser in the MVA than in the dijet-mass analysis in
order to maximise the information available to the final discriminant.

	\begin{table}[tb]
	\begin{center}
   \begin{footnotesize}
	\begin{tabular}{l|| c|c|c|c|c|| c|c}
	\hline\hline
	Variable    &  \multicolumn{5}{c||}{Dijet-mass analysis}                     & \multicolumn{2}{c}{Multivariate analysis} \\
	\hline\hline
	               \multicolumn{8}{c}{Common selection} \\
	\hline\hline
	\ptv\ [GeV]                &   0--90  & 90$^{(\ast)}$--120  & 120--160 & 160--200 & $>$ 200    & 0--120   & $>$ 120 \\
	$\Delta R(\mathrm{jet}_1,\mathrm{jet}_2)$ & 0.7--3.4 & 0.7--3.0 & 0.7--2.3 & 0.7--1.8 & $<$ 1.4 & \multicolumn{2}{c}{$>$ 0.7 (\ptv$<$200 \GeV)} \\
	\hline\hline
	              \multicolumn{8}{c}{0-lepton selection} \\
	\hline\hline
	        \mpt\ [GeV] & \multirow{6}{*}{NU} &  $>$ 30 & \multicolumn{3}{c||}{$>$ 30} & \multirow{6}{*}{NU} & $>$ 30 \\
	$\Delta\phi(\metvec,\mptvec)$ &                    &      $<$ $\pi/2$             & \multicolumn{3}{c||}{$<$ $\pi/2$} &    & $<$ $\pi/2$ \\
	${\rm min}[\Delta\phi(\metvec,{\rm jet})]$ &                    &    --               & \multicolumn{3}{c||}{$>$ 1.5} &    & $>$ 1.5 \\
	$\Delta \phi (\metvec, \mathrm{dijet})$       &                    &    $>$ 2.2               & \multicolumn{3}{c||}{$>$ 2.8} &    & -- \\
	$\sum\limits_{i=1}^{N_{\mathrm{jet}}=2(3)} \pt^{\mathrm{jet}_i}$ [GeV]  &        &     $>$ 120 (NU)              & \multicolumn{3}{c||}{$>$ 120 (150)} &    & $>$ 120 (150)\\
	\hline\hline
	               \multicolumn{8}{c}{1-lepton selection} \\
	\hline\hline
        \mtw\ [GeV]           & \multicolumn{5}{c||}{ $<$ 120 } & \multicolumn{2}{c}{--}     \\
	$H_{\mathrm{T}}$ [GeV]  &  \multicolumn{2}{c|}{$>$ 180} & \multicolumn{3}{c||}{--}      & $>$ 180 & --      \\
	\met\ [GeV] &  \multicolumn{2}{c|}{--} & \multicolumn{2}{c|}{$>$ 20} & $>$ 50  & --      & $>$ 20  \\
	\hline\hline
	               \multicolumn{8}{c}{2-lepton selection} \\
	\hline\hline
	\mll\ [GeV]  & \multicolumn{5}{c||}{ 83-99 }  & \multicolumn{2}{c}{71-121} \\
	\met\ [GeV]           & \multicolumn{5}{c||}{ $<$ 60 } & \multicolumn{2}{c}{--}     \\
	\hline\hline
	\end{tabular}
   \end{footnotesize}
	\end{center}
	\caption{\label{tab:selKinEvt}  Event topological and kinematic selections. 
NU stands for `Not Used'.
$(\ast)$ In the 0-lepton channel, the lower edge of the 
second \ptv\ interval is set at 100~GeV instead of 90~GeV.
For the 1-lepton channel, only the 1-muon sub-channel is used in the $\ptv < 120$~GeV 
intervals.}
	\end{table}

Further categorisation is performed according to the transverse momentum of the vector 
boson, \ptv, to take advantage of the better signal-to-background ratio at high \ptv.
The transverse momentum of the vector boson is reconstructed as the \met\ in the 0-lepton 
channel, the magnitude \ptw\ of the vector sum of the lepton transverse momentum 
and the \metvec\ in the 1-lepton channel, and the magnitude \ptz\ of the vector sum of the 
transverse momenta of the two leptons in the 2-lepton channel.
In the dijet-mass analysis, the events are categorised in five \ptv\ intervals, 
with boundaries at 0, 90, 120, 160, and 200~GeV. 
In the 0-lepton channel and for events fulfilling the condition on $\sum\pt^{\mathrm{jet}_i}$ 
mentioned in table~\ref{tab:selKinEvt}, the \met\ trigger is fully efficient for $\met > 160$~GeV, 
97\% efficient for $\met = 120$~GeV, and 80\% efficient for $\met = 100$~GeV, with
an efficiency that decreases rapidly for lower \met. Only four
intervals are therefore used in the 0-lepton channel, with a minimum \met\ value of 100~GeV.
In the 1-muon sub-channel, the \met\ trigger is used for $\ptw > 120$~GeV to recover events not 
selected by the single-muon trigger, thus increasing the signal acceptance 
in this channel by 8\%. 
In the MVA, only two intervals are defined, with \ptv\ below or above 120~GeV, but
the detailed \ptv\ information is used in the final discriminant. 

In the dijet-mass analysis, requirements are applied to the angular separation between the 
two jets of the dijet system, $\Delta R(\mathrm{jet}_1,\mathrm{jet}_2)$,
which depend on the \ptv\ interval. 
The requirement on the minimum value reduces the background from $V$+jet 
production, while the requirement on the maximum value, which reduces the 
background from \ttb\ production, is tightened with 
increasing \ptv\ to take advantage of the increasing collimation of the dijet system
for the signal. To increase the signal acceptance, the requirement on the minimum value is 
removed in the highest \ptv\ interval, where the amount of background is smallest. 
In the MVA, where the 
$\Delta R(\mathrm{jet}_1,\mathrm{jet}_2)$
information is used in the final discriminant, only a minimum
value is required, a requirement which is also removed for $\ptv > 200$~GeV. 

In the 0-lepton channel, the multijet (MJ) background is suppressed by
imposing requirements on the magnitude \mpt\ of the track-based
missing transverse momentum vector \mptvec, the azimuthal angle between \metvec\ and \mptvec, 
$\Delta \phi(\metvec, \mptvec)$, 
the azimuthal angle between \metvec\ and
the nearest jet, ${\rm min}[\Delta\phi(\metvec,{\rm jet})]$, and the
azimuthal angle between \metvec\ and the dijet system,
$\Delta \phi (\metvec, \mathrm{dijet})$. In addition, a minimum value is
required for the scalar sum of the jet transverse momenta, $\sum\pt^{\mathrm{jet}_i}$, which depends
on the jet multiplicity. 
Additional requirements are applied in the lowest \ptv\ interval of the 0-lepton channel, 
where the MJ background is largest: 
$N_{\mathrm{jet}} = 2$; $\met > 100$~GeV; $\dphi < 2.7$; ${\cal S} > 7$; and ${\cal L} > 0.5$. 
Here, \dphi\ is the azimuthal angle between the two jets, $\cal S$ is the \met\ significance,
defined as the ratio of \met\ to the square root of $\sum\pt^{\mathrm{jet}_i}$;  
and $\cal L$ is a likelihood ratio constructed to discriminate further against the MJ 
background.\footnote{The likelihood ratio uses the following inputs:
$\Delta \phi (\metvec, \mathrm{dijet})$; \dphi; the magnitude of the vector sum of the 
two jet transverse momenta, \mht; \mht\ divided by $\sum\pt^{\mathrm{jet}_i}$; and the
cosine of the helicity angle in the dijet rest frame as defined in ref.~\cite{Gallicchio:2010dq}.
For the MJ background, the probability density functions used in the likelihood ratio 
are constructed from data events selected with $\Delta \phi(\metvec, \mptvec) > \pi/2$.}

In the 1-lepton channel, a requirement is imposed on the transverse mass\footnote{
The transverse mass \mtw\ is calculated from the transverse momentum and the azimuthal 
angle of the charged lepton, $\pt^\ell$ and $\phi^\ell$, and from the missing transverse momentum 
magnitude, \met, and azimuthal angle, $\phi^{\mathrm{miss}}$: 
$\mtw = \sqrt{2\pt^\ell\met (1-\cos(\phi^\ell-\phi^{\mathrm{miss}}))}$.} 
\mtw\ in the dijet-mass analysis. This requirement reduces the contamination from 
the \ttb\ background. Requirements are also imposed 
on $H_{\mathrm{T}}$ (\met) for $\ptv < (>) 120$~GeV, where
$H_{\mathrm{T}}$ is the scalar sum of \met\ and the transverse momenta of 
the two leading jets and the lepton. This mainly reduces the MJ background.
As discussed in section~\ref{sec:multijet}, the MJ background is difficult to model
and remains substantial in the 1-electron sub-channel in the $\ptv < 120$~GeV intervals.
Therefore, only the 1-muon sub-channel is used in these intervals.

In the 2-lepton channel, criteria are imposed on the dilepton invariant
mass, \mll, which must be consistent with the mass of the $Z$ boson. 
In the dijet-mass analysis a requirement is imposed on \met; 
this variable is used in the final discriminant of the MVA. 

For events in which two jets are loosely $b$-tagged, 
these selection criteria define a set of ``2-tag signal regions'', categorised
in terms of channel (0, 1, or 2 leptons), \ptv\ interval, and jet multiplicity (2 or 3).
In the dijet-mass analysis, a further division is performed into the TT, MM and LL 
$b$-tagging categories. In the MVA, 
where the $b$-tagging information is used in the final discriminant, 
a similar subdivision is performed with the
difference that the TT and MM categories are merged in the 0- and 2-lepton channels.
Similarly defined 1-tag and 0-tag ``control regions'' are used in the analysis to constrain the 
main backgrounds. In the 1-lepton channel, the 2-tag signal regions with a third selected jet 
act in practice as control regions because they are largely dominated by \ttb\ events.
All 2-tag signal and 1-tag control regions are used simultaneously
in the global fit (described in section~\ref{sec:fit}) used to extract the results. 
The 0-tag control regions are used only for 
background modelling studies (reported in section~\ref{sec:bkg}).

After event selection, the energy calibration of the $b$-tagged jets is improved as follows.
The energy from muons within a jet is added
to the calorimeter-based jet energy after removing the energy 
deposited by the muon in the calorimeter (muon-in-jet correction), and a \pt-dependent 
correction is applied to account for biases in the response due to resolution effects
(resolution correction). This latter
correction is determined for the \pt\ spectrum of jets from the decay
of a Higgs boson with $\mh=125$~GeV in simulated $(W/Z)H$ events. The dijet mass resolution 
for the signal is improved by
14\% after these corrections 
and is typically 11\% (figure~\ref{fig:mass_resolution}(a)).
In the 2-lepton channel, wherein there is no true \met\ involved except possibly from 
semileptonic heavy-flavour decays, the energy calibration of the jets is further
improved by a kinematic likelihood fit, which includes a Breit--Wigner constraint
on the dilepton mass, Gaussian constraints on each of the transverse
components of the $\ell\ell bb$ system momentum 
(with a width of 9~GeV, as determined from $ZH$ simulated events), 
dedicated transfer functions relating the true jet transverse momenta to their 
reconstructed values (after the muon-in-jet correction, but without the resolution correction)
as well as a prior built from the expected true jet \pt\ spectrum in $ZH$ events
(playing a role similar to the resolution correction). Overall, the \bb\ mass resolution is 
improved by 30\% in the 2-lepton channel (figure~\ref{fig:mass_resolution}(b)).

\begin{figure}[tb!]
\centerline{
\subfigure[]{\includegraphics[width=0.49\textwidth]{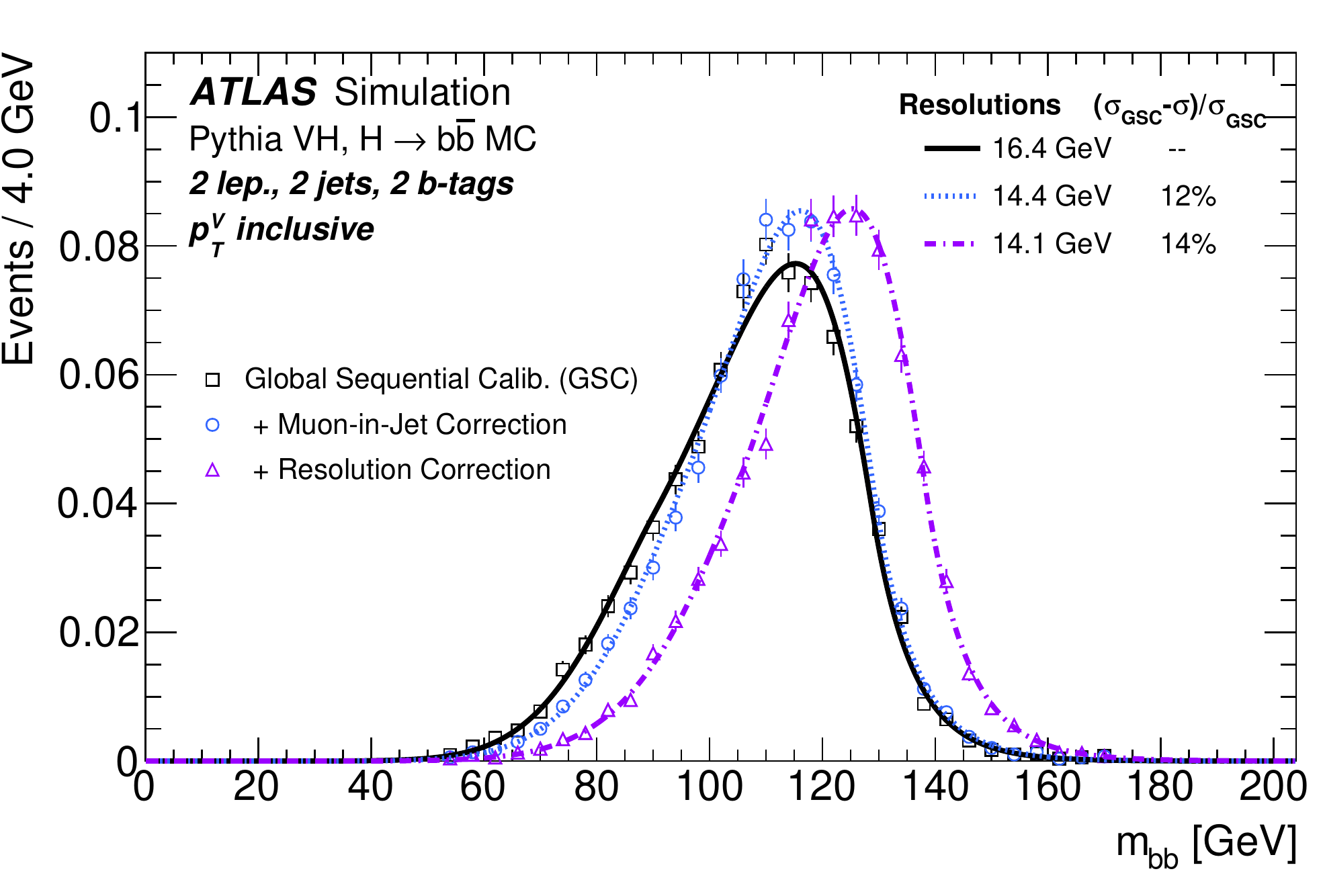}}
\subfigure[]{\includegraphics[width=0.49\textwidth]{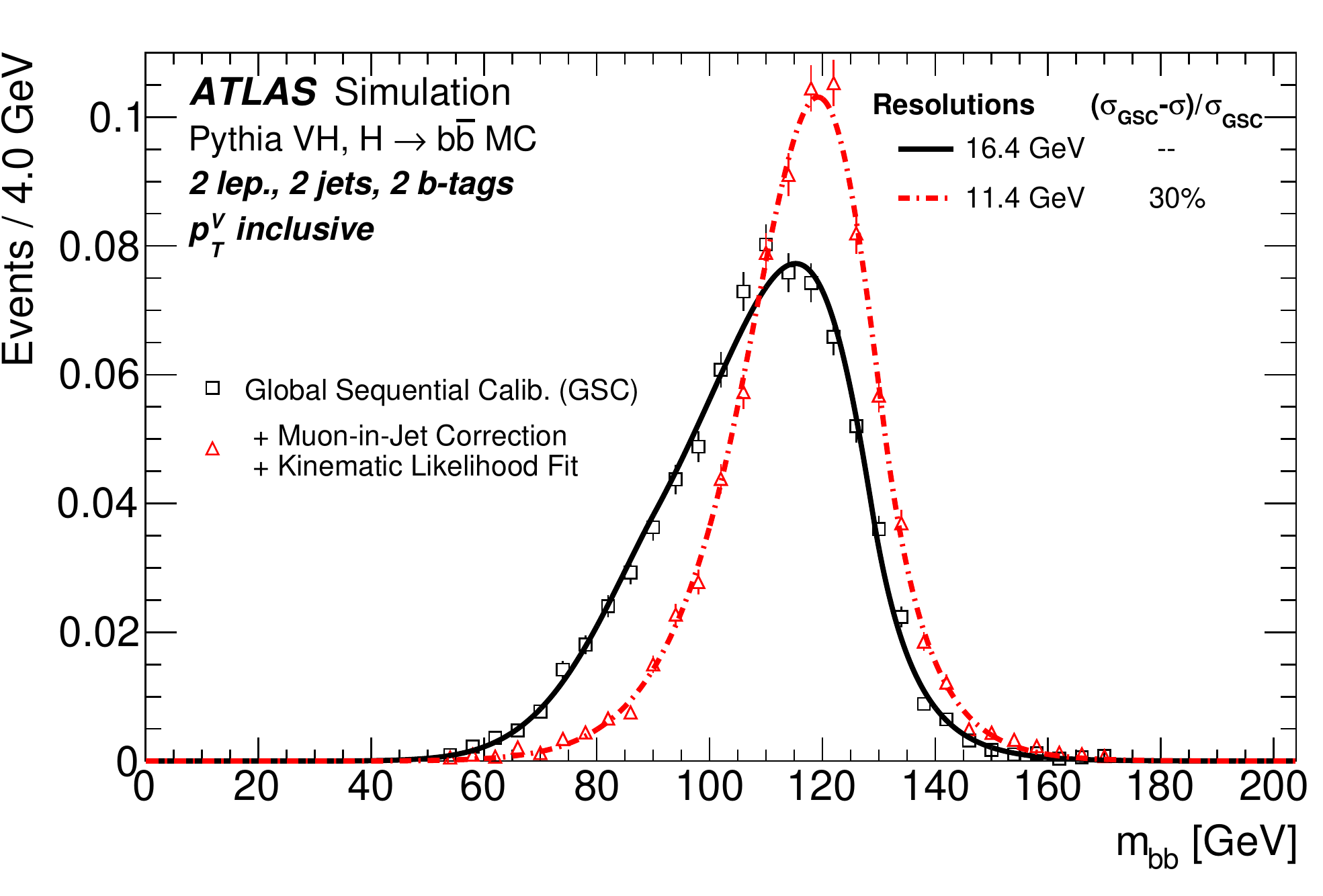}}}
\caption{Dijet-invariant-mass distribution for the decay products of a
  Higgs boson with $m_H = 125$~GeV in the 2-lepton MVA selection. 
The distributions are shown 
(a) using jets after global sequential calibration (GSC, solid), 
and after adding muons inside jets (dotted) 
and after correcting for resolution effects specific to the kinematics of
the decay of a Higgs boson with $m_H = 125$~GeV (dash-dotted); 
(b) using jets after global sequential calibration (GSC, solid), 
and after adding muons inside jets and applying the kinematic fit (dash-dotted). 
The distributions are fit to the Bukin function~\cite{Bukin} and the parameter representing 
the width of the core of the distribution is shown in the figures, 
as well as the relative improvement in the resolution with respect to jets after the global sequential calibration.
}
\label{fig:mass_resolution}
\end{figure}

The cross sections times branching ratios for \wzh\ with \wln, \zll, \znn,
and \hbb, as well as the acceptances in the three channels after full selection
are given in table~\ref{tab:crosssection_acc} for the MVA and the
dijet-mass analysis. 
The acceptance for other production and decay modes of the Higgs boson is negligible. 
The 0-lepton channel adds 7\% in acceptance for the \wln\ process 
with respect to the 1-lepton channel. Similarly, the 1-lepton channel adds 10\% in
acceptance for the \zll\ process with respect to the 2-lepton channel.  

\begin{table}[tb!]
\begin{center}
\begin{tabular}{l|ccccc}
\hline\hline
\multicolumn{5}{c}{$m_H = 125$~GeV at $\sqrt{s}=8$TeV} \\
\hline
\multirow{2}{*}{Process} & \multirow{2}{*}{Cross section $\times$ BR [fb]} & \multicolumn{3}{c}{Acceptance [\%]}\\
\cline{3-5}
&  & 0-lepton & 1-lepton & 2-lepton\\
\hline
$\qqb\to(Z\to\ell\ell)(\hbb)$ &  14.9  & --      & 1.3 (1.1) & 13.4 (10.9) \\ 
$gg\to(Z\to\ell\ell)(\hbb)$   &   1.3  & --      & 0.9 (0.7) & 10.5 (8.1) \\
$\qqb\to(W\to\ell\nu)(\hbb)$  & 131.7  & 0.3 (0.3) & 4.2 (3.7) & -- \\
$\qqb\to(Z\to\nu\nu)(\hbb)$ &  44.2  & 4.0 (3.8) & --        & -- \\
$gg\to(Z\to\nu\nu)(\hbb)$   &   3.8  & 5.5 (5.0) & --        & -- \\
\hline\hline
\end{tabular}
\end{center}
\caption{The cross section times branching ratio (BR) and acceptance for the three channels 
at 8 TeV. For $ZH$, the \qqb- and $gg$-initiated processes are shown separately.  
The branching ratios are calculated considering only decays to muons and electrons for \zll, 
decays to all three lepton flavours for \wln\ and decays to neutrinos for \znn. 
The acceptance is calculated as the fraction of events remaining in the combined 2-tag signal 
regions of the MVA (dijet-mass analysis) after the full event selection. 
\label{tab:crosssection_acc}}
\end{table}

\FloatBarrier

\clearpage
\section{Multivariate analysis}\label{sec:mva}

Although the dijet mass is the kinematic variable that provides the best discrimination between 
signal and backgrounds, the sensitivity of the search is improved by making use of additional 
kinematic, topological and $b$-tagging properties of the selected events in a multivariate analysis. 
The Boosted Decision Tree (BDT) technique~\cite{breiman,freund1} is used, which,
similarly to other multivariate methods, properly accounts for correlations between variables.


Dedicated BDTs are constructed, trained and evaluated in each of the 0-, 1- and 2-lepton
channels in the 2-tag regions (with the LL, MM and TT categories combined) 
and separately for the events with
two and three jets. In the 0-lepton channel, only events with $\ptv > 120$ GeV are 
used, whereas for the 1- and 2-lepton channels individual BDTs are used for 
$\ptv < 120$ GeV and $\ptv > 120$ GeV. Events in the electron
and muon sub-channels are combined since none of the variables used are lepton-flavour specific.
In the 0-lepton channel, the final results are obtained using the MVA
for $\ptv > 120$~GeV. For the small $100 < \ptv < 120$~GeV interval, which has reduced sensitivity, no
dedicated BDT is trained and only the dijet-mass distribution is used.

The BDTs are trained to separate the $(VH, H\to\bb)$ signal from the sum of 
the expected background processes. 
The input variables used to construct the BDTs are chosen in order
to maximise the separation, while avoiding the use of variables not
improving the performance significantly. Starting from the dijet mass, additional variables are 
tried one at a time and the one yielding the best separation gain is
kept. This procedure is repeated until adding more variables does not result
in a significant performance gain.  
The final sets of variables for the different channels are listed in table~\ref{tab:mvavars}. 
The $b$-tagged jets belonging to the dijet system
(with mass denoted $m_{bb}$) 
are labelled in decreasing \pt\ as $b_1$ and $b_2$, and
their separation in pseudorapidity is $|\Delta\eta({b}_1,{b}_2)|$. 
The $b$-tagging information is provided by the outputs of the MV1c neural network,
$MV1c({b}_1)$ and $MV1c({b}_2)$.
The angular separation, in the transverse plane,
of the vector boson and the dijet system of $b$-tagged jets and
their pseudorapidity separation 
are denoted $\Delta\phi(V,{bb})$ and $|\Delta\eta(V,{bb})|$, respectively.
In the 0-lepton channel, $H_{\mathrm{T}}$ is defined as the scalar sum 
of the transverse momenta of all jets and \met.
In the 1-lepton channel, the angle 
between the lepton and the closest $b$-tagged jet in the transverse plane is denoted
$\mathrm{min}[\Delta\phi(\ell,{b})]$. 
The other variables were defined in the previous sections.
In 3-jet events, the third jet is labelled as jet$_3$ and the mass of the 3-jet system is denoted $m_{bbj}$.
\begin{table}[tb!]
\begin{center}
\begin{tabular}{l|ccc}
\hline\hline
Variable                                      &  0-Lepton   & 1-Lepton    & 2-Lepton    \\
\hline                                      
\ptv                                          &             & $\times$  & $\times$  \\
\met                                          &  $\times$ & $\times$  & $\times$  \\
$\pt^{{b}_1}$                                 &  $\times$ & $\times$  & $\times$  \\
$\pt^{{b}_2}$                                 &  $\times$ & $\times$  & $\times$  \\
$m_{{bb}}$                                    &  $\times$ & $\times$  & $\times$  \\
$\Delta R({b}_1,{b}_2)$                      &  $\times$ & $\times$  & $\times$  \\
$|\Delta\eta({b}_1,{b}_2)|$                  &  $\times$ &           & $\times$  \\ 
$\Delta\phi(V,{bb})$                         &  $\times$ & $\times$  & $\times$  \\
$|\Delta\eta(V,{bb})|$                       &             &             & $\times$  \\
$H_{\mathrm{T}}$                               &  $\times$ &             &             \\
$\mathrm{min}[\Delta\phi(\ell,{b})]$         &             & $\times$  &          \\
\mtw                                          &             & $\times$  &             \\
\mll                                          &             &             & $\times$  \\
$MV1c({b}_1)$                        &  $\times$ & $\times$  & $\times$  \\
$MV1c({b}_2)$                        &  $\times$ & $\times$  & $\times$  \\
\hline
  & \multicolumn{3}{c}{Only in 3-jet events} \\
\hline
$\pt^{\mathrm{jet}_3}$                          &  $\times$ & $\times$  & $\times$  \\
$m_{bbj}$                                     &  $\times$ & $\times$  & $\times$  \\
\hline
\hline\hline
\end{tabular}
\end{center}
\caption{\label{tab:mvavars} Variables used in the multivariate analysis for the 
 0-, 1- and 2-lepton channels.}
\end{table}

%
%
\begin{sidewaysfigure}[p]
\begin{center}
%
%
\subfigure[]{\includegraphics[width=0.30\textwidth]{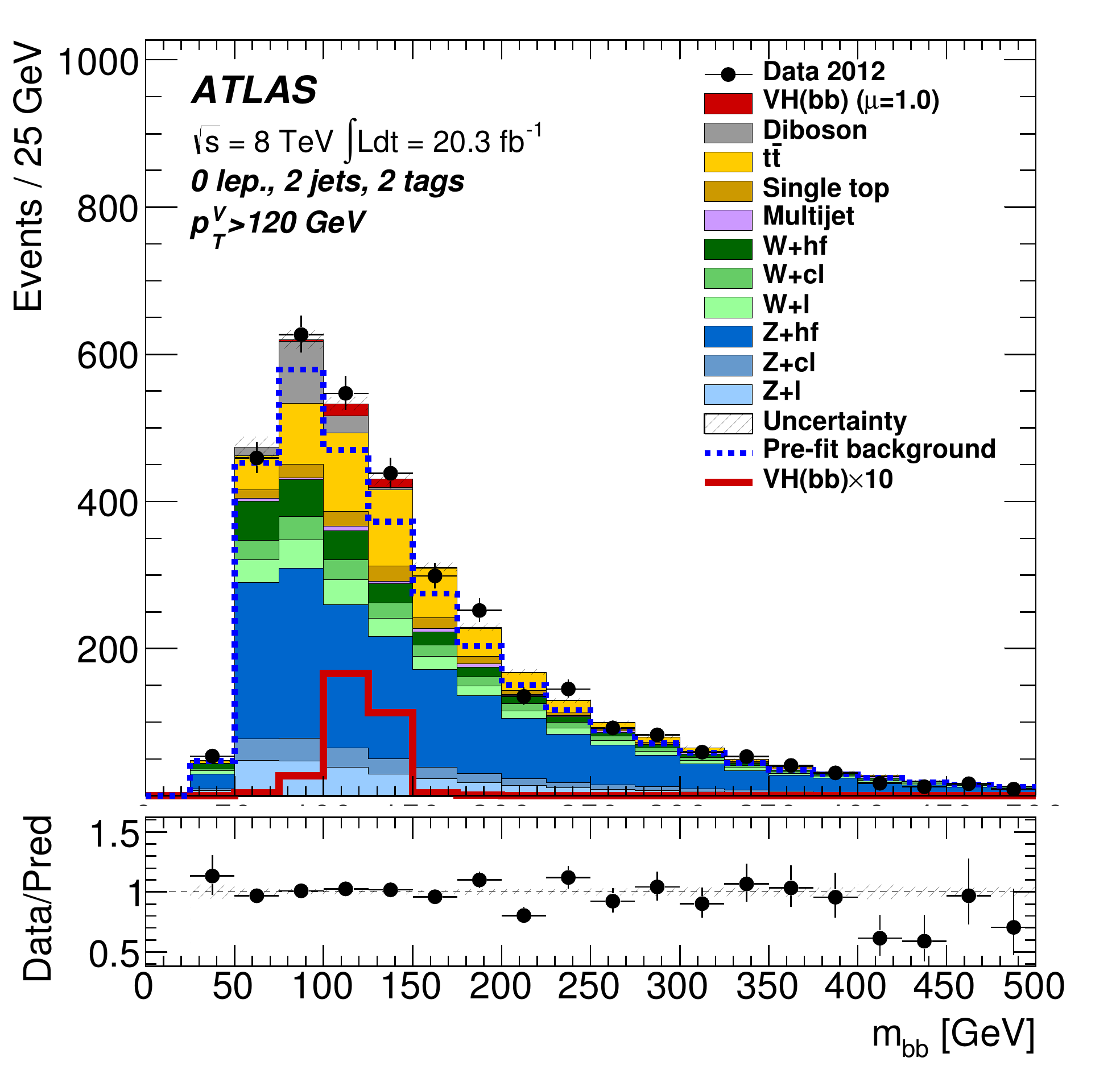}}
\hfill
\subfigure[]{\includegraphics[width=0.30\textwidth]{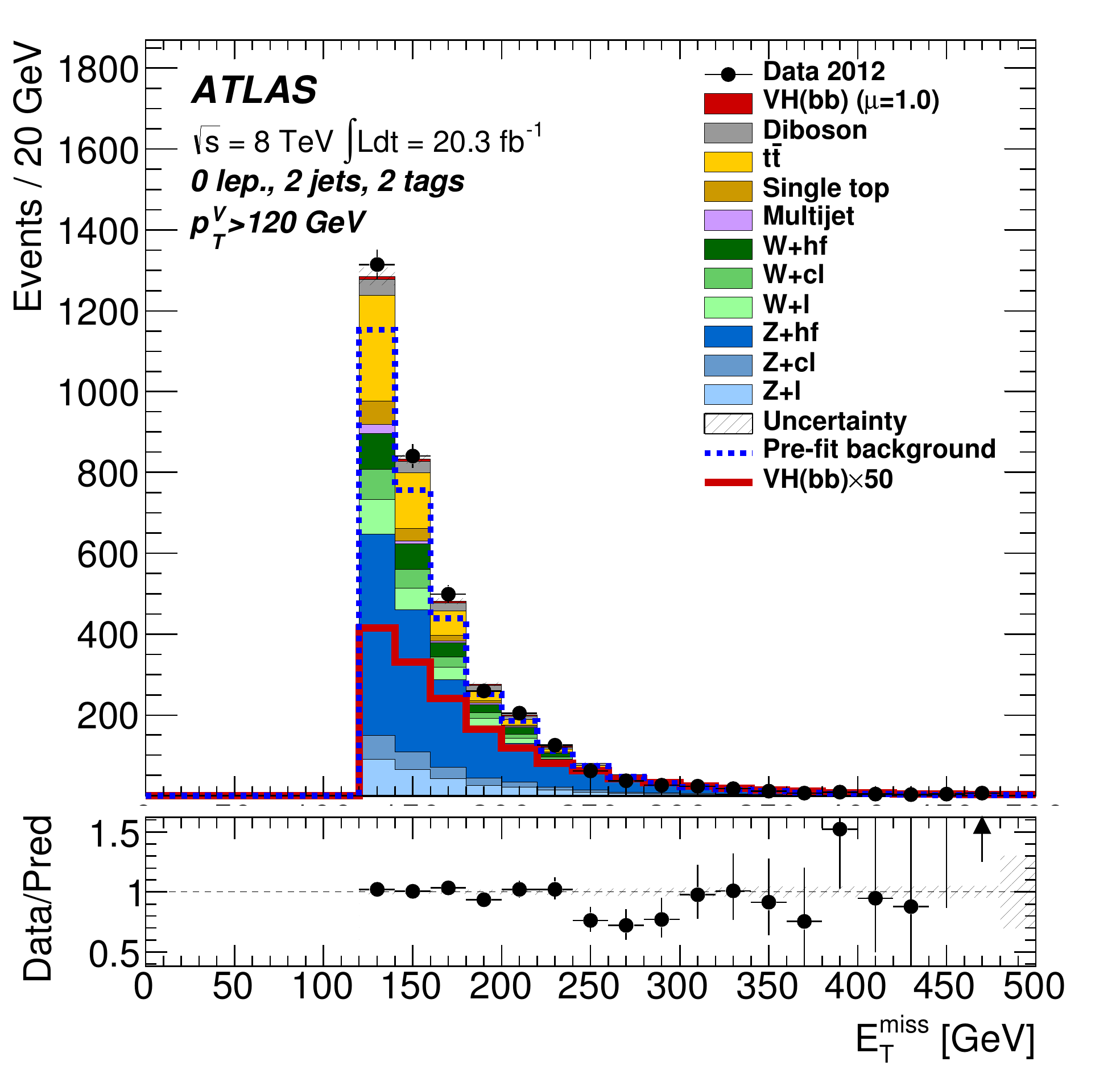}}
\hfill
\subfigure[]{\includegraphics[width=0.30\textwidth]{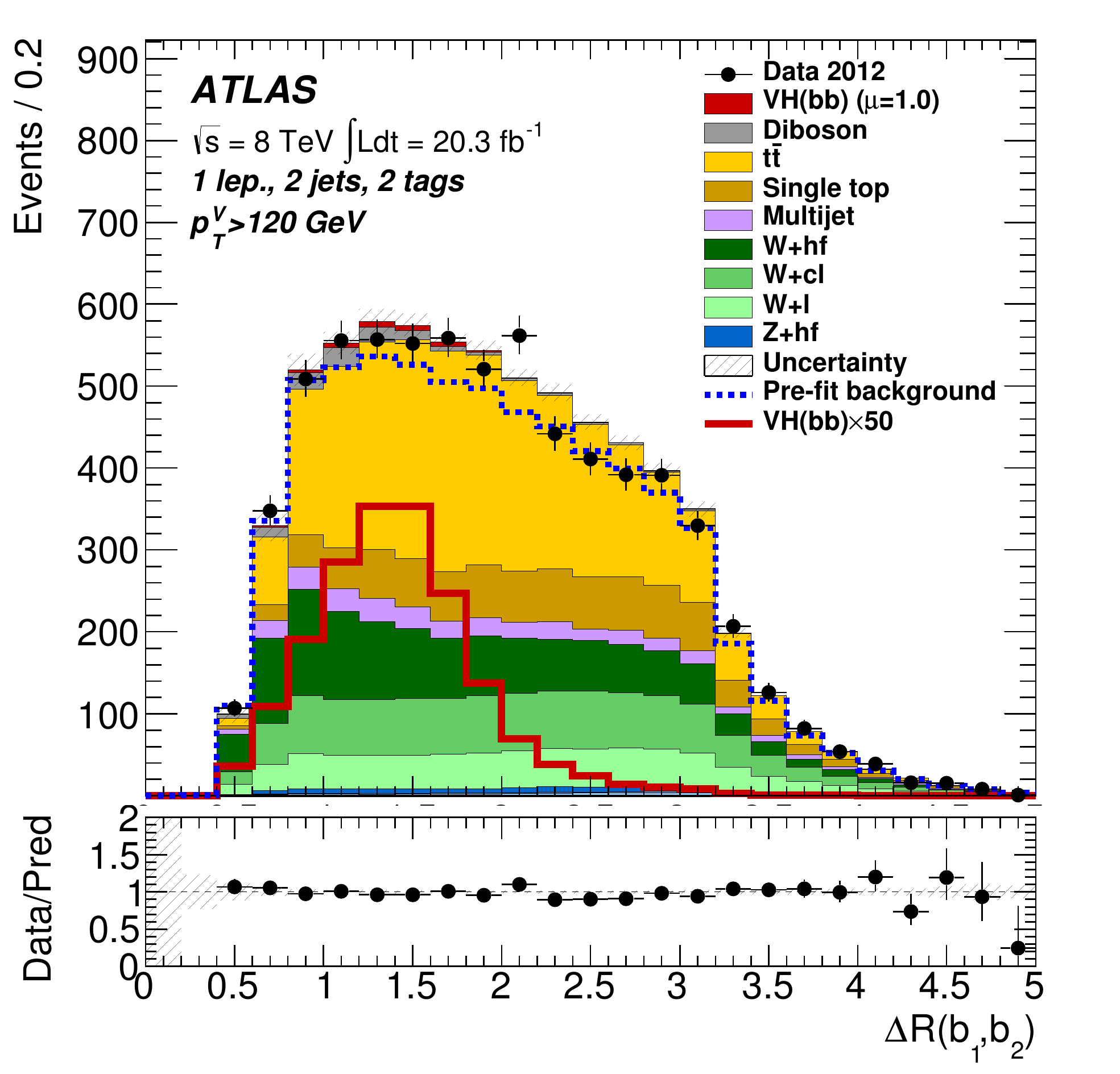}}
\subfigure[]{\includegraphics[width=0.30\textwidth]{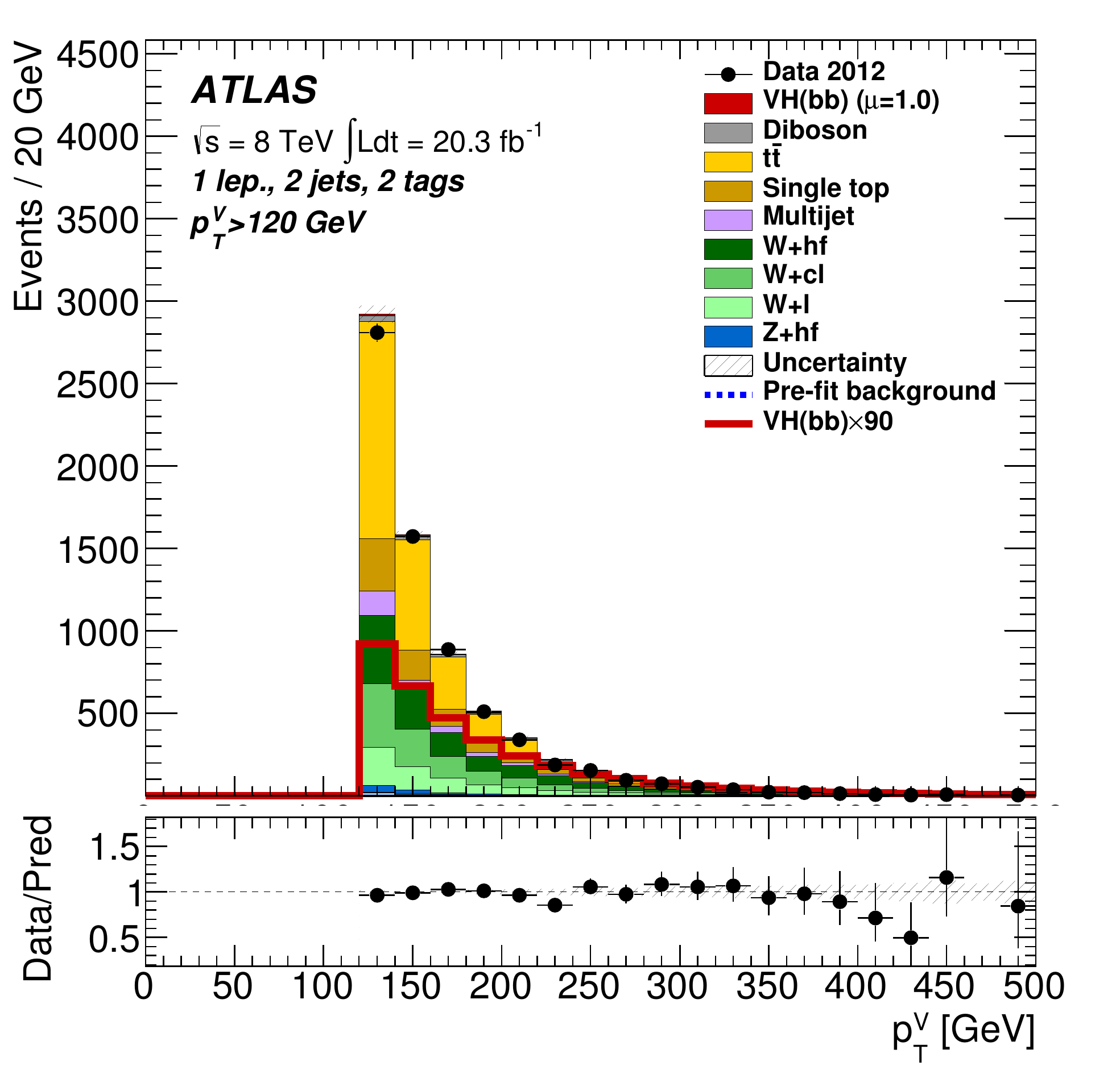}}
\hfill
\subfigure[\textbf{}]{\includegraphics[width=0.30\textwidth]{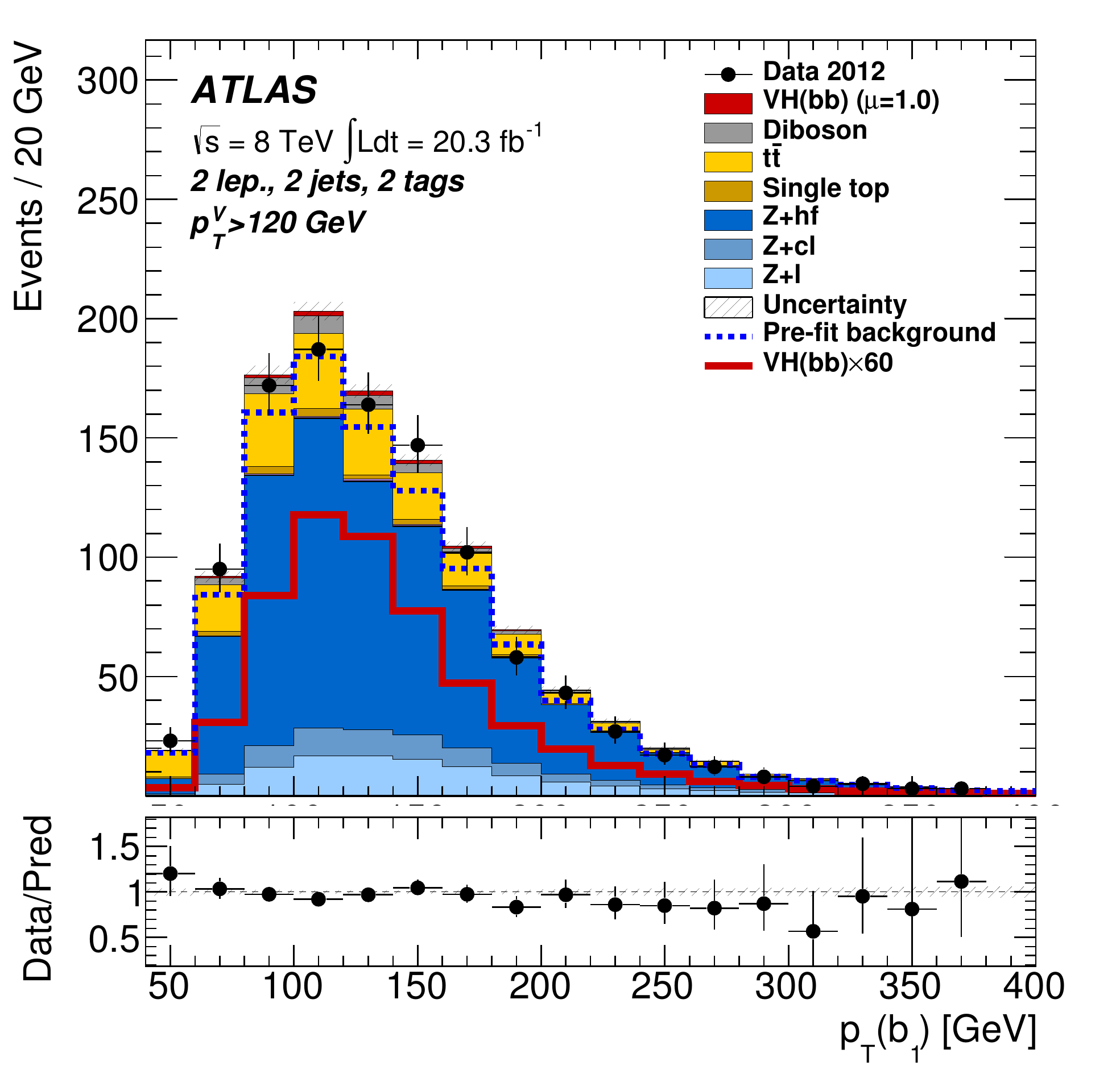}}
\hfill
\subfigure[\textbf{}]{\includegraphics[width=0.30\textwidth]{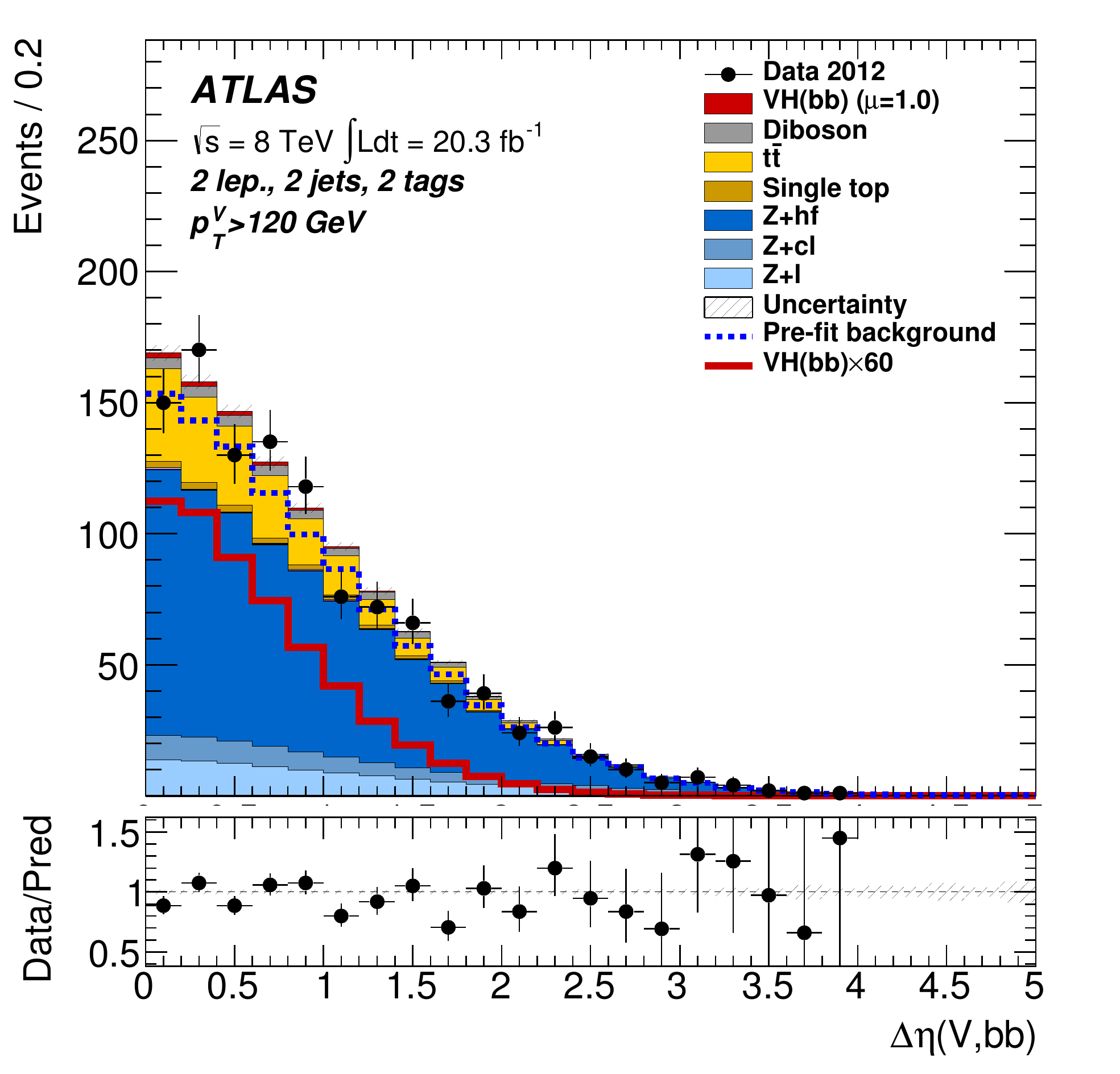}}
\caption{
Examples of variables input to the BDT in the 2-jet 2-tag category (LL, MM and TT combined) for $\ptv > 120 \GeV$:
(a) 0-lepton channel, dijet mass;    
(b) 0-lepton channel, \met;          
(c) 1-lepton channel, $\Delta R({b}_1,{b}_2)$;
(d) 1-lepton channel, $\ptW$;
(e) 2-lepton channel, $\pt^{{b}_1}$;
(f) 2-lepton channel, $|\Delta\eta(V,{bb})|$.
The distributions for the 2-lepton channel in (e) and (f) are shown after having applied the kinematic fit
as described in section~\ref{sec:selec}.       
The background contributions after the global fit of the MVA are shown 
as filled histograms.
The Higgs boson signal ($\mh = 125$~GeV) is shown as a filled histogram on top of 
the fitted backgrounds, 
as expected from the SM (indicated as $\mu=1.0$),
and, unstacked as an unfilled histogram, scaled by the factor indicated in the legend. 
The dashed histogram shows the total background as expected from the pre-fit 
MC simulation. The entries in overflow are included in the last bin.
The size of the combined statistical and systematic uncertainty on the
sum of the signal and fitted background is indicated by the hatched band. The ratio
of the data to the sum of the signal and fitted background is shown in the lower panel.
\label{fig:inputs_2jet_2tag}}
\end{center}
\end{sidewaysfigure}

The input variables of the BDTs are compared between data and simulation, 
and good agreement is found within the assessed uncertainties.
Selected input-variable distributions are shown in figure~\ref{fig:inputs_2jet_2tag}.\footnote{In 
this and all similar figures, all backgrounds are taken into account, 
but those contributing less than 1\% are omitted from the legend.}
In this figure, as for all figures in this section, the MJ background is estimated 
as described in section~\ref{sec:multijet}, corrections to the simulation as explained 
in section~\ref{sec:corrections} are applied, and background normalisations and shapes 
are adjusted by the global fit of the MVA as outlined at the beginning of section~\ref{sec:bkg} 
and presented in more detail in section~\ref{sec:fit}.
A similarly good agreement is found for the correlations between pairs of input variables,
as can be seen in figure~\ref{fig:correlations_inputs_2jet_2tag_0l}.

%
%
\begin{sidewaysfigure}[tb!]
\begin{center}
\subfigure[]{\includegraphics[width=0.55\textwidth]{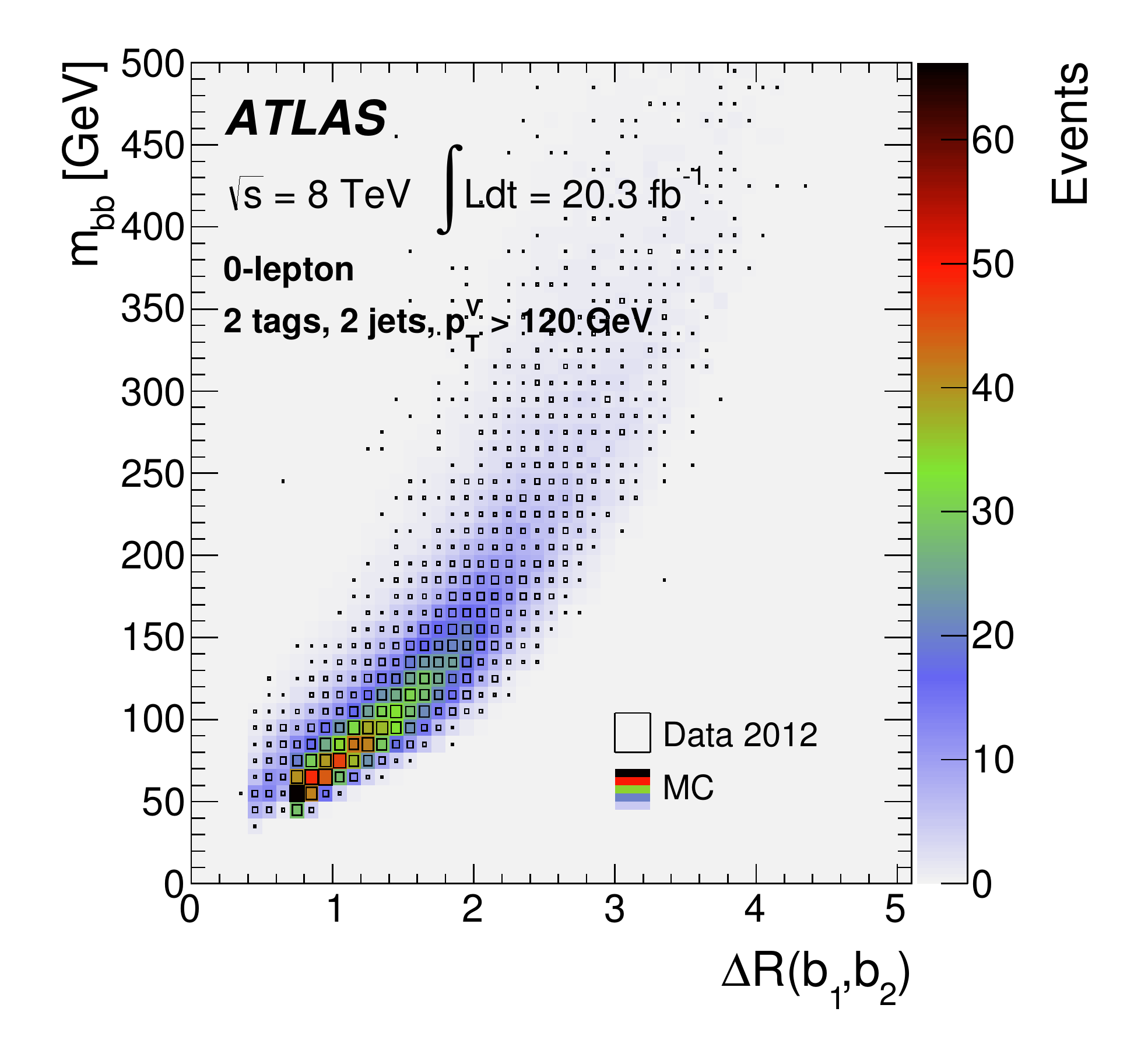}}
\hfill
\parbox[b]{0.44\textwidth}{
\subfigure[]{\includegraphics[width=0.43\textwidth]{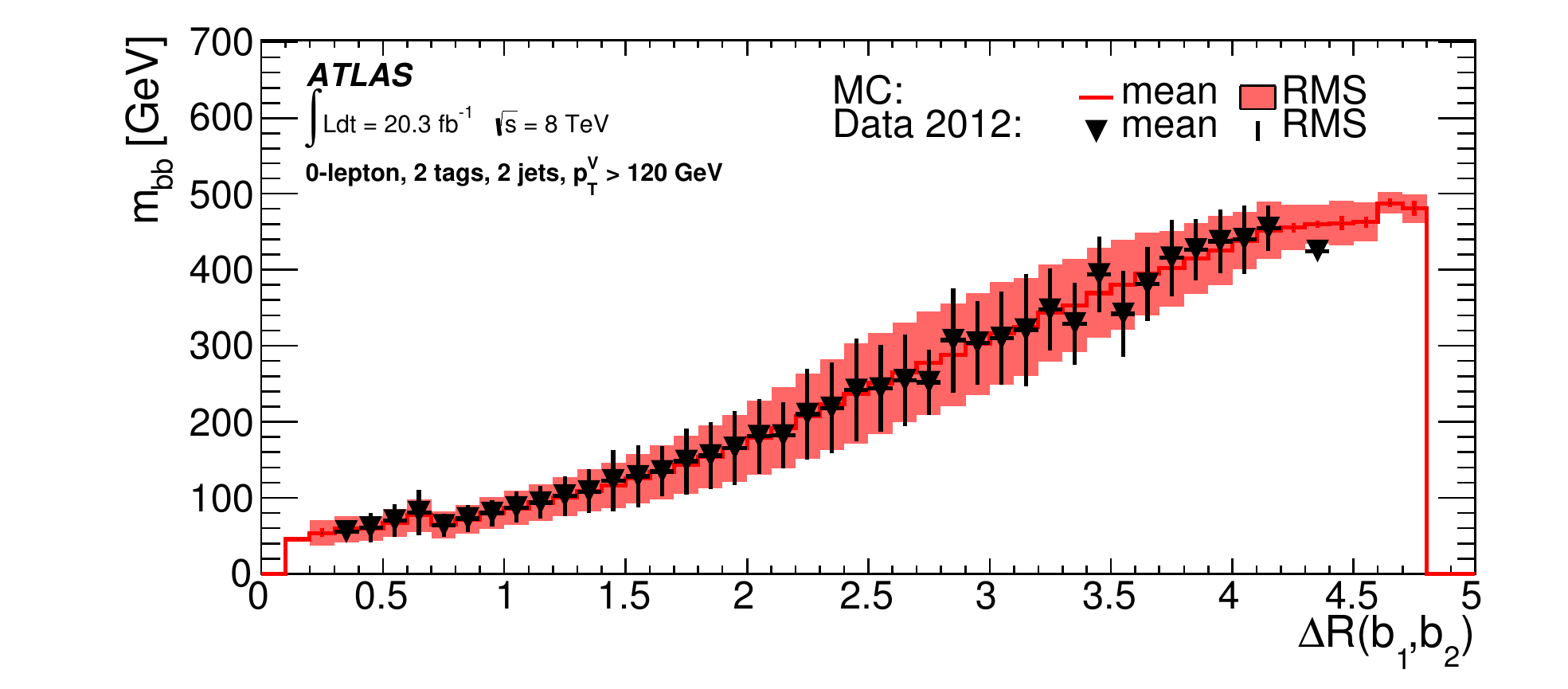}}
\subfigure[]{\includegraphics[width=0.43\textwidth]{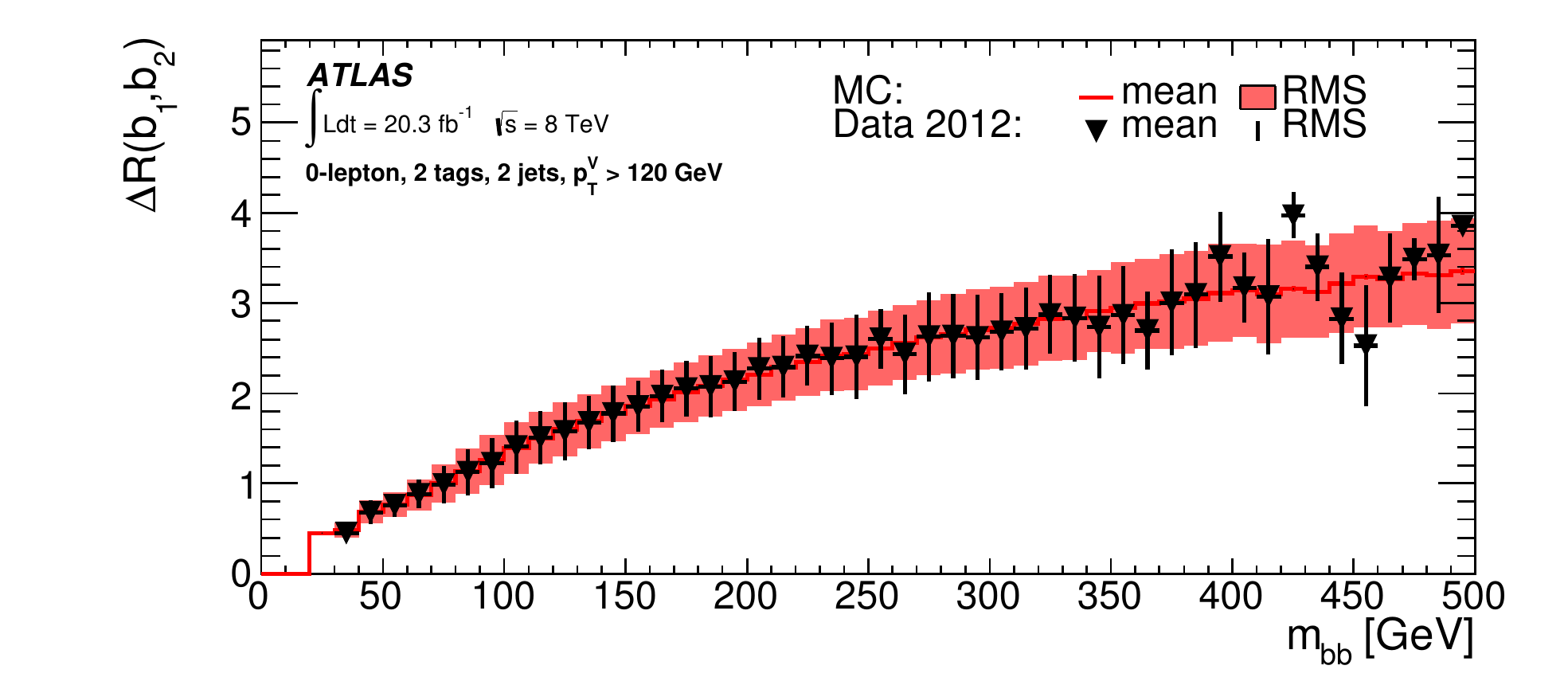}}
\vspace*{0.4cm}
\vfill
}
\caption{
Correlations between the dijet mass and $\Delta R({b}_1,{b}_2)$ input variables to the BDT
in the 2-jet 2-tag category (LL, MM and TT combined) of the 0-lepton channel for $\ptv > 120 \GeV$:
(a) dijet mass versus $\Delta R({b}_1,{b}_2)$ for the total expected background 
    (shadings indicating the numbers of events)
    and the data (open boxes with the box size being proportional to the number of events),
(b) and (c) show the mean values and RMS of the projections 
onto the $\Delta R({b}_1,{b}_2)$ and dijet-mass axes, respectively, 
for the total expected background 
after the global fit of the MVA
and the data.
\label{fig:correlations_inputs_2jet_2tag_0l}}
\end{center}
\end{sidewaysfigure}

The Toolkit for Multivariate Data Analysis, TMVA~\cite{tmva}, is used to train the BDTs.
The values for the training parameters are found by determining the 
configuration with the best separation between signal and background in a coarsely binned multi-dimensional training parameter 
space, followed by more finely grained one-dimensional scans of individual training
parameters.    
%
In order to make use of the complete set of simulated MC events for the 
BDT training and 
evaluation in an unbiased way, 
the MC events are split into two samples of equal size, $A$ and $B$. 
The performance of the BDTs trained on sample $A$ ($B$) is evaluated with sample $B$ ($A$) in order
to avoid using identical events for both training and evaluation of the same BDT. 
Half of the data are analysed with the BDTs trained on sample $A$, and the other half with the BDTs 
trained on sample $B$.
At the end, the output distributions of the BDTs trained on samples $A$ and $B$ are
merged for both the simulated and data events.

The values of the BDT outputs do not have a well-defined interpretation.
A dedicated procedure is applied to transform the BDT-output distributions
to obtain a smoother distribution for the background processes and a finer binning
in the regions with the largest signal contribution, while at the same time preserving a sufficiently 
large number of background events in each bin.
Starting from a very fine-binned histogram of the BDT-output distribution,
the procedure merges histogram bins, from high to low BDT-output values, until
a certain requirement, based on the fractions of signal and background events in
the merged bin, is satisfied.   
To limit the number of bins and to reduce the impact of statistical fluctuations,
a further condition is that the statistical uncertainty of the expected total background
contribution has to be smaller than 10\% in each merged bin.
The free parameters of the transformation algorithm
are optimised to maximise the expected signal sensitivity.
For simplicity, these transformed outputs, which are used for the analysis, 
are called ``BDT$_{VH}$ discriminants'' in the following.
An optimisation of the number of bins and bin boundaries is also
performed for the $m_{bb}$ distribution used in the dijet-mass analysis in
a similar way, where 
the free parameters of the transformation algorithm are optimised 
separately for the different analysis regions.
The effect of the transformation on the BDT-output and dijet-mass distributions can be seen
in figure~\ref{fig:transformation} for the 1-lepton channel and one signal region.
The transformation groups into few bins the \mbb\ regions that are far from the signal on each
of the low and high mass sides, while it expands the region close to the signal mass, where the
signal-to-background ratio is largest. The effect on the BDT output is similar, but simpler to 
visualise because the signal and the background accumulate initially on the high and the low sides 
of the distribution, respectively.
\begin{figure}[tb!]
\begin{center}
\centerline{
\hfill
\subfigure[]{\includegraphics[width=0.46\textwidth]{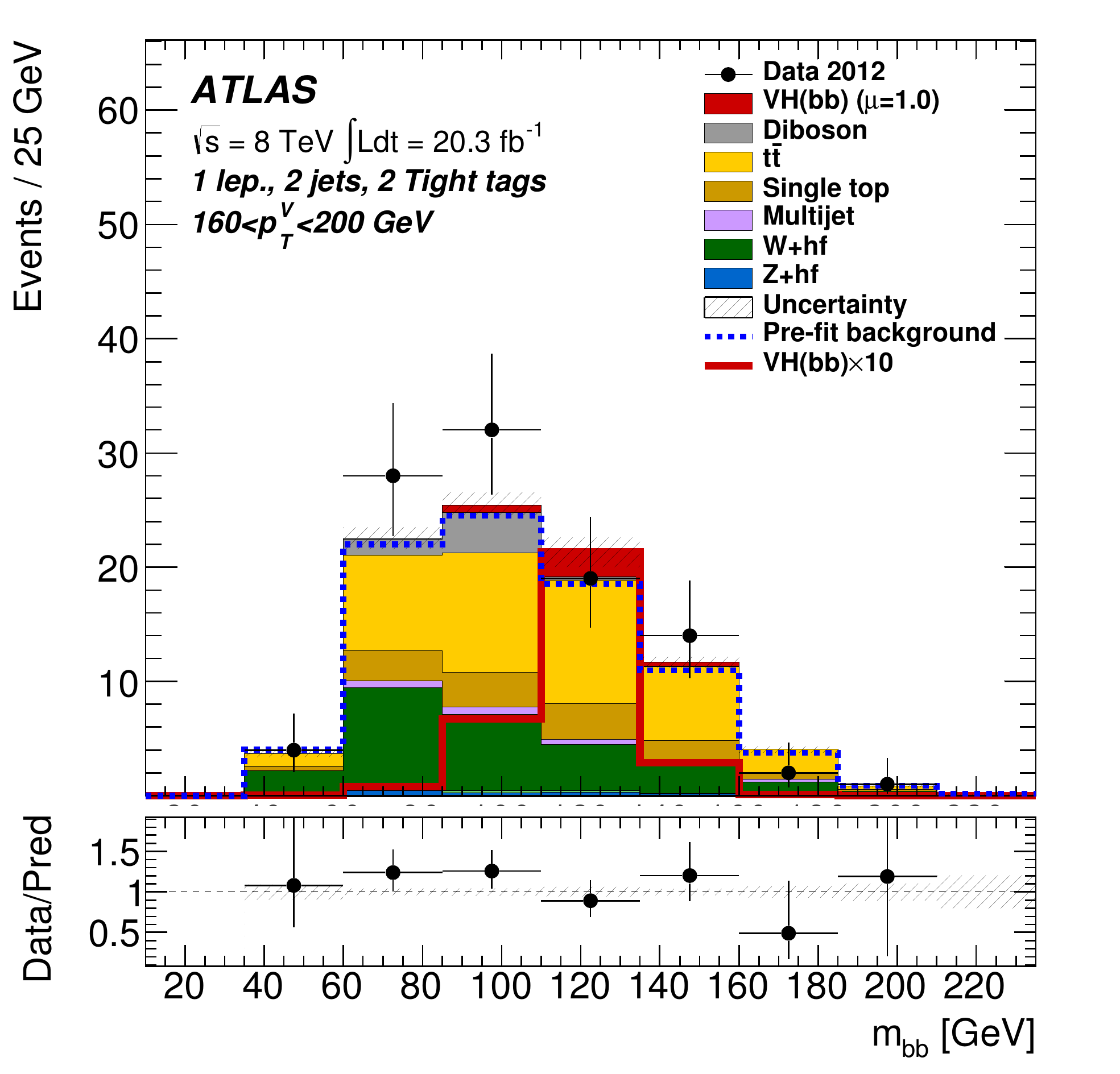}}
\hfill
\subfigure[]{\includegraphics[width=0.46\textwidth]{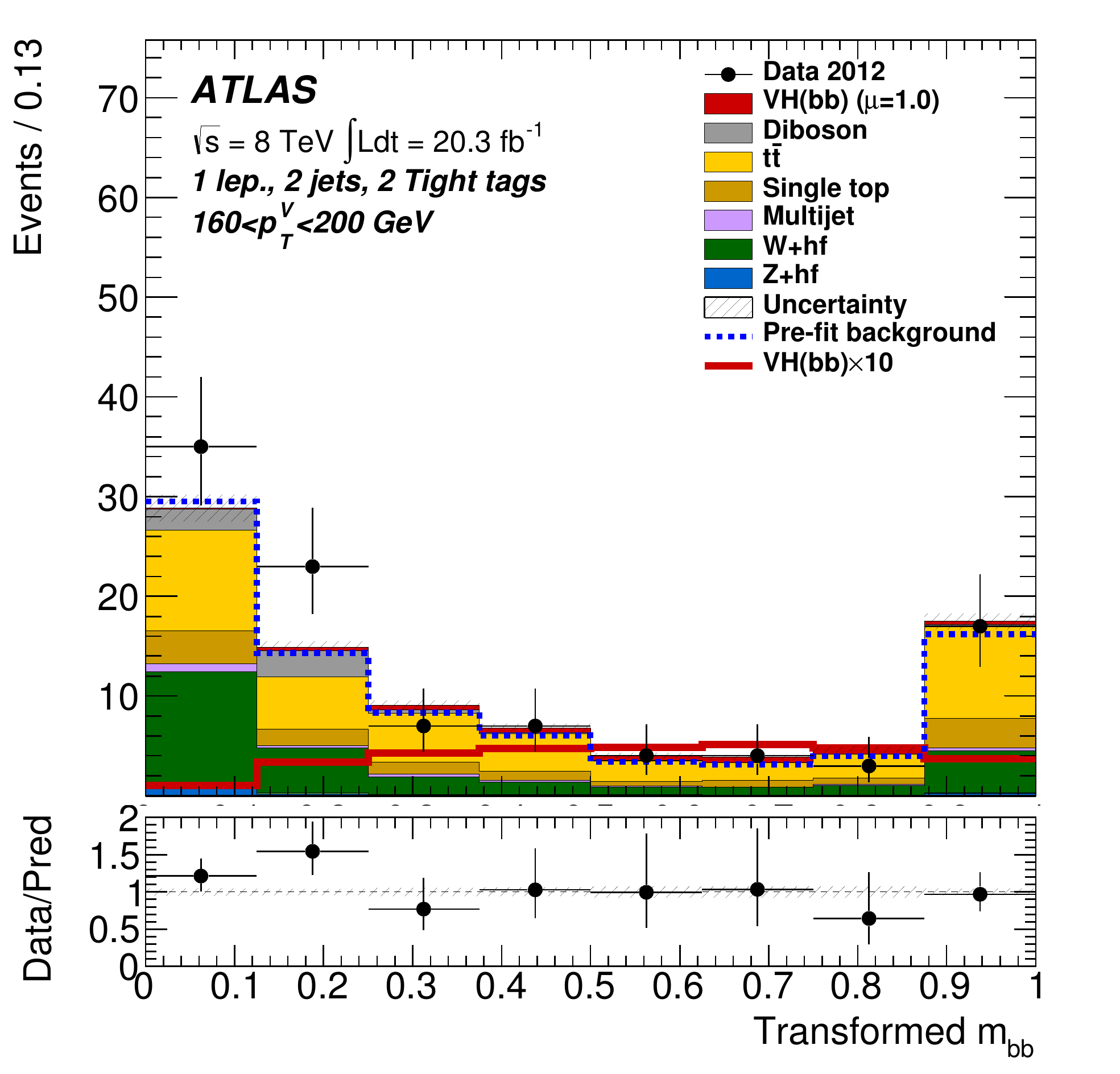}}
\hfill
}
\centerline{
\hfill
\subfigure[]{\includegraphics[width=0.46\textwidth]{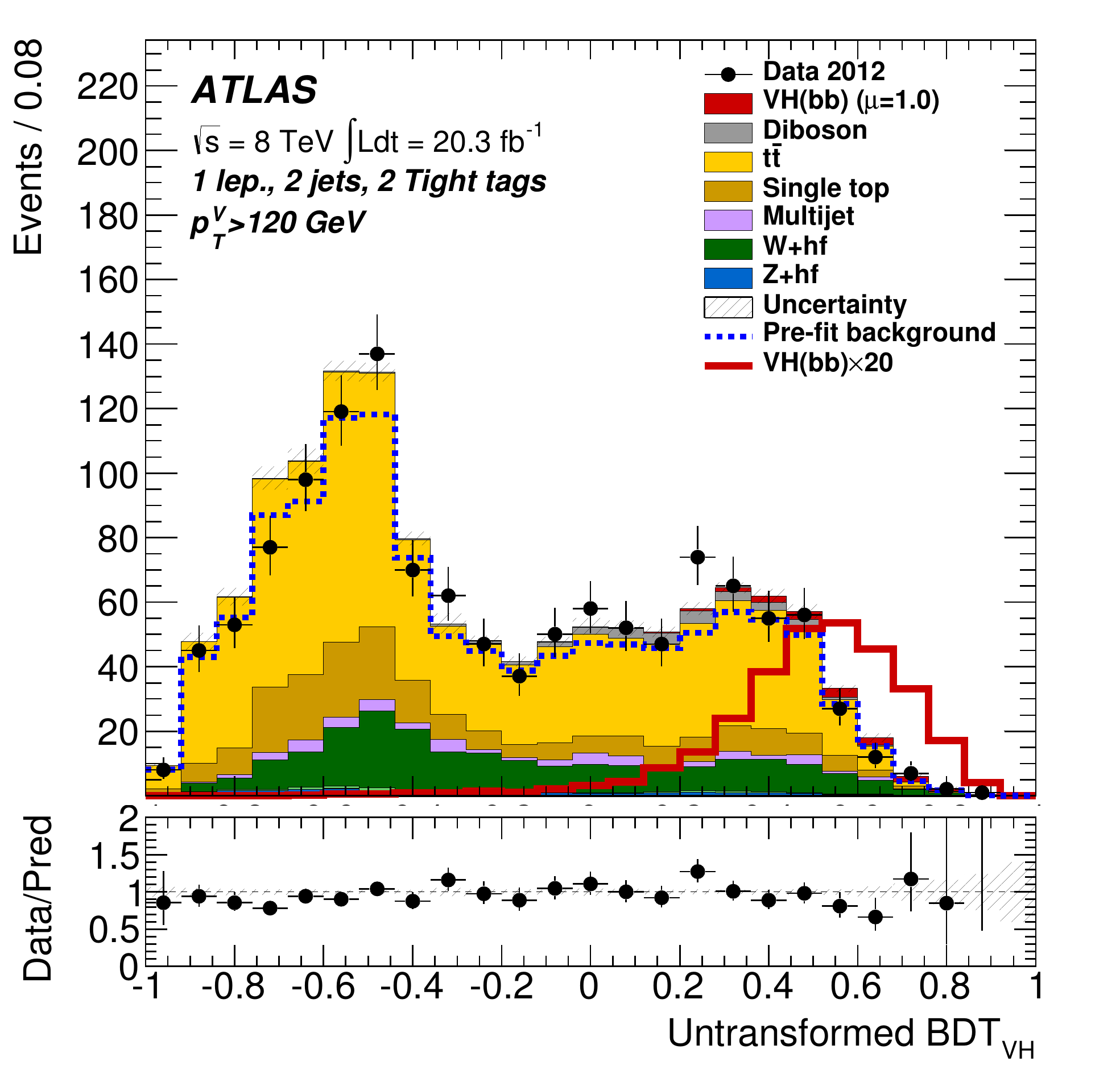}}
\hfill
\subfigure[]{\includegraphics[width=0.46\textwidth]{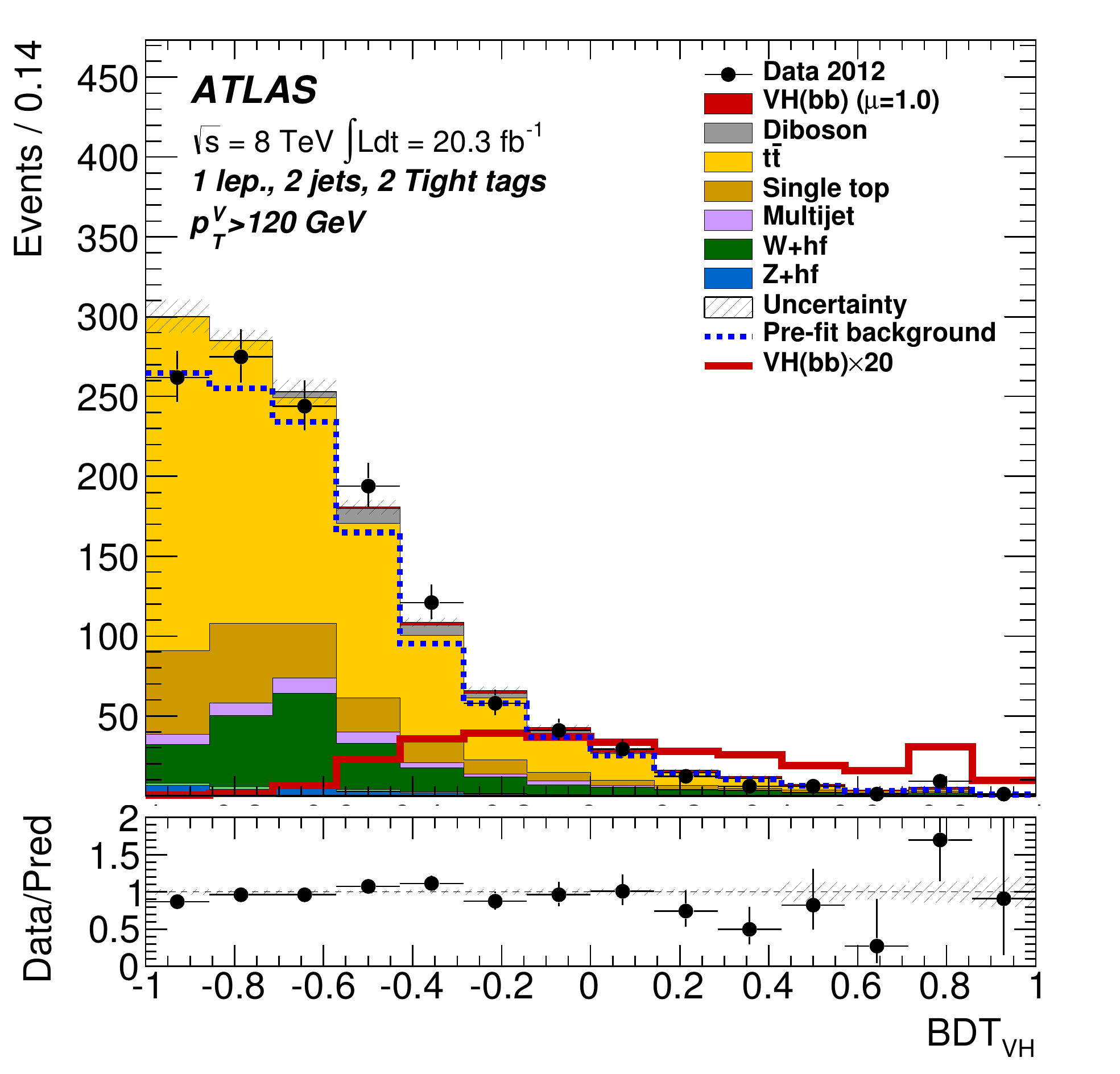}}
\hfill
}
\caption{
Top: The dijet-mass distributions for the expected background and signal contributions
     in the 1-lepton channel and the 2-jet 2-tag TT category for $160 \GeV < \ptw \leq 200 \GeV$
     (a) before and (b) after applying the transformation of the histogram bins.
Bottom: The BDT-output distribution for the expected background and signal contributions
     in the 1-lepton channel and the 2-jet 2-tag TT category for $\ptw > 120 \GeV$
     (c) before and (d) after applying the transformation of the histogram bins.
The background contributions after the relevant global fit 
(of the dijet-mass analysis in (a) and (b) and of the MVA in (c) and (d))
are shown as filled histograms.
The Higgs boson signal ($\mh = 125$~GeV) is shown as a filled histogram on top of 
the fitted backgrounds, 
as expected from the SM (indicated as $\mu=1.0$),
and, unstacked as an unfilled histogram, scaled by the factor indicated in the legend. 
The dashed histogram shows the total background as expected from the pre-fit 
MC simulation. The entries in overflow are included in the last bin.
The size of the combined statistical and systematic uncertainty on the
sum of the signal and fitted background is indicated by the hatched band. The ratio
of the data to the sum of the signal and fitted background is shown in the lower panel. 
\label{fig:transformation}}
\end{center}
\end{figure}

Correlations between input variables and the BDT$_{VH}$ discriminant can provide information on the impact 
of individual variables on the classification.
Figure~\ref{fig:correlations_mbb_output_2tag_0lep} 
shows such correlations for the dijet mass, which is the BDT input that
provides the best single-variable discriminating power. 

\begin{sidewaysfigure}[tb!]
\begin{center}
\subfigure[]{\includegraphics[width=0.55\textwidth]{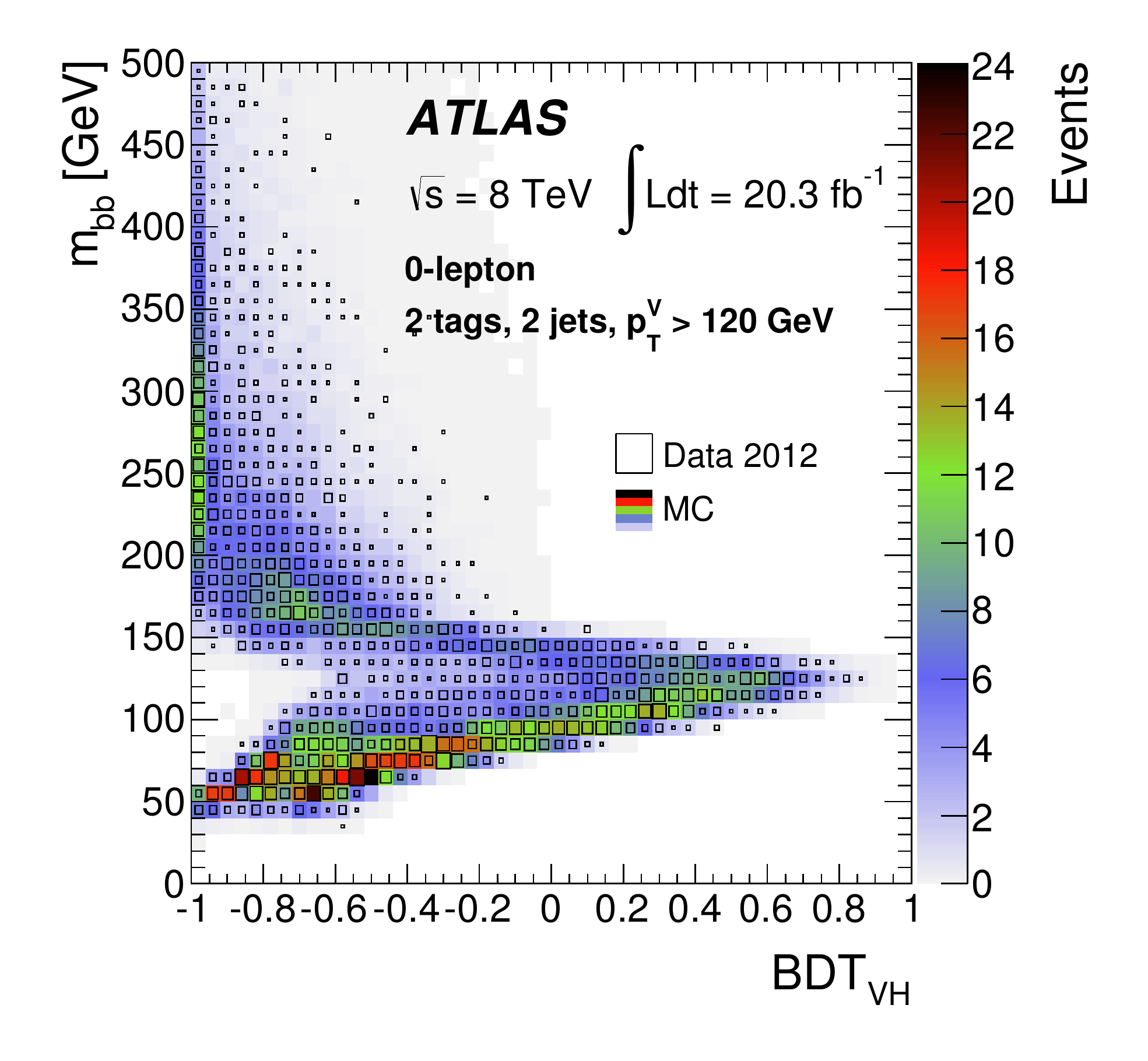}}
\hfill
\parbox[b]{0.44\textwidth}{
\subfigure[]{\includegraphics[width=0.43\textwidth]{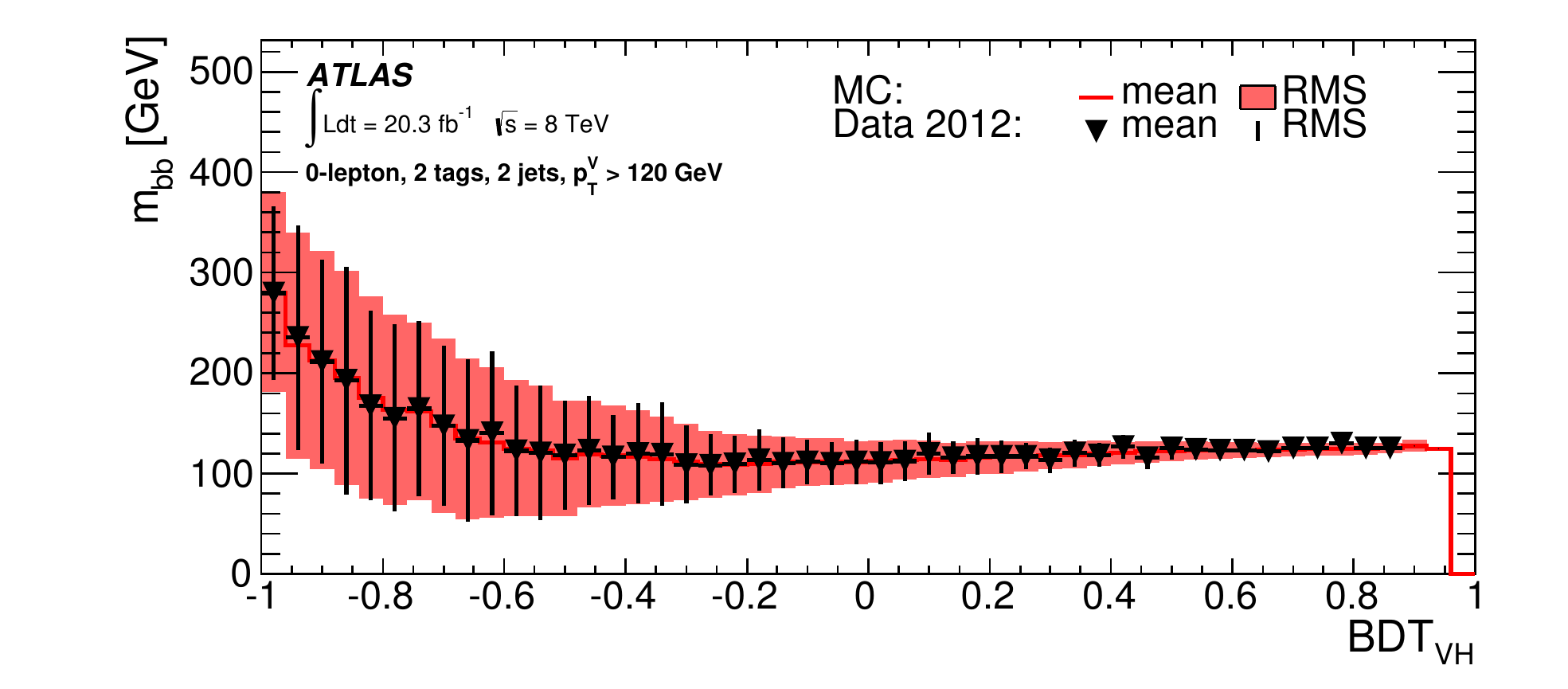}}
\subfigure[]{\includegraphics[width=0.43\textwidth]{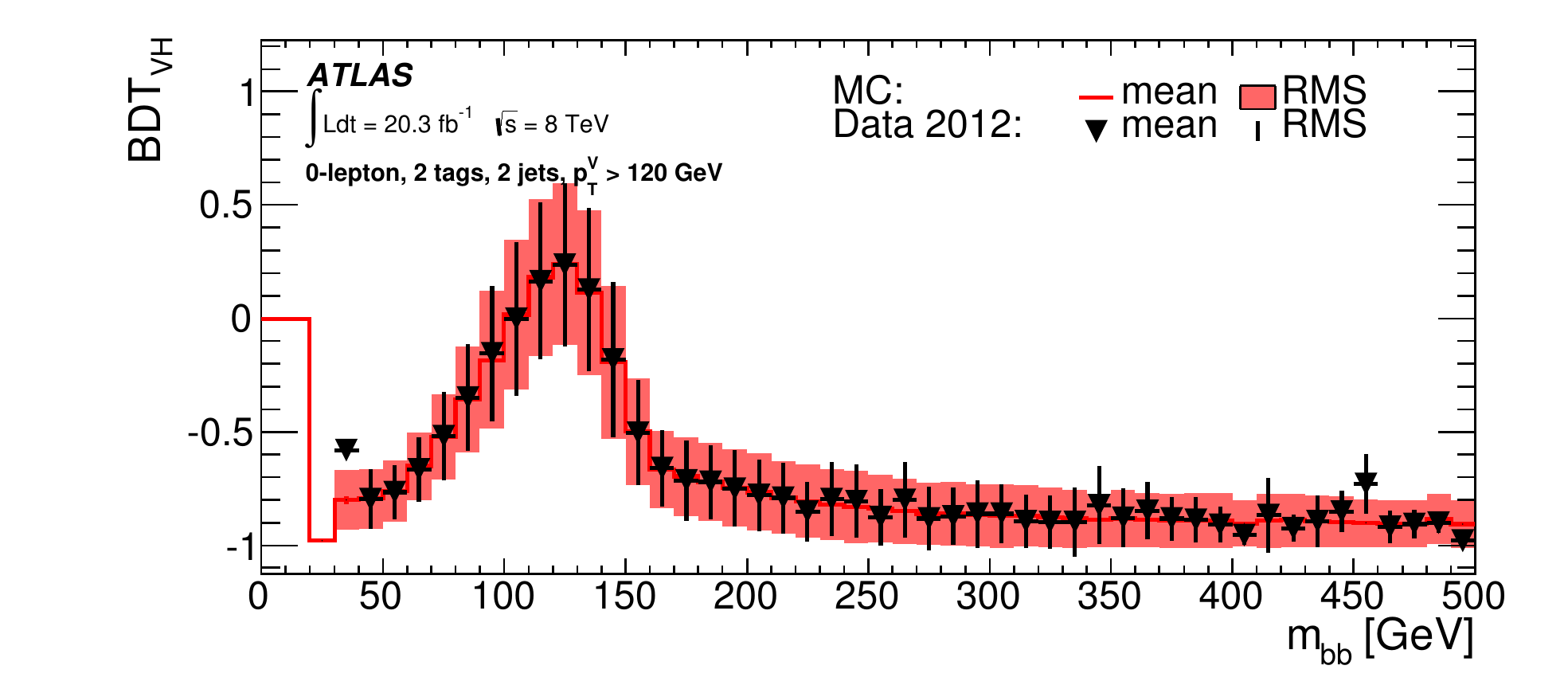}}
\vspace*{0.4cm}
\vfill
}
\caption{
Correlations between 
the dijet mass and the BDT$_{VH}$ discriminant 
in the 2-jet 2-tag category (LL, MM and TT combined) of the 
0-lepton channel for $\ptv > 120$~GeV:
(a) the dijet mass versus the BDT$_{VH}$ discriminant for the total expected background
    (shadings indicating the numbers of events)
    and the data (open boxes with the box size being proportional to the number of events),
(b) and (c) show the mean values and RMS of the projections onto the axes
of the BDT$_{VH}$ discriminant and dijet mass, respectively,
for the total expected background 
after the global fit of the MVA
and the data. 
\label{fig:correlations_mbb_output_2tag_0lep}}
\end{center}
\end{sidewaysfigure}

\FloatBarrier

\section{Background composition and modelling}\label{sec:bkg}

This section describes the modelling of individual backgrounds. In many cases, the data
are able to constrain the normalisations and shapes better than the a priori estimates.
A likelihood fit (also called ``global fit'') 
is used to simultaneously extract both the signal yield and constraints on the 
background normalisations and shapes.
The distributions used by the fit are those of the dijet mass or BDT$_{VH}$ discriminant in the 2-tag
signal regions, as appropriate, as well as those of the $MV1c$ value of the $b$-tagged jet in 
the 1-tag control regions. More details are provided in section~\ref{sec:fit}.

For the multijet (MJ) backgrounds, the normalisations and shapes
provided as inputs to the fit are estimated from data, as explained below.
For the other backgrounds the inputs are taken from the simulation, except for 
the normalisations of the $V$+jets and \ttb\ backgrounds that are left free
to float in the fit. The corrections to these two backgrounds, described below,
are applied prior to the fit.

In all distributions presented in this section, unless otherwise specified,
the normalisations of the various backgrounds are those extracted from the 
global fit for the dijet-mass or multivariate 
analysis, as appropriate. The fit also adjusts the background shapes in those 
distributions within the constraints from the systematic uncertainties discussed in 
section~\ref{sec:sys}.

\subsection{Multijet background}\label{sec:multijet}

Multijet events are produced with a huge cross section via the strong interaction, and therefore give
rise to potentially large backgrounds.
A first class of MJ background arises from jets or photon conversions misidentified as electrons, 
or from semileptonic heavy-flavour decays;
the 1- and 2-lepton channels
are especially sensitive to this class of background. 
Another class, which affects mostly the 0-lepton 
channel, arises from large fluctuations in jet energy measurements in the calorimeters,
which create ``fake'' \met. 
These MJ backgrounds 
cannot be determined reliably by simulation, and are estimated from data in each of the 0-, 1-, 
and 2-lepton channels, and in each of the 2- and 3-jet, 0-, 1-, and 2-tag regions.
 
The MJ background is estimated in the 0-lepton channel using an ``ABCD method'',
within which the data are divided into four regions based on the
${\rm min}[\Delta \phi(\metvec,{\rm jet})]$ and 
$\Delta\phi(\metvec, \mptvec)$ variables, such that three of the regions are dominated 
by background.
(In the 100--120~GeV \ptv\ interval, 
the likelihood ratio $\cal L$ designed to suppress the MJ background is used instead 
of ${\rm min}[\Delta \phi(\metvec,{\rm jet})]$.)
For events with real \met, it is expected that the directions of 
the calorimeter-based and track-based missing transverse momenta,
\metvec\ and \mptvec, are similar.
In events with fake \met\ arising from a jet energy fluctuation, it is expected that the direction of 
\metvec\ is close to the direction of the poorly measured jet.
The signal region (A) is therefore selected with 
${\rm min}[\Delta \phi(\metvec,{\rm jet})] > 1.5$ and
$\Delta\phi(\metvec, \mptvec) < \pi/2$.
In region C, the requirement on $\Delta\phi(\metvec, \mptvec)$ is reversed.
In regions B and D, ${\rm min}[\Delta \phi(\metvec,{\rm jet})] < 0.4$ is required,
with requirements on $\Delta\phi(\metvec, \mptvec)$ as in regions A and C, respectively.
A comparison of the ${\rm min}[\Delta \phi(\metvec,{\rm jet})]$ distributions
for $\Delta\phi(\metvec, \mptvec)$ above and below $\pi/2$ shows that these two variables 
are only weakly correlated, and this observation is confirmed in a multijet event sample
simulated with {\sc pythia8}. 
An MJ template in region A is obtained using events in region C after subtracting
the contribution of other backgrounds, taken from simulation. The template is normalised
by the ratio of the number of events in region B to that in region D, again after subtracting other
backgrounds from those regions. The populations of events in the various regions suffer from low statistical precision after
the 2-tag requirement. The $b$-tagging requirement is therefore dropped in regions B, C
and D, and an additional $b$-tagging normalisation factor is applied to the resulting
template, taken as the fraction of 2-tag events in region D. 
The MJ background in the signal regions is found to amount to $\sim 1$\% 
of the total background.

In the 1-lepton channel, the MJ background is determined separately for
the electron and muon sub-channels. For each signal or control region, an MJ-background 
template is obtained in an MJ-dominated region 
after subtracting the small remaining contribution from the other backgrounds. 
The other backgrounds are taken from a simulation improved by scale factors 
for the various contributions obtained from a preliminary global fit.
The MJ-dominated region is obtained by modifying the nominal selection to use medium, instead of tight, 
leptons and loosening both the track and calorimeter-based isolation criteria. 
The track-based isolation is changed to the intervals 5\%--12\% and 7\%--50\% for electrons and muons respectively, 
instead of $< 4$\%; and the calorimeter-based isolation is loosened to $<7$\% from $<4$\%.
The sample sizes of the MJ-templates are however rather low in the 2-tag regions.
Since it is observed that the kinematic properties of the 1-tag and 2-tag events 
in the MJ-dominated regions are similar, 1-tag events are used to enrich the 2-tag MJ 
templates. Events in the 1-tag category are promoted to the 2-tag category by assigning 
to the untagged jet an emulated $MV1c$ value drawn from the appropriate $MV1c$ distribution 
observed in the corresponding 2-tag MJ template. This distribution depends on the rank (leading or 
sub-leading) of the untagged jet and on the $MV1c$ value of the tagged jet. 
To cope with residual differences observed in some distributions between these pseudo-2-tag 
MJ events and the actual 2-tag MJ events, a reweighting is applied according to the $MV1c$ 
of the tagged jet and, for the electron sub-channel, according to  
$\Delta R(\mathrm{jet}_1,\mathrm{jet}_2)$ and $\ptw$. This procedure is applied in each
of the 2- and 3-jet, LL, MM and TT categories.
The normalisations of the MJ templates are 
then obtained from ``multijet fits'' to the \met\ distributions in the 2- and 3-jet, 
1- and 2-tag (LL, MM and TT combined) categories, with floating normalisations for the 
templates of the other background processes. The templates for these other background processes
are taken from the improved simulation mentioned above.

The MJ background in the 1-lepton channel is concentrated at low \ptw, 
and in the 2-jet 2-tag sample with $\ptw < 120$~GeV it ranges from 11\% 
of the total background in the LL category to 6\% in the TT category.
The main purpose of including the $\ptw < 120$~GeV intervals is to provide 
constraints on the largest backgrounds ($V$+jets and $\ttb$) in the
global fit. Since the MJ background is twice as large for $\ptw < 120$~GeV 
in the 1-electron sub-channel than in the 1-muon sub-channel, 
only the 1-muon sub-channel is kept for $\ptw < 120$~GeV so as
to provide the most reliable constraints on the non-MJ backgrounds.
The resulting loss in sensitivity is 0.6\%.
For $\ptw > 120$~GeV, the MJ background is much smaller:  
4\% and 2\% in the LL and TT categories, respectively, for 2-jet events. 

A template for the MJ background in the 2-electron sub-channel is obtained in a
similar way, by loosening identification and isolation requirements. 
The normalisation is performed by a fit to the dilepton-mass 
distribution, where the $Z$+jets and MJ components are free parameters, while
the other backgrounds (mostly \ttb) are taken from the simulation. The MJ
normalisation factors are found to be consistent in the 0-, 1- and 2-tag 
regions. To cope with the reduced size of the 2-tag MJ event sample, a procedure similar to
that used in the 1-lepton channel is used, wherein the pretag MJ sample is
weighted by its 2-tag fraction and combinations of $MV1c$ values are randomly 
assigned to the jets according to their distribution in the 2-tag MJ template. 
In the 2-muon sub-channel, the MJ background is found to be negligible from a 
comparison between data and MC prediction in the sidebands of the $Z$ mass peak.
Altogether, the MJ background amounts to $<$1\% 
of the total background in the 2-lepton channel. 

\subsection{Corrections to the simulation}\label{sec:corrections}

The large number of events in the 0-tag samples allows for detailed investigations 
of the modelling of the $V$+jet backgrounds by the version of the {\sc sherpa} 
generator used in this analysis. 
Given that the search is performed in intervals of \ptv, with the higher 
intervals providing most of the sensitivity, an accurate modelling of the 
\ptv\ distribution is important. 

Figure~\ref{fig:ptw_0tag_1lep}(a) shows
that the \ptw\ spectrum for $W$+jets production in the 1-muon sub-channel  
is softer in the data than in the simulation. 
It is found that this mismodelling is strongly correlated with a mismodelling 
of the \dphi\ distribution,\footnote{
It has indeed been observed that the shape of the \dphi\ distribution in data 
is better reproduced by NLO generators than by the baseline {\sc sherpa} 
generator used in this analysis~\cite{Hoeche:2012yf}.}
shown in figure~\ref{fig:dphi_0tag_1lep}(a).\footnote{The peak around \dphi=0.7
comes from the combination of two effects: a rise towards low \dphi\ due to gluon splitting, 
and a drop towards low \dphi\ due to the two jets becoming unresolved.} 
\begin{figure}[t!]
\begin{center}
\hfill
\subfigure[]{\includegraphics[width=0.46\textwidth]{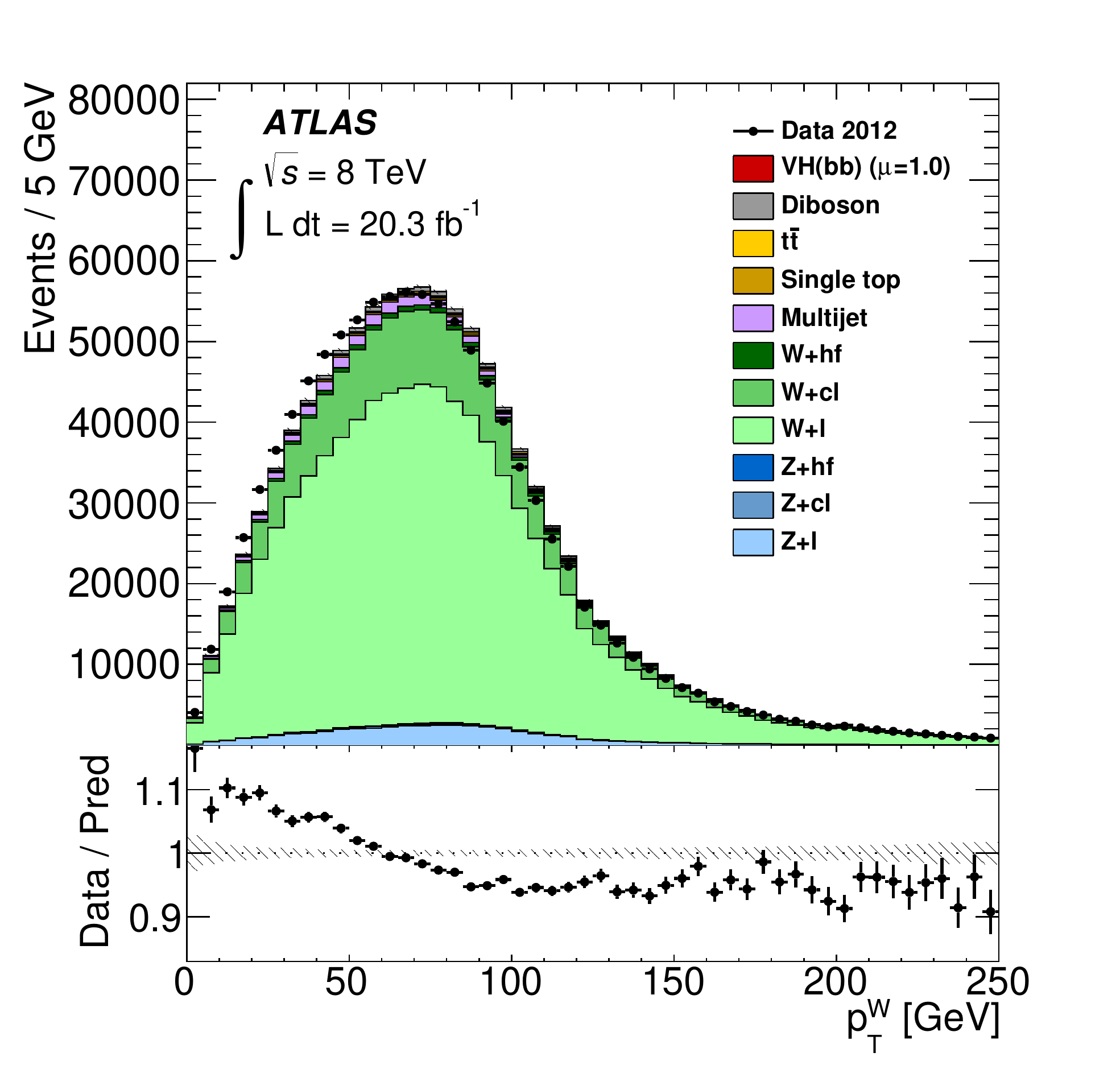}}
\hfill
\subfigure[]{\includegraphics[width=0.46\textwidth]{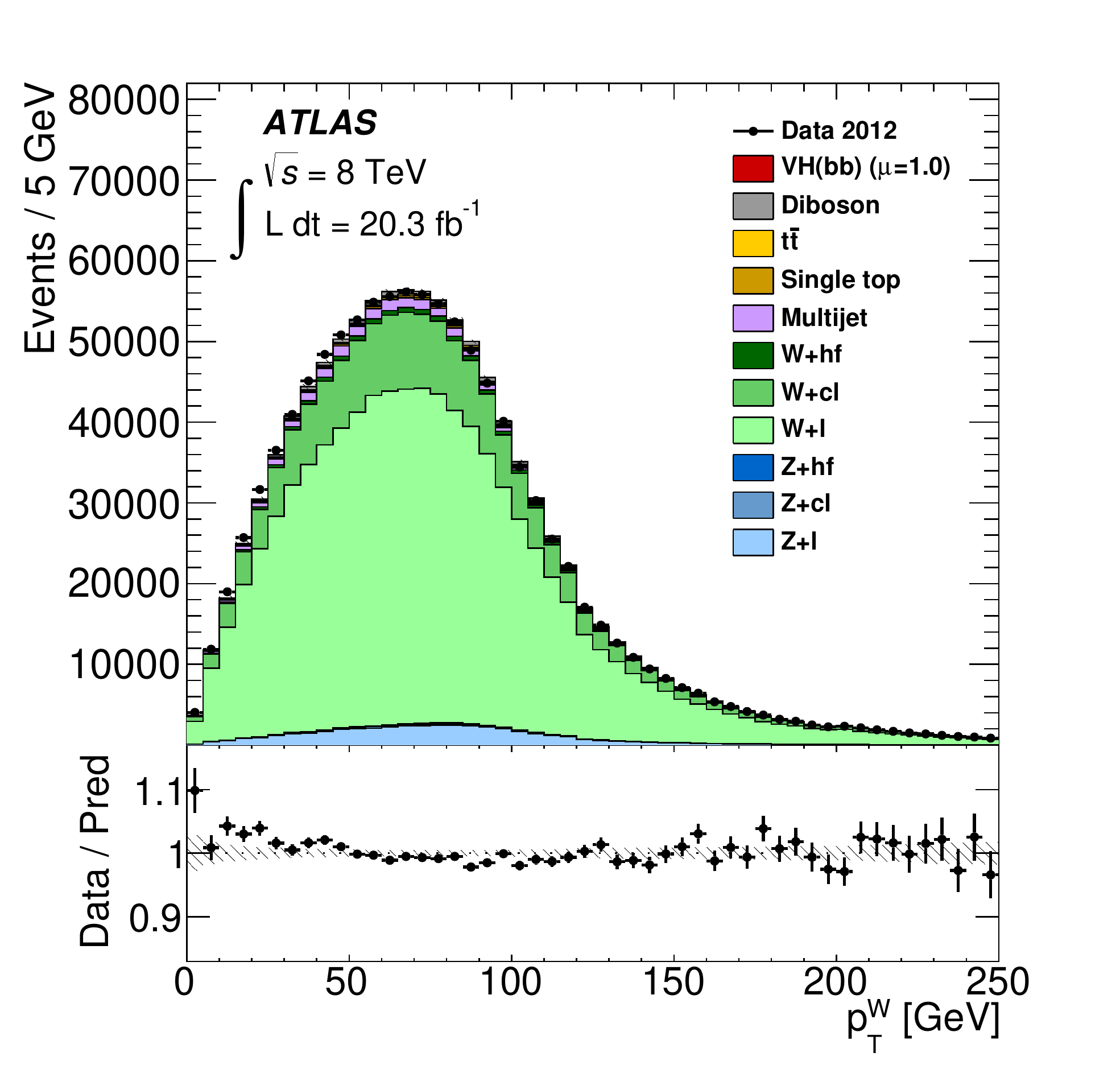}}
\hfill
\end{center}
\caption{
The \ptw\ distribution observed in data (points with error bars) and expected (histograms) for the 
2-jet 0-tag control region of the 1-muon sub-channel (MVA selection), (a) before and (b) after 
\dphi\ reweighting.
The multijet and simulated-background normalisations are provided by the multijet fits.
The size of the statistical uncertainty is indicated by the shaded band. 
The data-to-background ratio is shown in the lower panel.
  \label{fig:ptw_0tag_1lep}}
\end{figure}
\begin{figure}[hb!]
\begin{center}
\hfill
\subfigure[]{\includegraphics[width=0.45\textwidth]{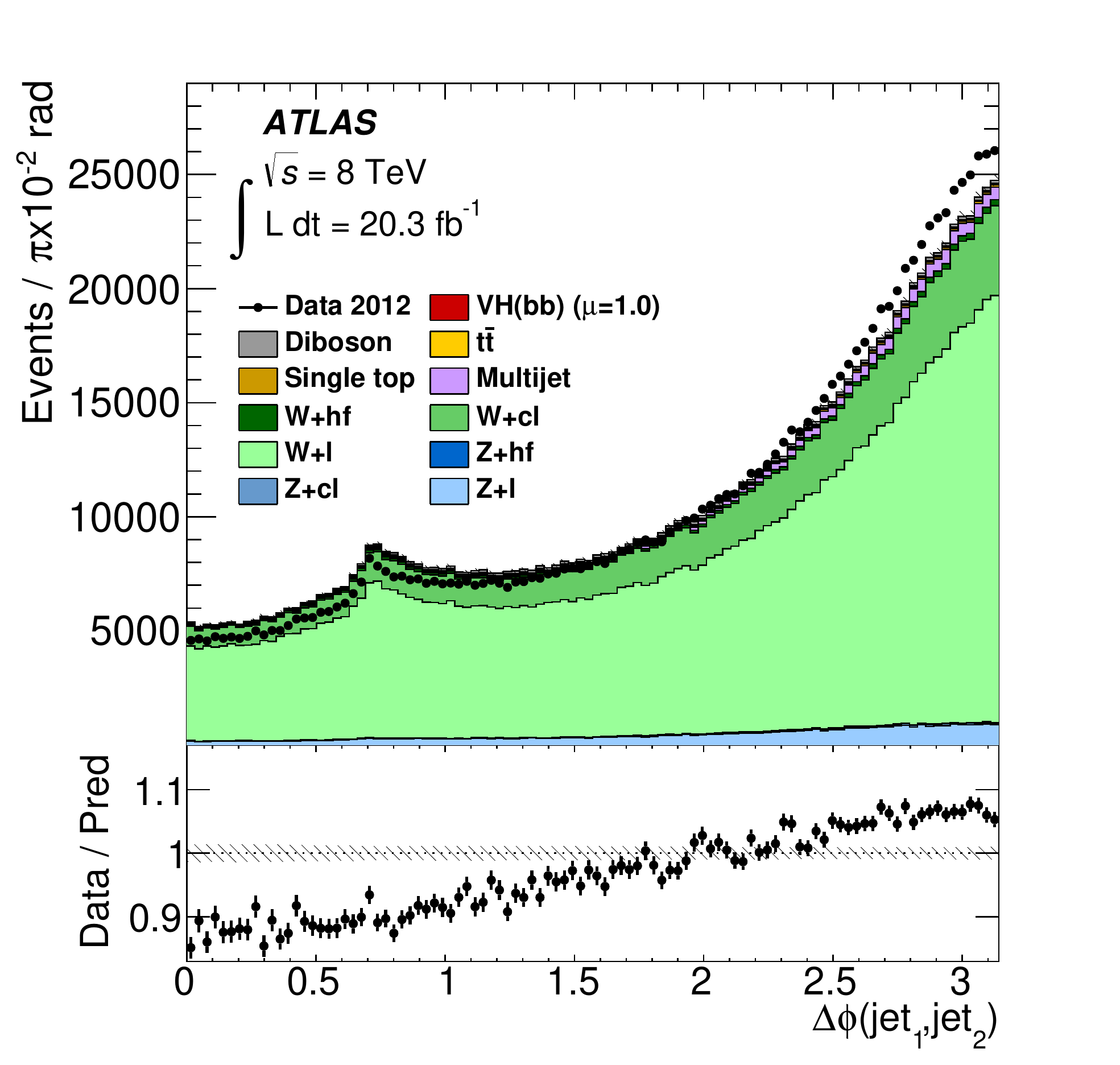}}
\hfill
\subfigure[]{\includegraphics[width=0.45\textwidth]{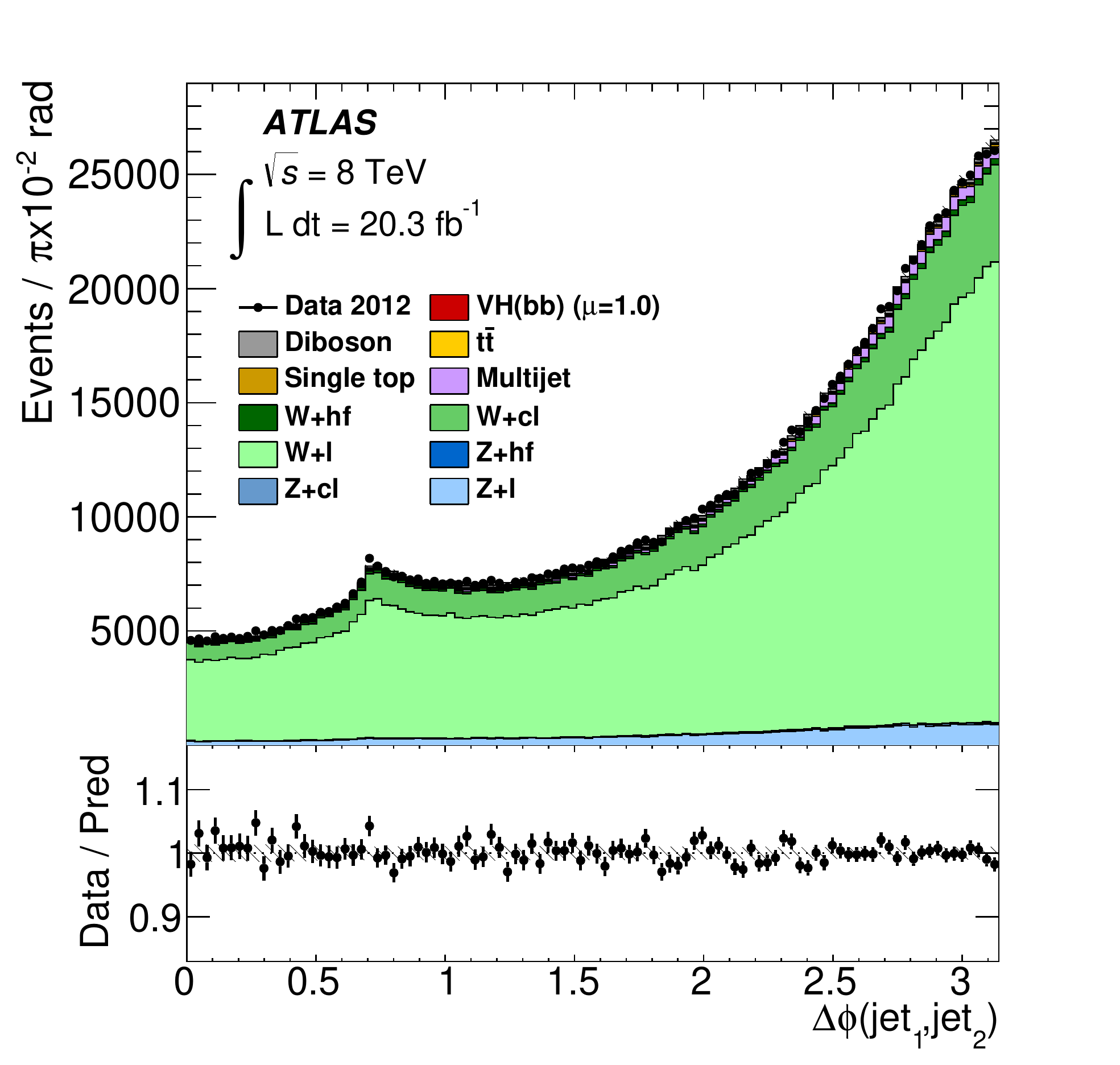}}
\hfill
\end{center}
\caption{
The \dphi\ distribution observed in data (points with error bars) and expected (histograms) for the 
2-jet 0-tag control region of the 1-muon sub-channel (MVA selection), (a) before and (b) after 
reweighting. All \ptw\ intervals are combined.
The multijet and simulated-background normalisations are provided by the multijet fits.
The size of the statistical uncertainty is indicated by the shaded band.
The data-to-background ratio is shown in the lower panel.
  \label{fig:dphi_0tag_1lep}}
\end{figure}
In order to address this mismodelling, the $Wl$ and $Wcl$
simulations are reweighted based on parameterised fits to the ratio of data to simulation 
in the \dphi\ variable in the 0-tag region, where these backgrounds dominate. 
Four separate functions are derived: for the 2- and 3-jet categories and for
$\ptw$ above and below $120$~GeV. 
The reweighted \dphi\ distributions show good agreement between data and simulation
(figure~\ref{fig:dphi_0tag_1lep}(b)).
This reweighting increases (reduces) by 0.7\% (5.6\%) 
the normalisation of the $\ptw\ <$ ($>$)~120~GeV region. 
After this reweighting, the modelling of the whole \ptw\ 
distribution is greatly improved, as can be seen 
in figure~\ref{fig:ptw_0tag_1lep}(b). 
This reweighting also improves the modelling of other distributions, most 
notably the dijet mass. It also improves the modelling in the 1-tag 
control regions and is therefore applied to the $Wl$ and $Wcl$ backgrounds 
in all regions of all channels. The numbers of $Wcc$ and $Wb$ background events 
in the 0- and 1-tag regions are too small to allow conclusive studies of their modelling, 
so no reweighting is applied to these backgrounds, 
but an associated systematic uncertainty is assessed instead, 
as explained in section~\ref{sec:sys}.

A similar, but not identical, procedure is used for the $Z$+jet events in the 
2-lepton channel. A \dphi\ reweighting is found to improve the modelling of the \ptz\ 
distribution in the 0-tag regions. In the signal-depleted 2-tag regions obtained by the exclusion 
of the 100--150~GeV dijet mass interval, there is no evidence of a need for a \dphi\ 
correction, but the \ptz\ distribution is mismodelled. A dedicated \ptz\ reweighting
is therefore determined in the 2-tag regions. 
Applying the \dphi\ reweighting to the $Zl$ component and the \ptz\ 
reweighting to the $Zc$ and $Zb$ components leads to good modelling also in the 1-tag 
regions. This procedure is therefore used in all regions of all channels.
 
It has been observed in an unfolded measurement of the \pt\ distribution of 
top quarks from pair production that the {\sc powheg} generator interfaced to {\sc pythia} 
predicts too hard a 
spectrum~\cite{Aad:2014zka}.
A correction accounting for this discrepancy is therefore applied at the level of
generated top quarks in the \ttb\ production process.

\subsection{Distributions in the dijet-mass analysis}\label{sec:dijet}

Distributions of \ptv\ and dijet mass are shown in 
figure~\ref{fig:ptv_all} and in figures~\ref{fig:mbb_0lep} and~\ref{fig:mbb_12lep}, respectively,
for a selection of 2-tag signal regions
of the dijet-mass analysis. 
It can be seen 
that the background composition in the signal regions varies 
greatly from channel to channel, with the \ptv\ interval, with the jet multiplicity,
and with the $b$-tagging category considered. 
The signal-to-background ratio is larger in the 2-jet and tighter $b$-tagging categories, 
and lower in the 3-jet and loose $b$-tagging categories.

In the 2-lepton channel, the dominant background is always $Zbb$. There is also a
significant contribution from \ttb\ in the lower \ptz\ intervals, and 
the relative diboson contribution increases with \ptz.

For the 1-lepton channel and in the 2-jet samples
the combination of $Wbb$ and \ttb\ accounts for most 
of the background in the most sensitive MM and TT categories, with the relative 
contribution of $Wbb$ and dibosons being largest in the tighter $b$-tagging categories 
and increasing with \ptw.
The flavours of the two selected jets from \ttb\ depend on the reconstructed 
\ptw\ interval. In particular, at high \ptw, when the $b$-quark and the $W$
from a top-quark decay are collimated, there is a large $bc$ contribution,
where the $c$-quark comes from the $W\to cs$ decay.
A significant contribution from single-top-quark production processes is also seen.
In the 3-jet category, the \ttb\ contribution 
is in general dominant, but there are significant contributions  
from single-top-quark production (mostly in the $Wt$ channel) and from $Wbb$, 
the latter increasing with \ptw. A non-negligible contribution of MJ background can be
seen in the lowest \ptw\ intervals of the 2-jet category.

In the 0-lepton channel, the main backgrounds arise from $Zbb$ and \ttb, 
but the $Wbb$ background is also significant. 
The relative \ttb\ contribution is largest in the lowest \ptv\ intervals, 
and larger in the 3-jet than in the 2-jet category.

The variations in the background composition between categories allow the global fit
to disentangle the rates of the various background sources. The
non-negligible contributions from the $Vcl$ and, to a 
lesser extent, the $Vl$ backgrounds are constrained in the global fit
by the LL $b$-tagging categories, and also by the $MV1c$ distributions
of the $b$-tagged jet in the 1-tag control regions. 
The 0-tag control regions are not taken into account in the global fit,
but are mainly used to improve the modelling of the $V$+jets backgrounds, 
as explained in section~\ref{sec:corrections}.

\begin{sidewaysfigure}[tb!]
\begin{center}
\centerline{
\subfigure[]{\includegraphics[width=0.3\textwidth]{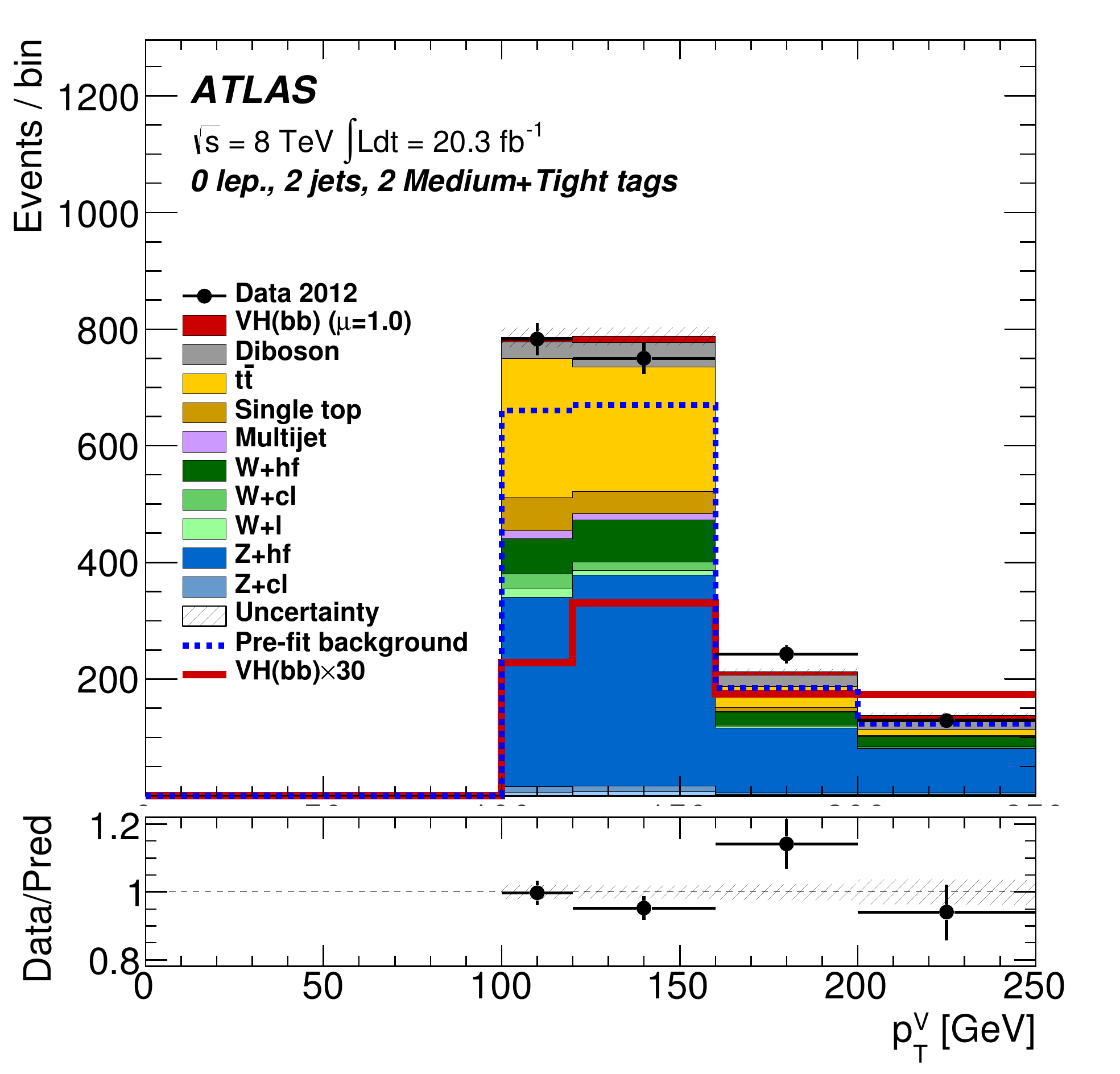}}
\subfigure[]{\includegraphics[width=0.3\textwidth]{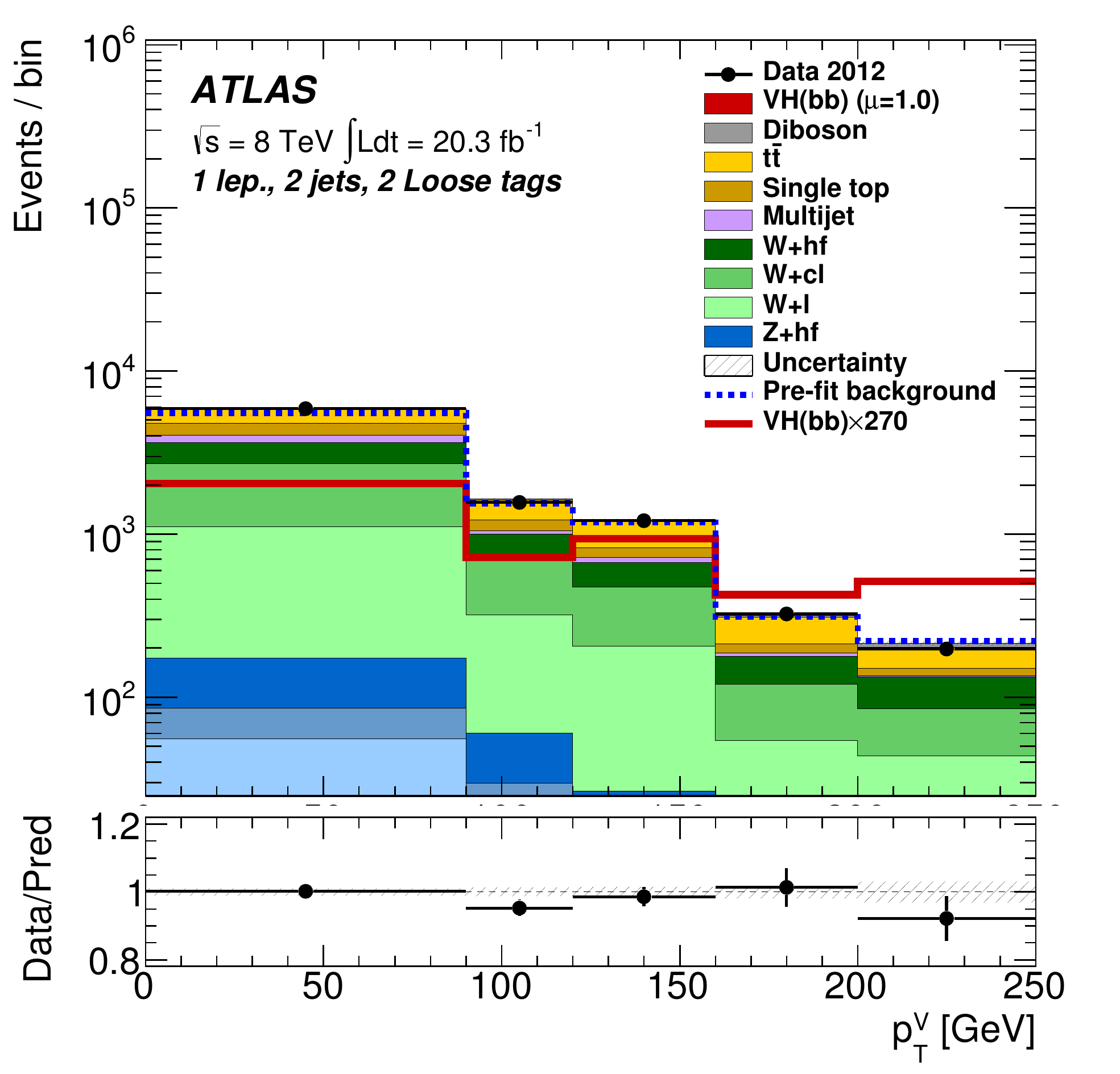}}
\subfigure[]{\includegraphics[width=0.3\textwidth]{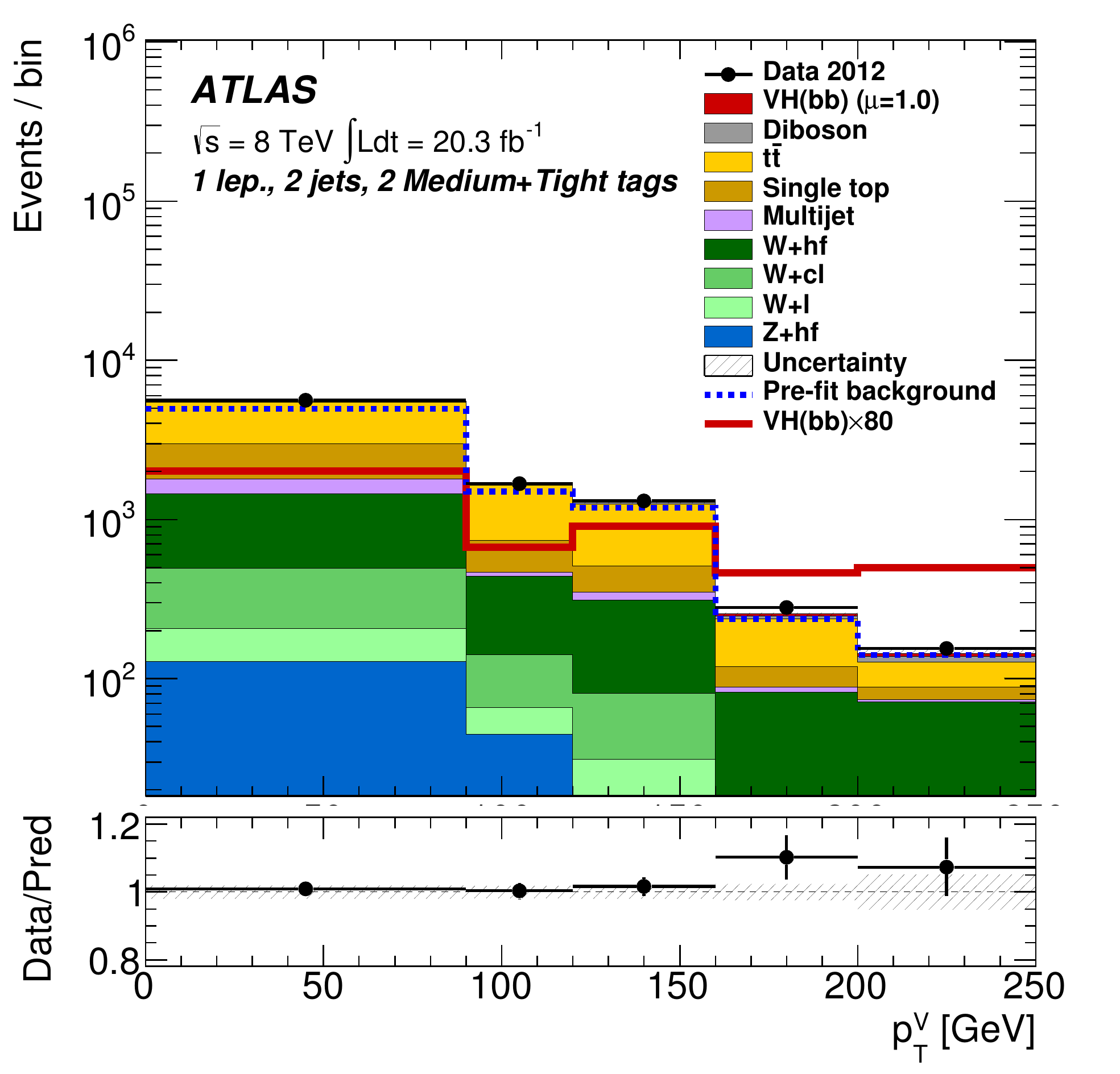}}
}
\centerline{
\subfigure[]{\includegraphics[width=0.3\textwidth]{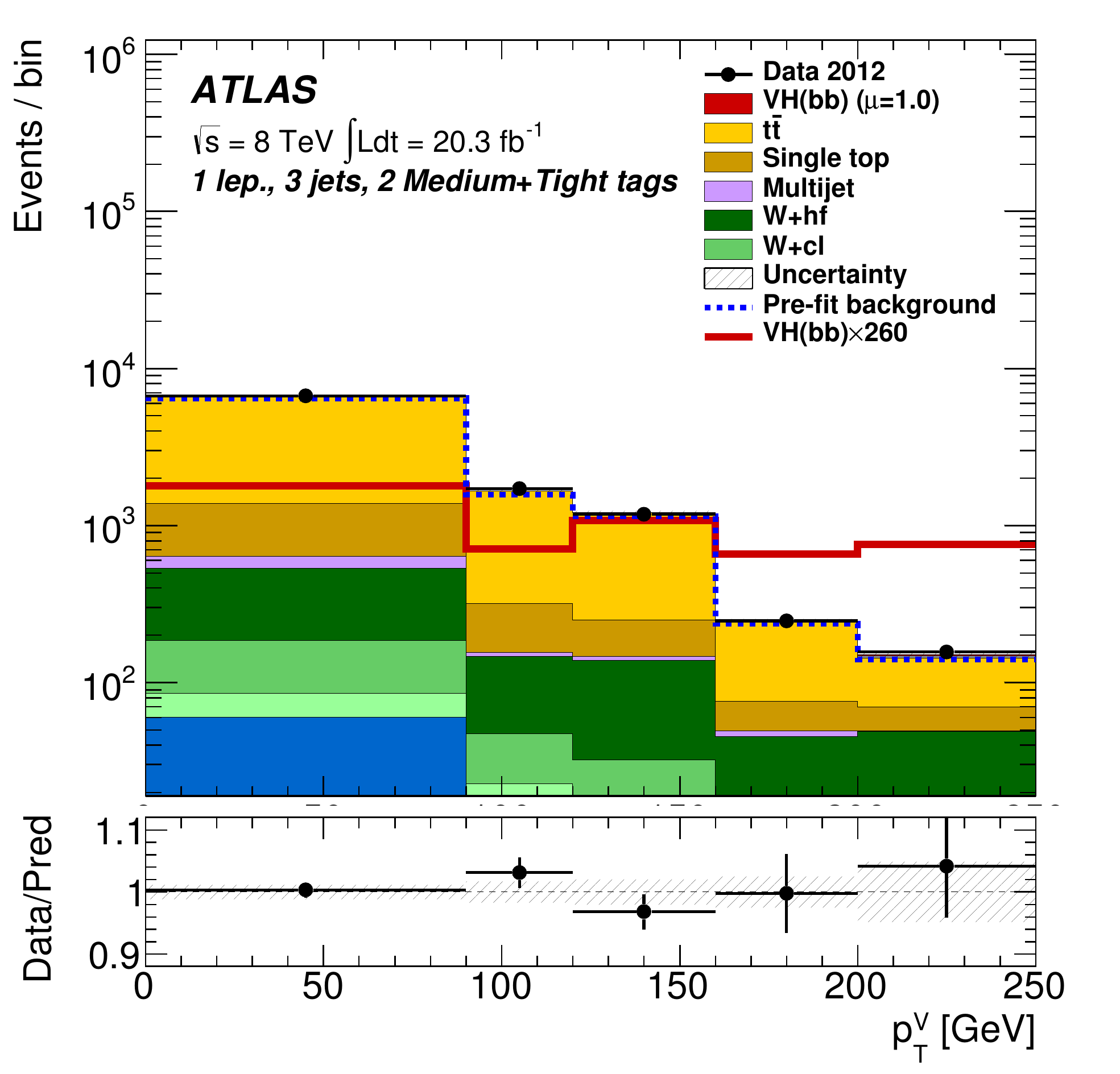}}
\subfigure[]{\includegraphics[width=0.3\textwidth]{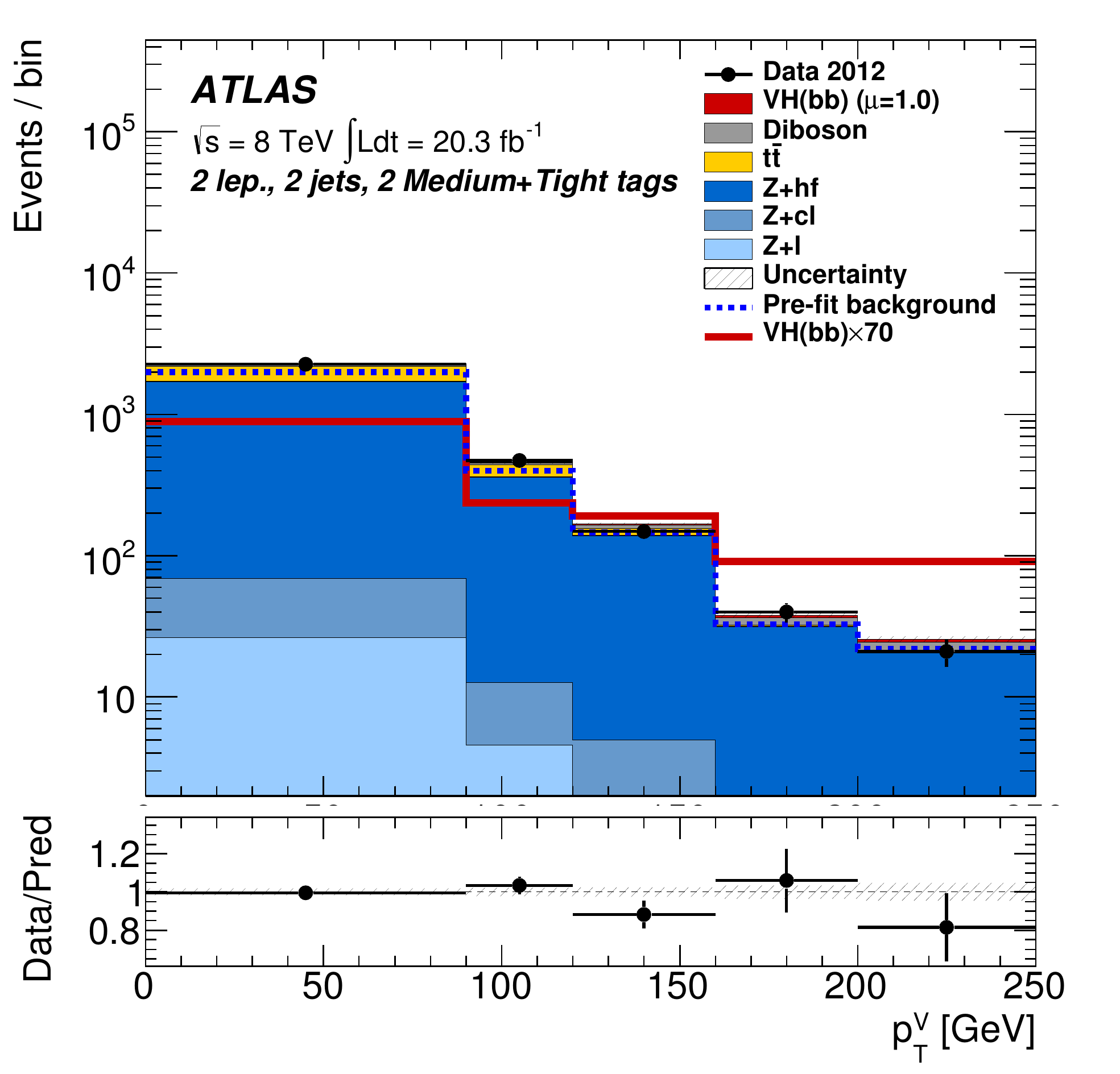}}
}
\hfill
\caption{
The \ptv\ distribution observed in data (points with error bars) and expected (histograms) for 
(a) the 2-jet signal regions of the 0-lepton channel with the Medium and Tight 
$b$-tagging categories (also referred to as MM and TT in the text) combined,
(b) the 2-jet signal regions of the 1-lepton channel for the LL category,
(c) the 2-jet signal regions of the 1-lepton channel with the MM and TT categories combined,
(d) the 3-jet signal regions of the 1-lepton channel with the MM and TT categories combined, and
(e) the 2-jet signal regions of the 2-lepton channel with the MM and TT categories combined.
The background contributions after the global fit of the dijet-mass analysis are shown as filled histograms.
The Higgs boson signal ($\mh = 125$~GeV) is shown as a filled histogram on top of 
the fitted backgrounds, 
as expected from the SM (indicated as $\mu=1.0$),
and, unstacked as an unfilled histogram, scaled by the factor indicated in the legend. 
The dashed histogram shows the total background as expected from the pre-fit 
MC simulation. Overflow entries are included in the last bin.
The size of the combined statistical and systematic uncertainty on the
sum of the signal and fitted background is indicated by the hatched band. The ratio
of the data to the sum of the signal and fitted background is shown in the lower panel.
\label{fig:ptv_all}}
\end{center}
\end{sidewaysfigure}

\begin{sidewaysfigure}[tb!]
\begin{center}
\subfigure[]{\includegraphics[width=0.32\textwidth]{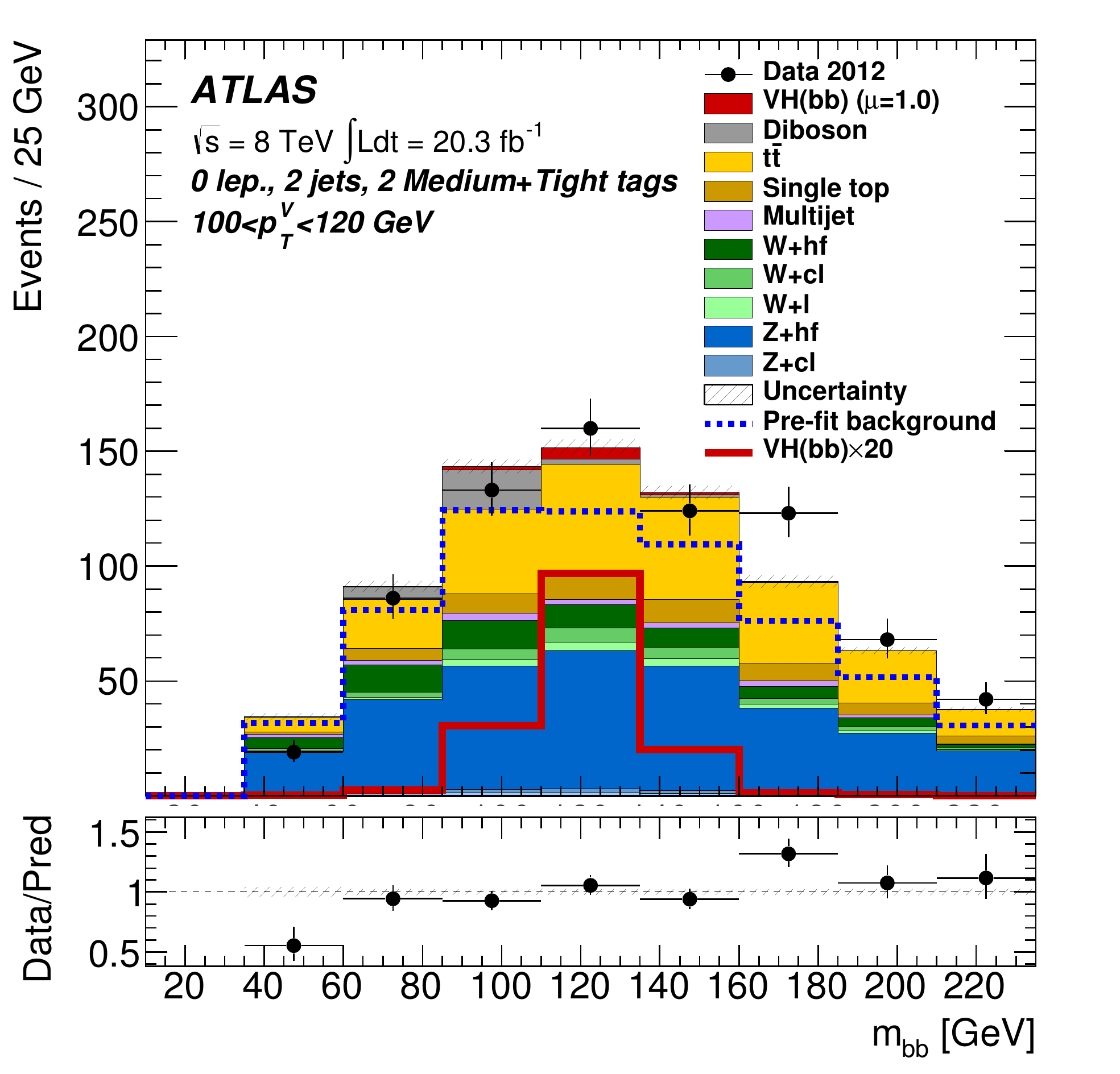}}
\hfill
\subfigure[]{\includegraphics[width=0.32\textwidth]{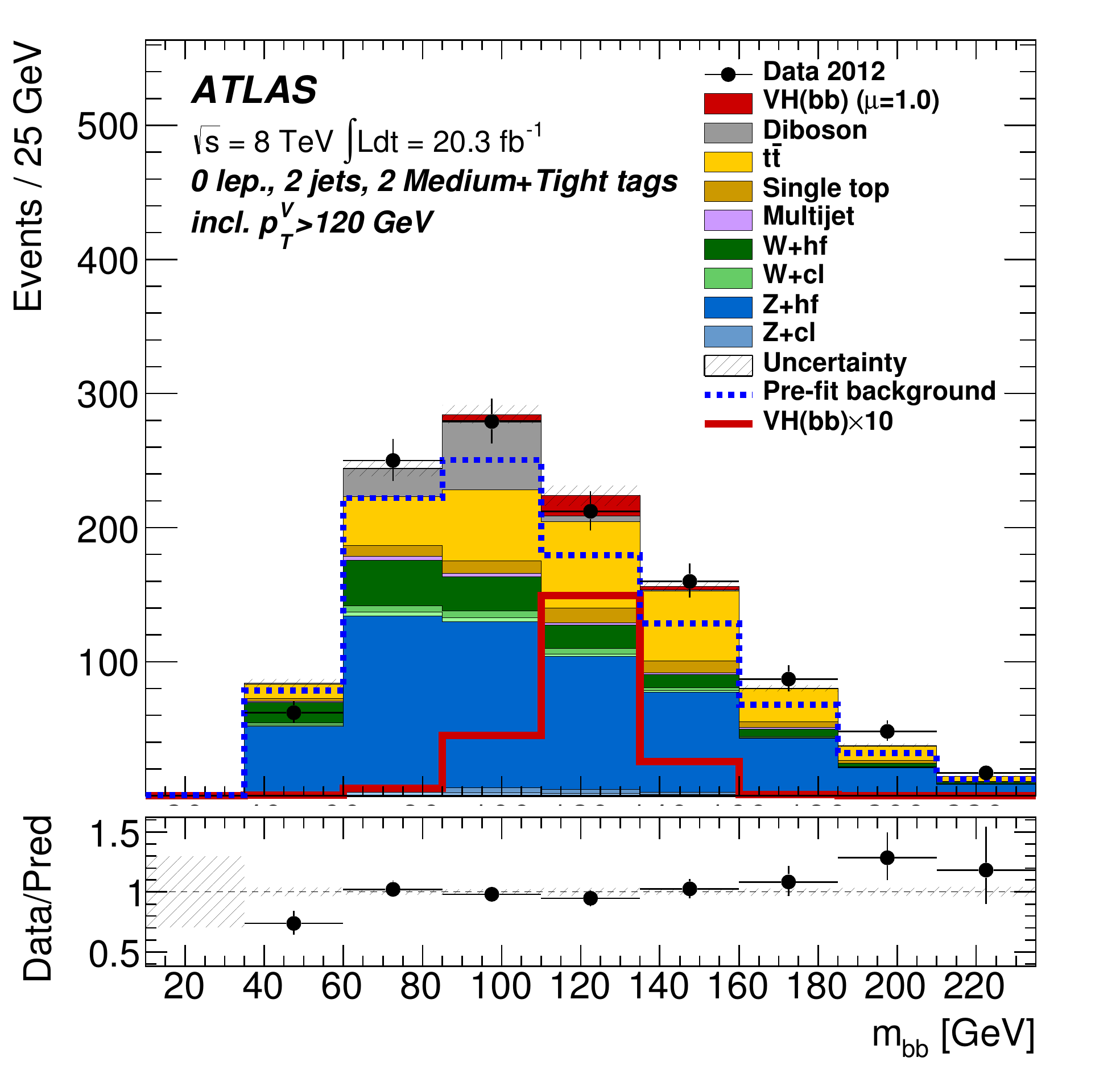}}
\hfill
\subfigure[]{\includegraphics[width=0.32\textwidth]{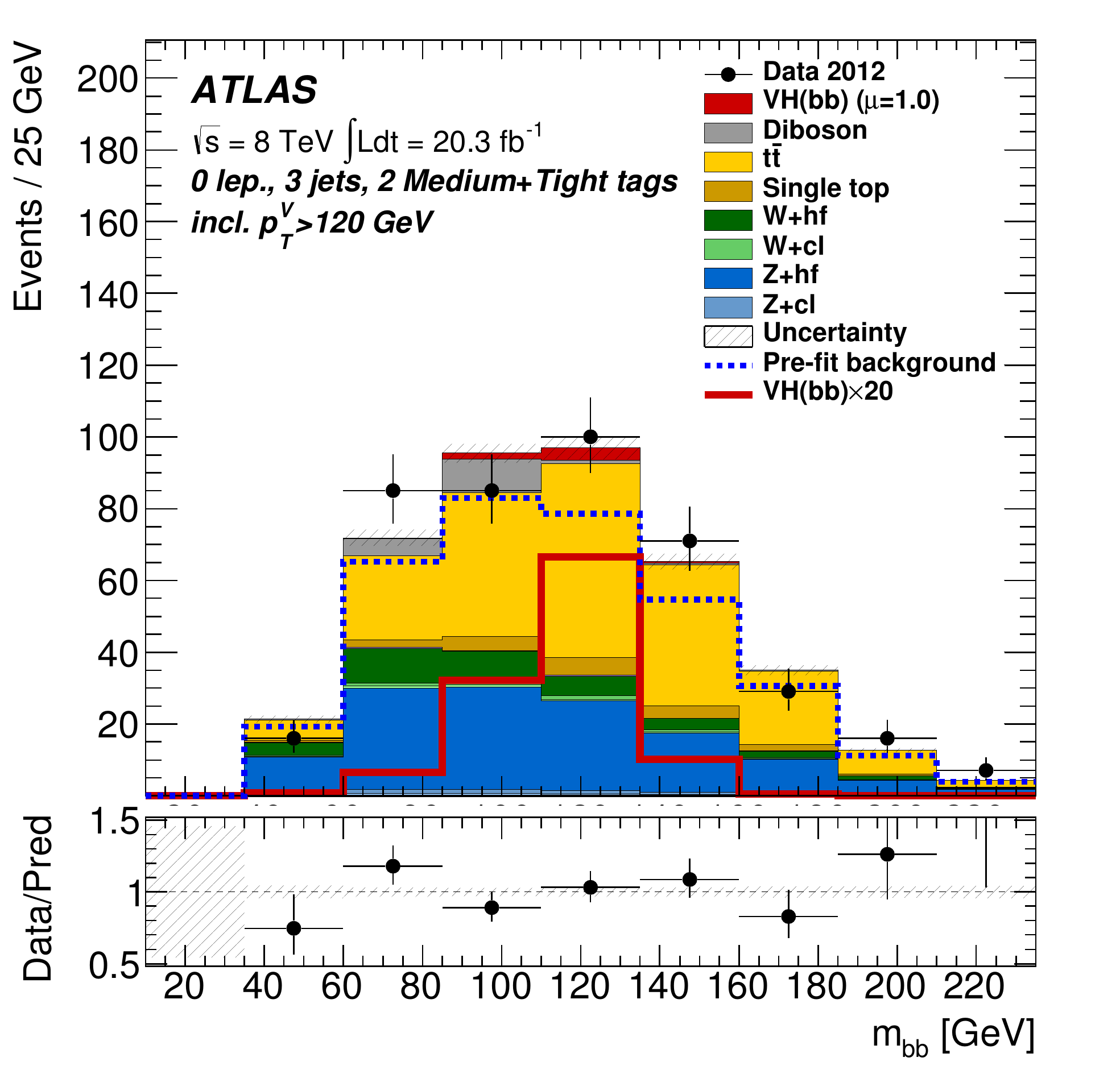}}
\end{center}
\caption{
The dijet-mass distribution observed in data (points with error bars) and expected (histograms) 
in the 0-lepton channel with the Medium and Tight 
$b$-tagging categories (also referred to as MM and TT in the text) combined for
(a) the 2-jet signal region in the $100<\ptv<120$~GeV interval,
(b) the 2-jet signal regions in the three intervals with $\ptv>120$~GeV combined, and
(c) the 3-jet signal regions in the three intervals with $\ptv>120$~GeV combined.
The background contributions after the global fit of the dijet-mass analysis are shown as filled histograms.
The Higgs boson signal ($\mh = 125$~GeV) is shown as a filled histogram on top of 
the fitted backgrounds, 
as expected from the SM (indicated as $\mu=1.0$),
and, unstacked as an unfilled histogram, scaled by the factor indicated in the legend. 
The dashed histogram shows the total background as expected from the pre-fit 
MC simulation. The entries in overflow are included in the last bin.
The size of the combined statistical and systematic uncertainty on the
sum of the signal and fitted background is indicated by the hatched band. The ratio
of the data to the sum of the signal and fitted background is shown in the lower panel.
\label{fig:mbb_0lep}}
\end{sidewaysfigure}

\begin{figure}[tb!]
\centerline{
\hfill
\subfigure[]{\includegraphics[width=0.495\textwidth]{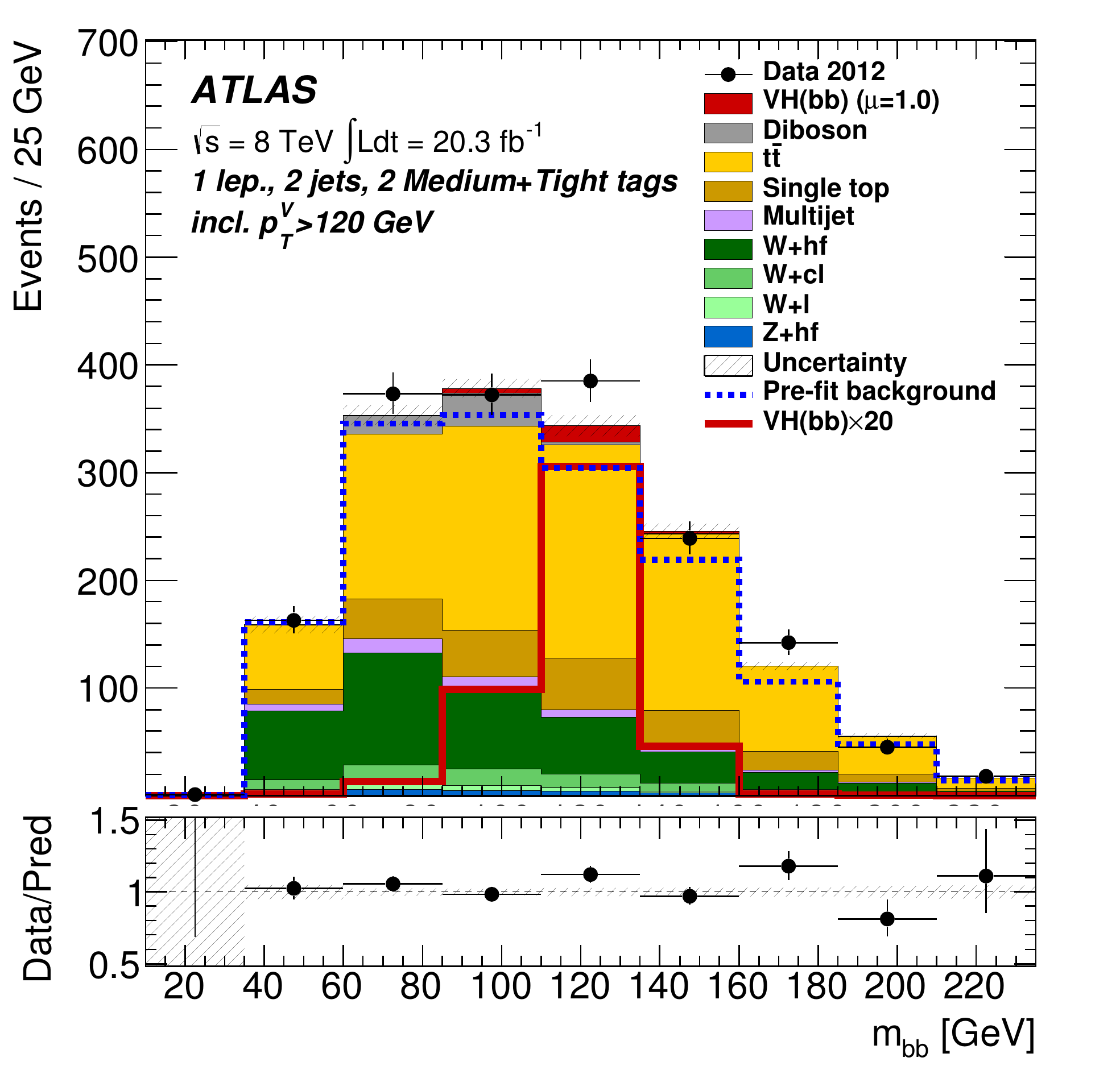}}
\hfill
\subfigure[]{\includegraphics[width=0.495\textwidth]{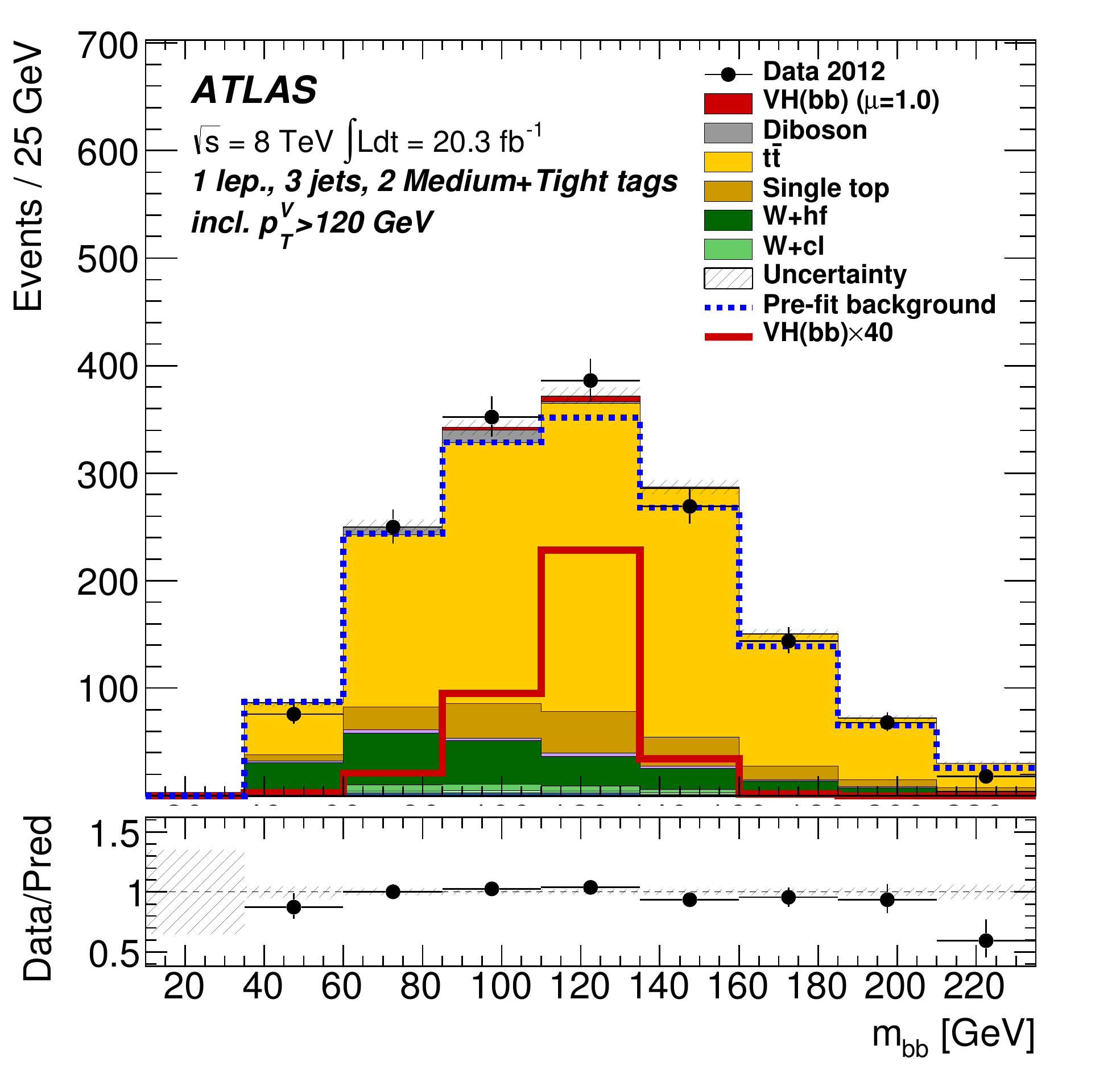}}
\hfill
}
\centerline{
\hfill
\subfigure[]{\includegraphics[width=0.495\textwidth]{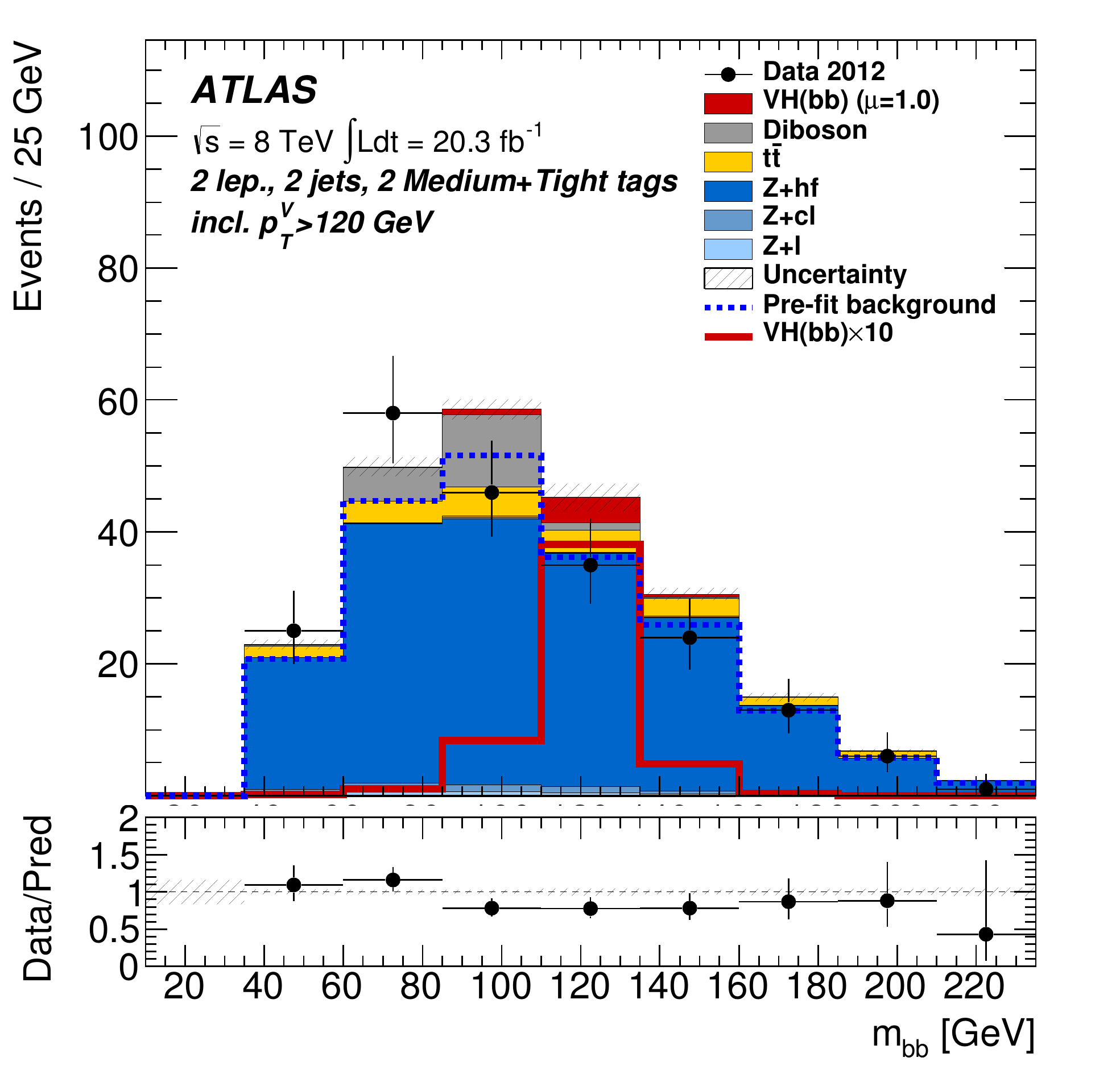}}
\hfill
\subfigure[]{\includegraphics[width=0.495\textwidth]{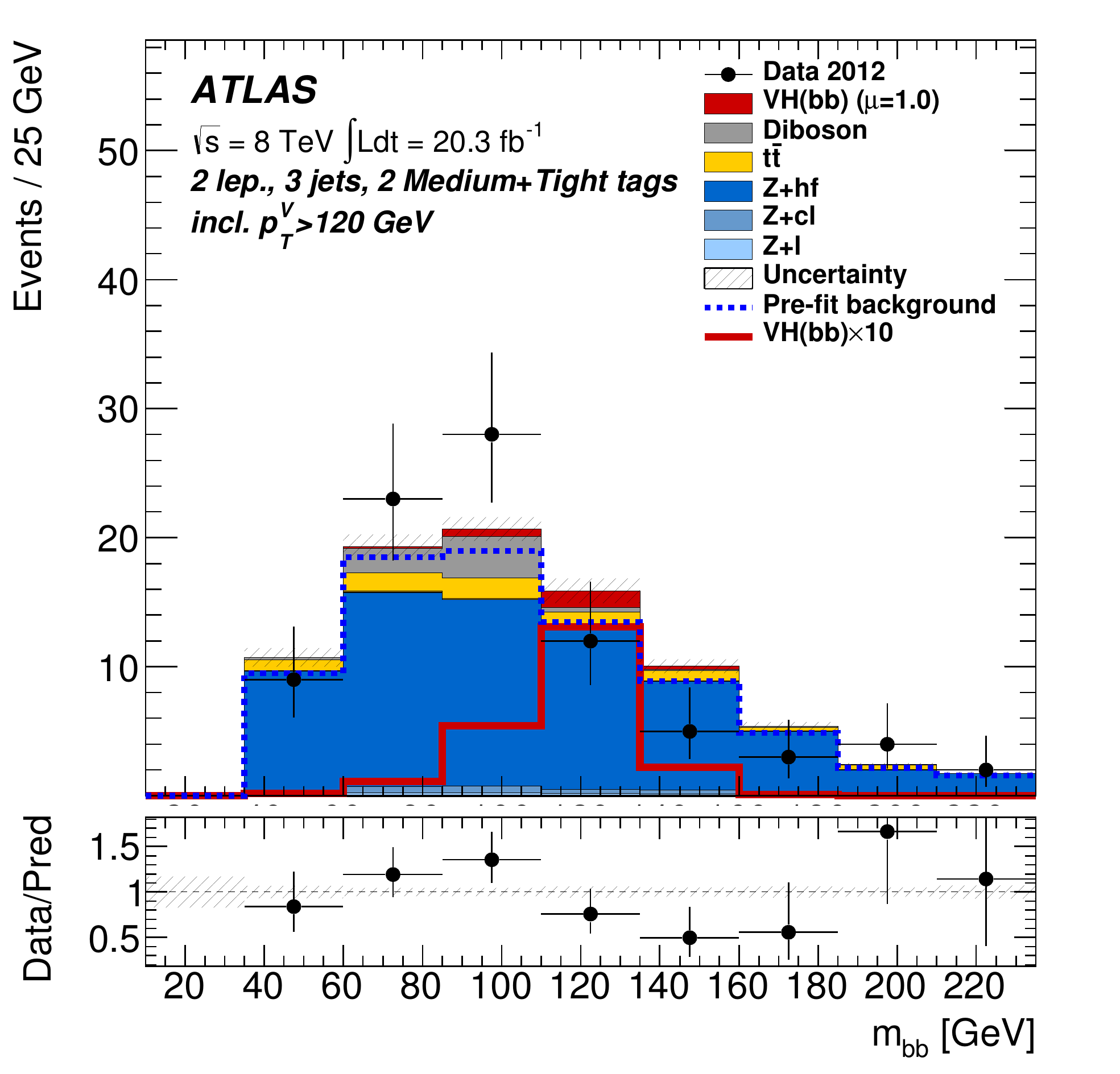}}
\hfill
}
\caption{
The dijet-mass distribution observed in data (points with error bars) and expected (histograms) 
with the Medium and Tight $b$-tagging categories (also referred to as MM and TT in the text) 
combined and the three intervals with $\ptv>120$~GeV combined for
(a) the 2-jet signal regions of the 1-lepton channel,
(b) the 3-jet signal regions of the 1-lepton channel,
(c) the 2-jet signal regions of the 2-lepton channel, and
(d) the 3-jet signal regions of the 2-lepton channel.
The background contributions after the global fit of the dijet-mass analysis are shown as filled histograms.
The Higgs boson signal ($\mh = 125$~GeV) is shown as a filled histogram on top of 
the fitted backgrounds, 
as expected from the SM (indicated as $\mu=1.0$),
and, unstacked as an unfilled histogram, scaled by the factor indicated in the legend. 
The dashed histogram shows the total background as expected from the pre-fit 
MC simulation. The entries in overflow are included in the last bin.
The size of the combined statistical and systematic uncertainty on the
sum of the signal and fitted background is indicated by the hatched band. The ratio
of the data to the sum of the signal and fitted background is shown in the lower panel.
\label{fig:mbb_12lep}}
\end{figure}

\FloatBarrier

\subsection{Distributions in the multivariate analysis}\label{sec:BDT}

Distributions of the BDT$_{VH}$ discriminants of the MVA are shown in figures~\ref{fig:BDT_2tag_0lep}
to~\ref{fig:BDT_2tag_2lep} for 2-tag signal regions in the 2- and 3-jet categories of 
the 0-, 1- and 2-lepton channels.
It can be seen that the backgrounds dominated by light jets and, to a lesser extent, $c$-jets 
accumulate at lower values of the BDT$_{VH}$ discriminants, due to the inclusion of the MV1c information
as inputs to the BDTs. The composition of the dominant backgrounds accumulating at higher 
values of the BDT$_{VH}$ discriminant is similar to what was already observed in the 2-tag signal regions 
of the dijet-mass analysis, namely $Vbb$ and \ttb, however with a larger contribution of the
latter due to the looser requirement on $\Delta R(\mathrm{jet}_1,\mathrm{jet}_2)$ 
in the MVA selection.

Distributions of the output of the MV1c $b$-tagging algorithm are shown in figure~\ref{fig:MV1c_1tag}
for the $b$-tagged jet in the 1-tag control regions of the MVA, in the 2-jet category and for $\ptv>120$~GeV. 
In these distributions, the four bins correspond to the four 
$b$-tagging operating points and are ordered from left to right in increasing $b$-jet purity.
It can be seen that these distributions, which are used in the global fit, provide strong 
constraints on the $Vc$ and $Vl$ backgrounds. As in the dijet-mass analysis, the 0-tag control 
regions are not used in the global fit.

\begin{figure}[b!]
\centerline{
\hfill
\subfigure[]{\includegraphics[width=0.495\textwidth]{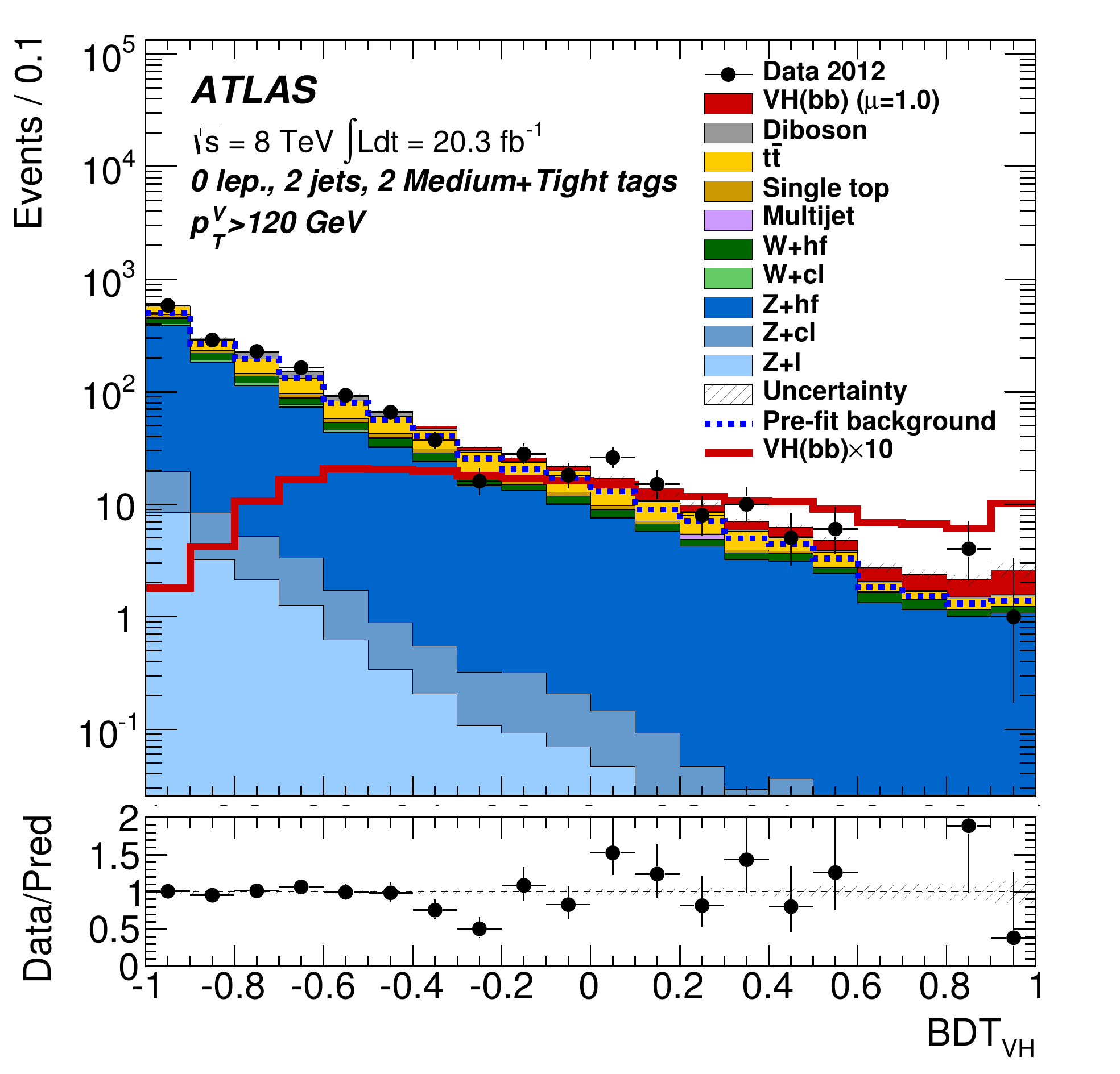}}
\hfill
\subfigure[]{\includegraphics[width=0.495\textwidth]{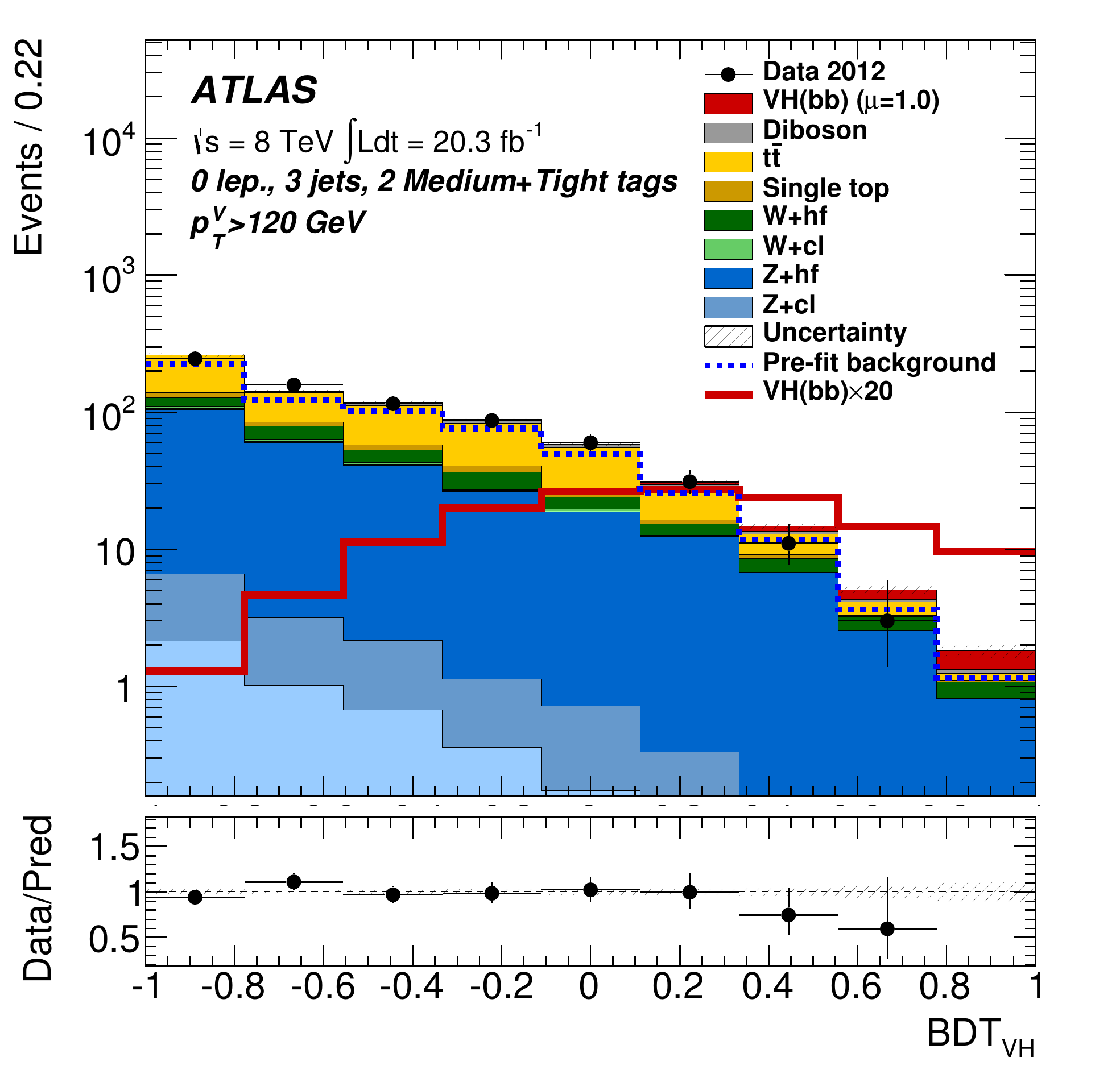}}
\hfill
}
\caption{
The BDT$_{VH}$-discriminant distribution observed in data (points with error bars) and expected (histograms) 
for the 0-lepton channel combining the 2-tag Medium and Tight $b$-tagging categories 
(also referred to as MM and TT in the text) for $\ptv > 120$~GeV for
(a) 2-jet events and (b) 3-jet events. 
The background contributions after the global fit of the MVA are shown 
as filled histograms.
The Higgs boson signal ($\mh = 125$~GeV) is shown as a filled histogram on top of 
the fitted backgrounds, 
as expected from the SM (indicated as $\mu=1.0$),
and, unstacked as an unfilled histogram, scaled by the factor indicated in the legend. 
The dashed histogram shows the total background as expected from the pre-fit 
MC simulation. 
The size of the combined statistical and systematic uncertainty on the
sum of the signal and fitted background is indicated by the hatched band. The ratio
of the data to the sum of the signal and fitted background is shown in the lower panel.
\label{fig:BDT_2tag_0lep}}
\end{figure}

\begin{figure}[tb!]
\centerline{
\hfill
\subfigure[]{\includegraphics[width=0.495\textwidth]{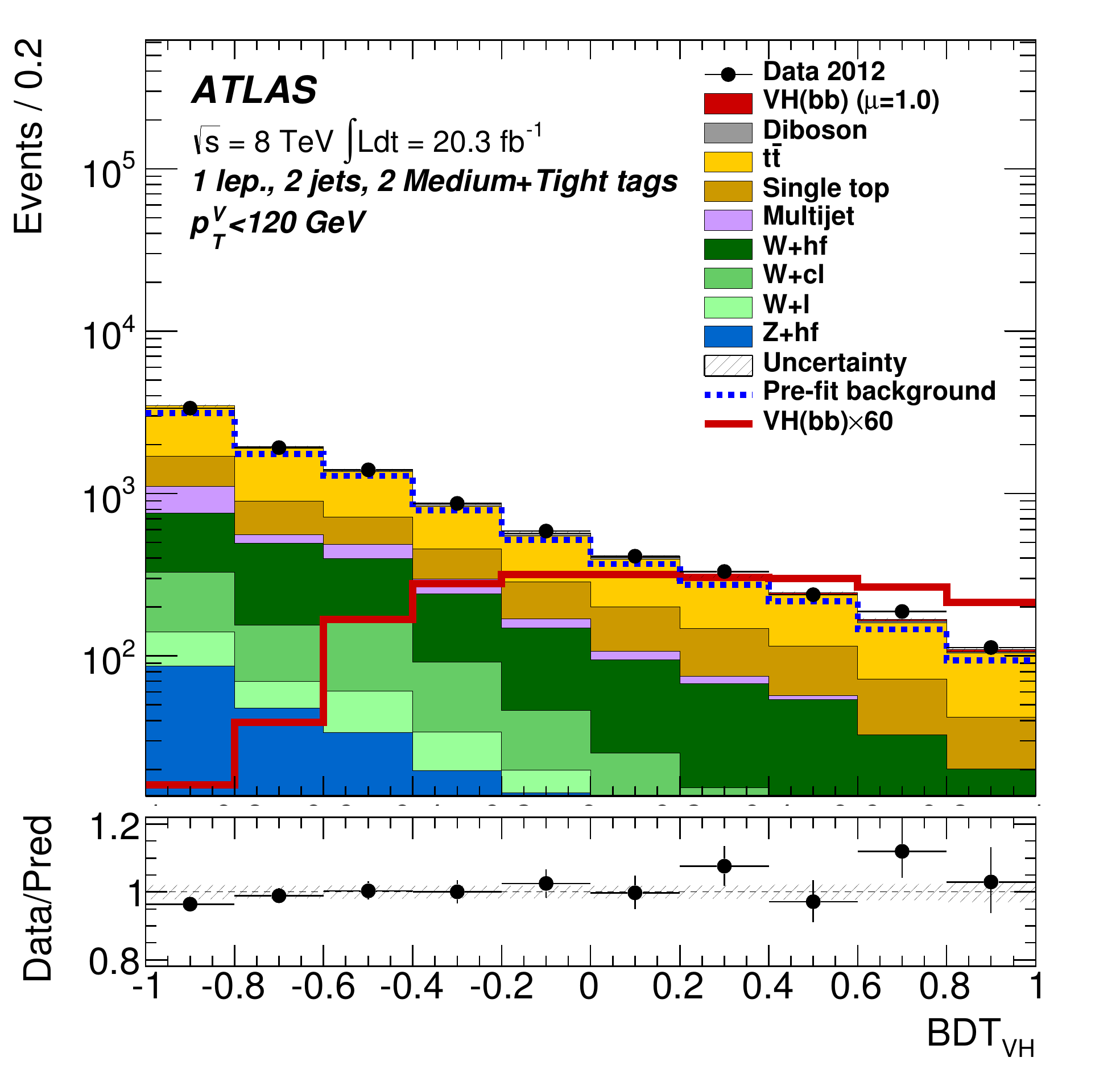}}
\hfill
\subfigure[]{\includegraphics[width=0.495\textwidth]{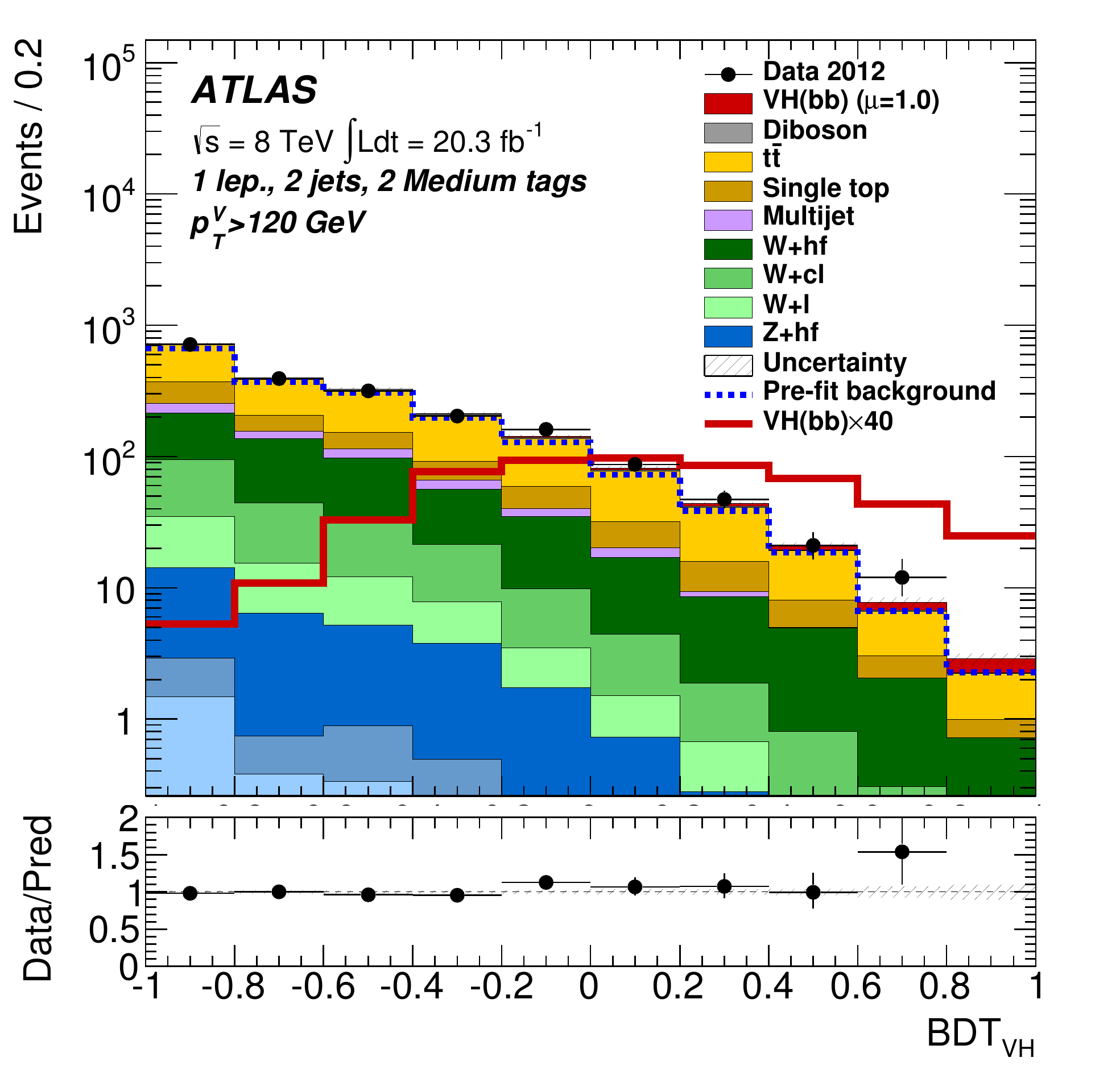}}
\hfill
}
\centerline{
\hfill
\subfigure[]{\includegraphics[width=0.495\textwidth]{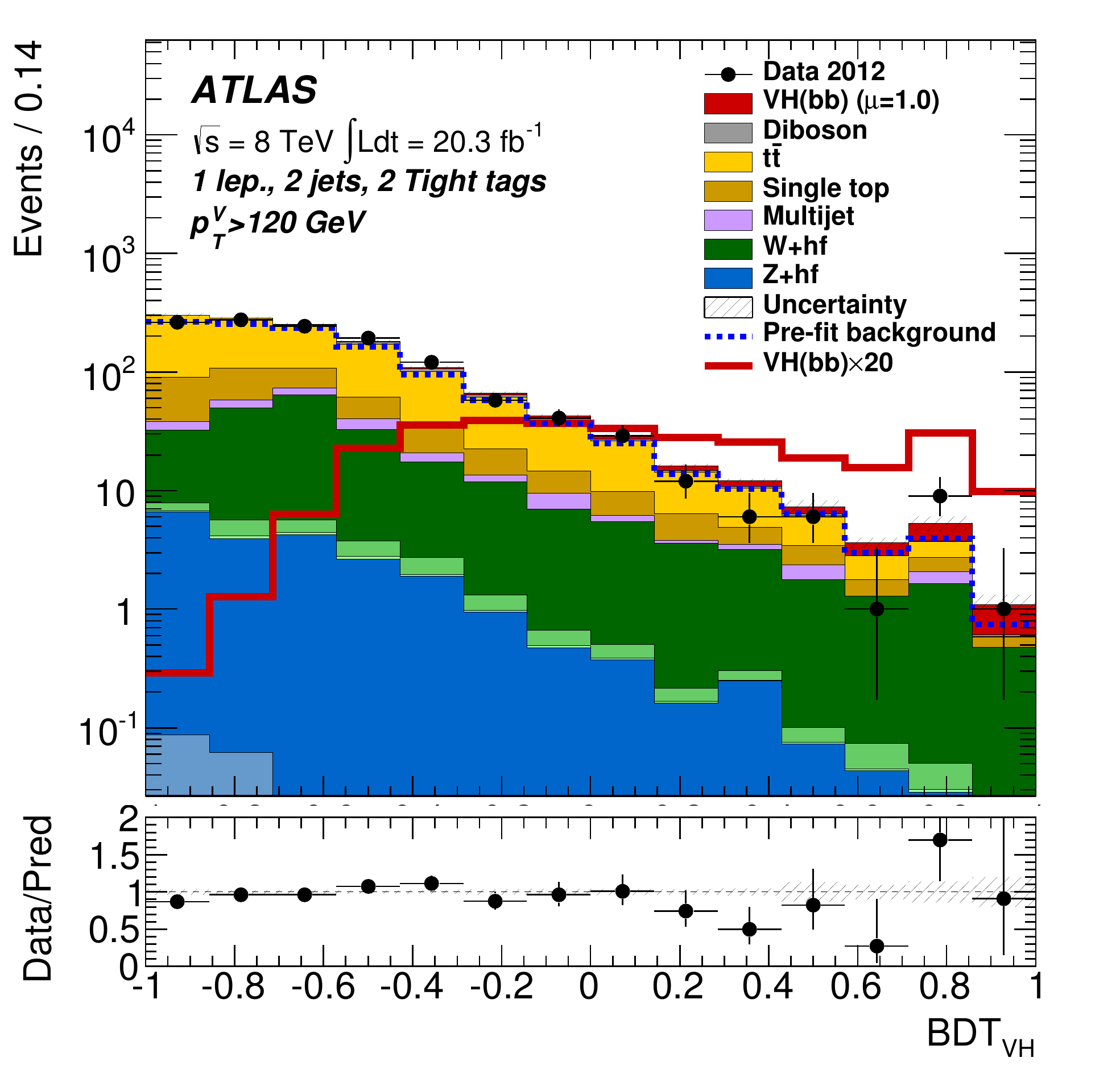}}
\hfill
\subfigure[]{\includegraphics[width=0.495\textwidth]{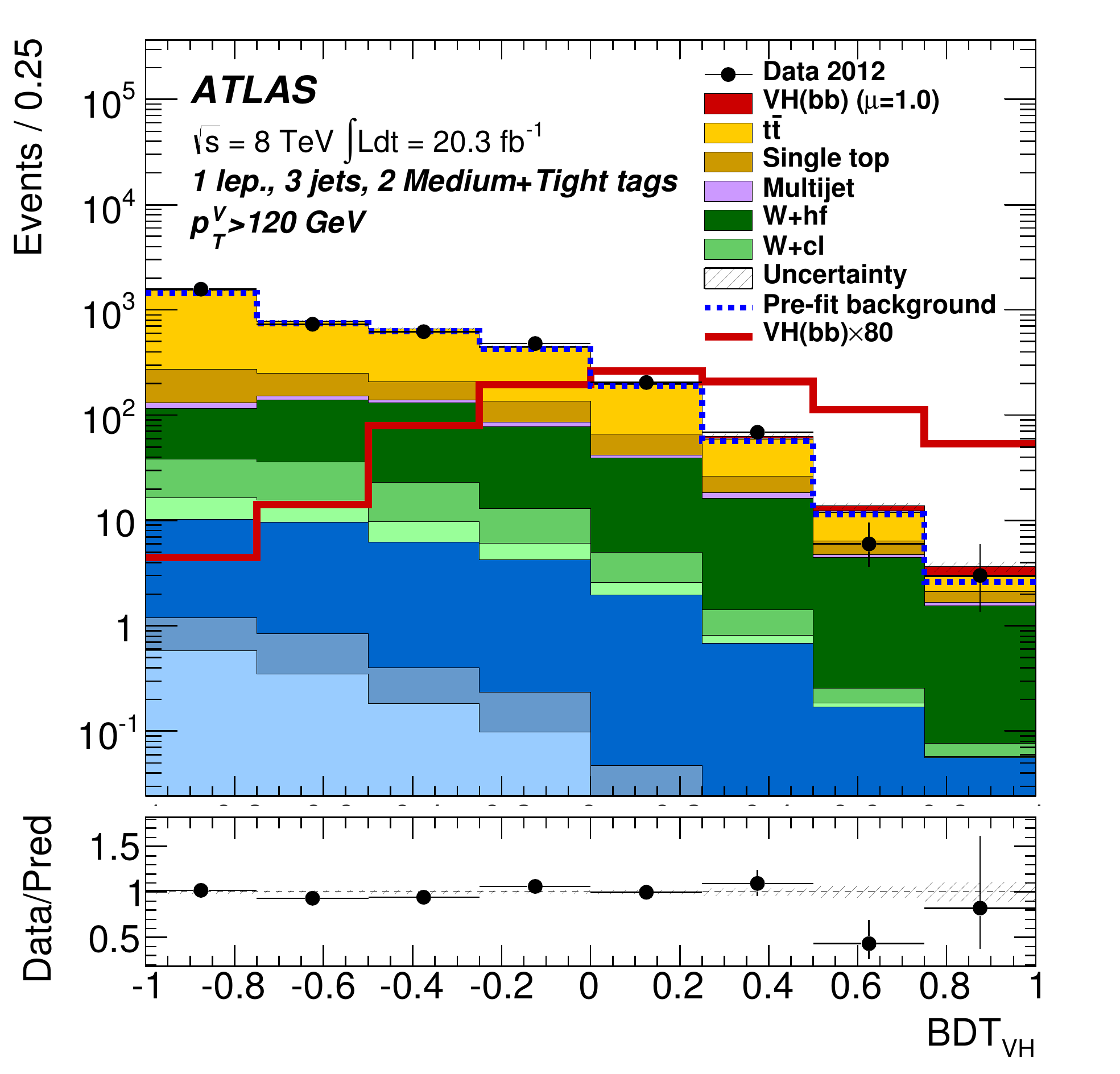}}
\hfill
}
\caption{
The BDT$_{VH}$-discriminant distribution observed in data (points with error bars) and expected (histograms) 
for the 2-tag signal regions of the 1-lepton channel for 
(a) 2-jet events with the Medium and Tight $b$-tagging categories 
(also referred to as MM and TT in the text) combined 
and with $\ptw \leq 120$~GeV, 
(b) MM 2-jet events with $\ptw > 120$~GeV, 
(c) TT 2-jet events with $\ptw > 120$~GeV, and 
(d) MM and TT combined 3-jet events with $\ptw > 120$~GeV. 
The background contributions after the global fit of the MVA are shown 
as filled histograms.
The Higgs boson signal ($\mh = 125$~GeV) is shown as a filled histogram on top of 
the fitted backgrounds, 
as expected from the SM (indicated as $\mu=1.0$),
and, unstacked as an unfilled histogram, scaled by the factor indicated in the legend. 
The dashed histogram shows the total background as expected from the pre-fit 
MC simulation.
The size of the combined statistical and systematic uncertainty on the
sum of the signal and fitted background is indicated by the hatched band. The ratio
of the data to the sum of the signal and fitted background is shown in the lower panel.
\label{fig:BDT_2tag_1lep}}
\end{figure}

\begin{sidewaysfigure}[tb!]
\centerline{
\subfigure[]{\includegraphics[width=0.32\textwidth]{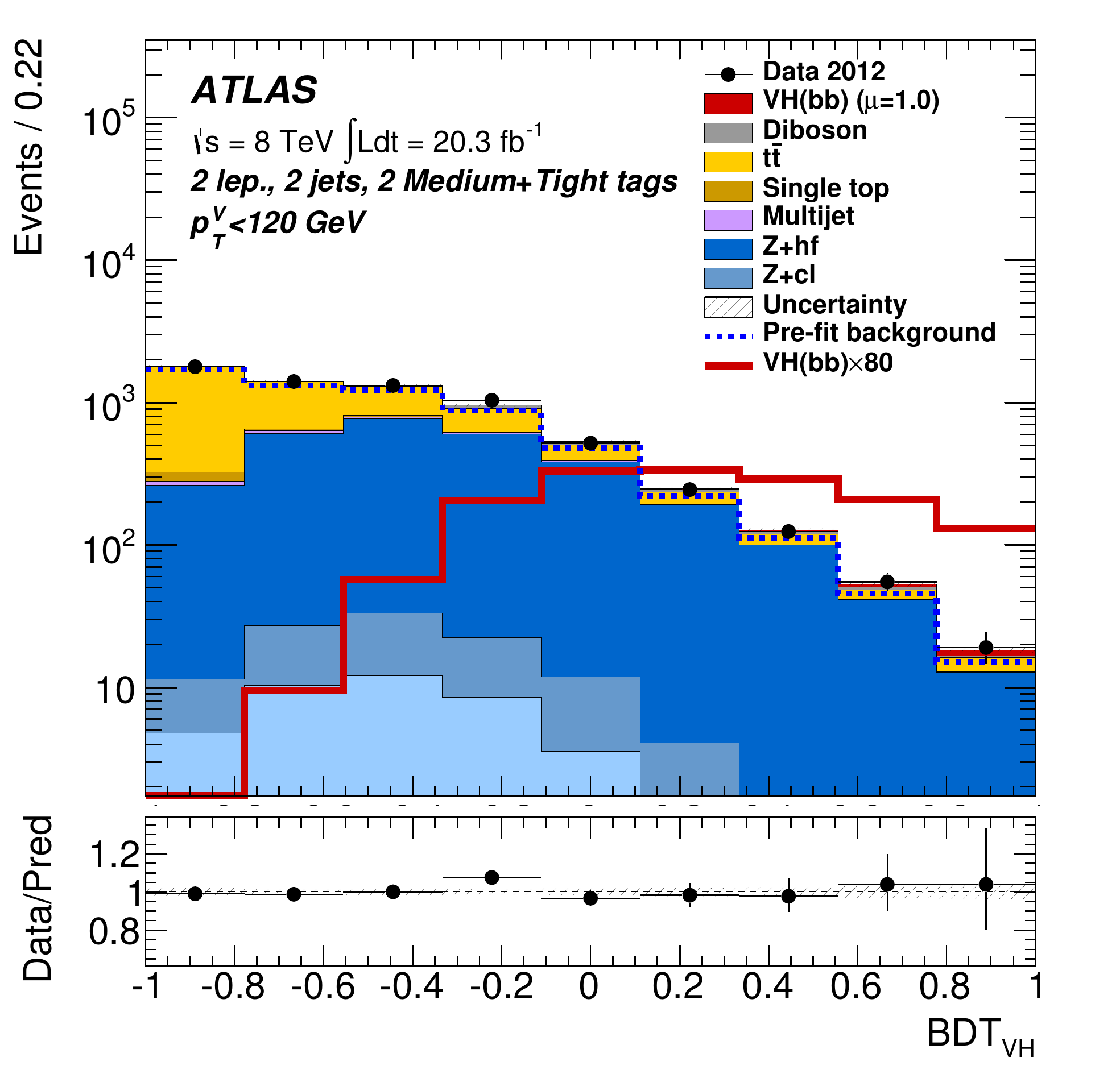}}
\hfill
\subfigure[]{\includegraphics[width=0.32\textwidth]{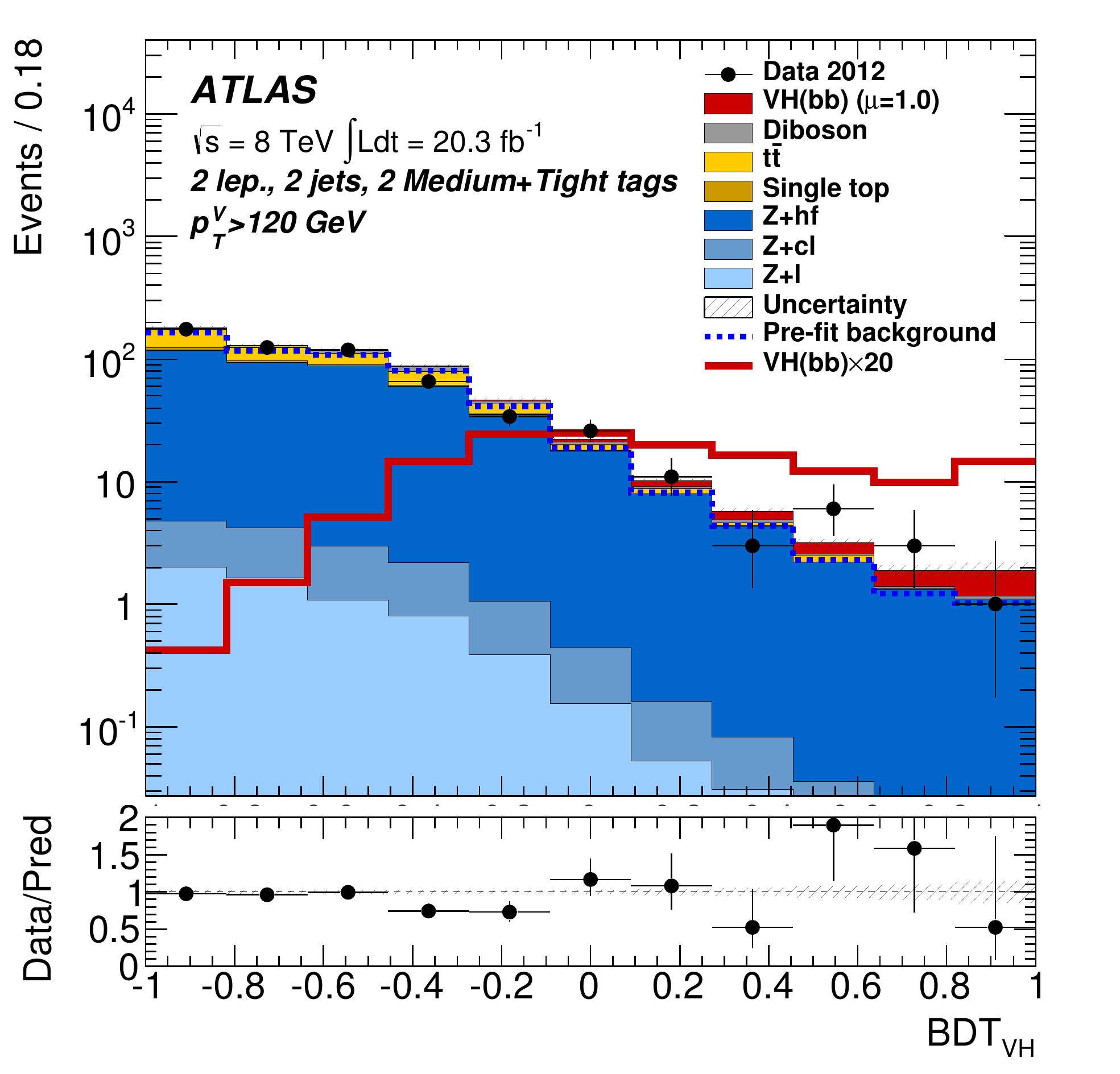}}
\hfill
\subfigure[]{\includegraphics[width=0.32\textwidth]{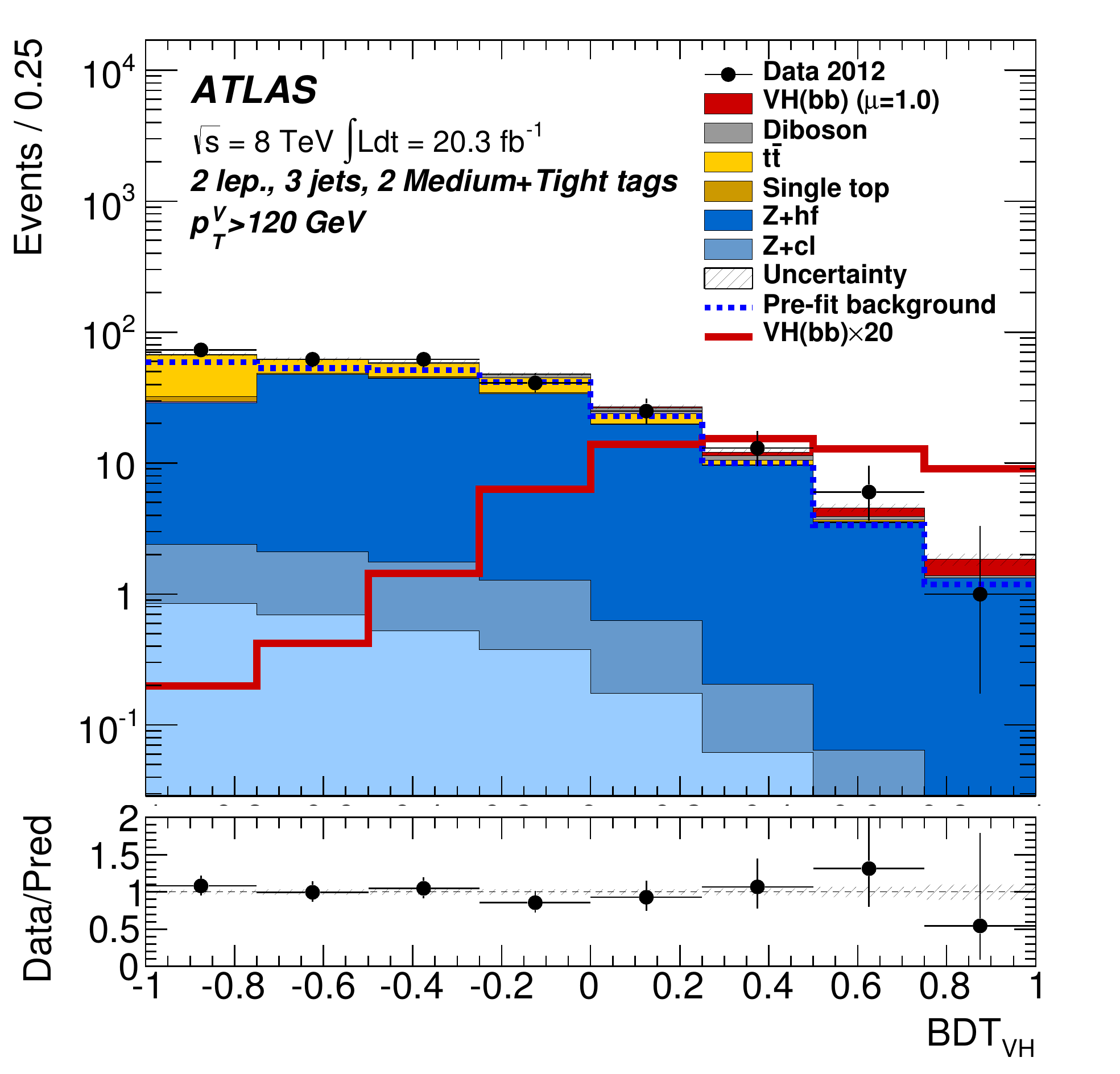}}
}
\caption{
The BDT$_{VH}$-discriminant distribution observed in data (points with error bars) and expected (histograms)
for the 2-lepton channel combining the 2-tag Medium and Tight $b$-tagging categories 
(also referred to as MM and TT in the text) for 
(a) 2-jet events with $\ptz \leq 120$~GeV, 
(b) 2-jet events with $\ptz > 120$~GeV, and 
(c) 3-jet events with $\ptz > 120$~GeV. 
The background contributions after the global fit of the MVA are shown 
as filled histograms.
The Higgs boson signal ($\mh = 125$~GeV) is shown as a filled histogram on top of 
the fitted backgrounds, 
as expected from the SM (indicated as $\mu=1.0$),
and, unstacked as an unfilled histogram, scaled by the factor indicated in the legend. 
The dashed histogram shows the total background as expected from the pre-fit 
MC simulation.
The size of the combined statistical and systematic uncertainty on the
sum of the signal and fitted background is indicated by the hatched band. The ratio
of the data to the sum of the signal and fitted background is shown in the lower panel.
\label{fig:BDT_2tag_2lep}}
\end{sidewaysfigure}

\begin{sidewaysfigure}[tb!]
\begin{center}
\subfigure[]{\includegraphics[width=0.32\textwidth]{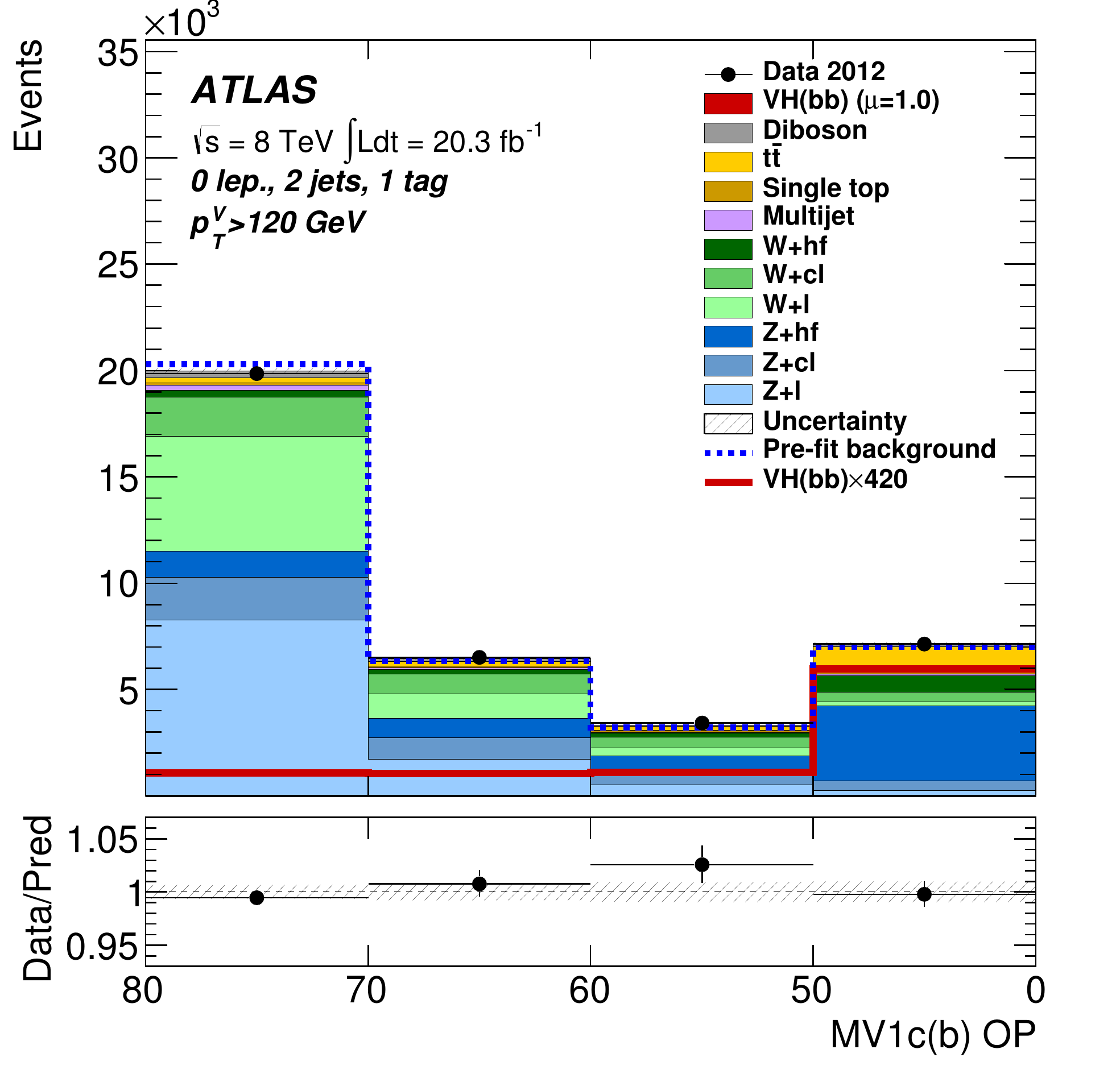}}
\hfill
\subfigure[]{\includegraphics[width=0.32\textwidth]{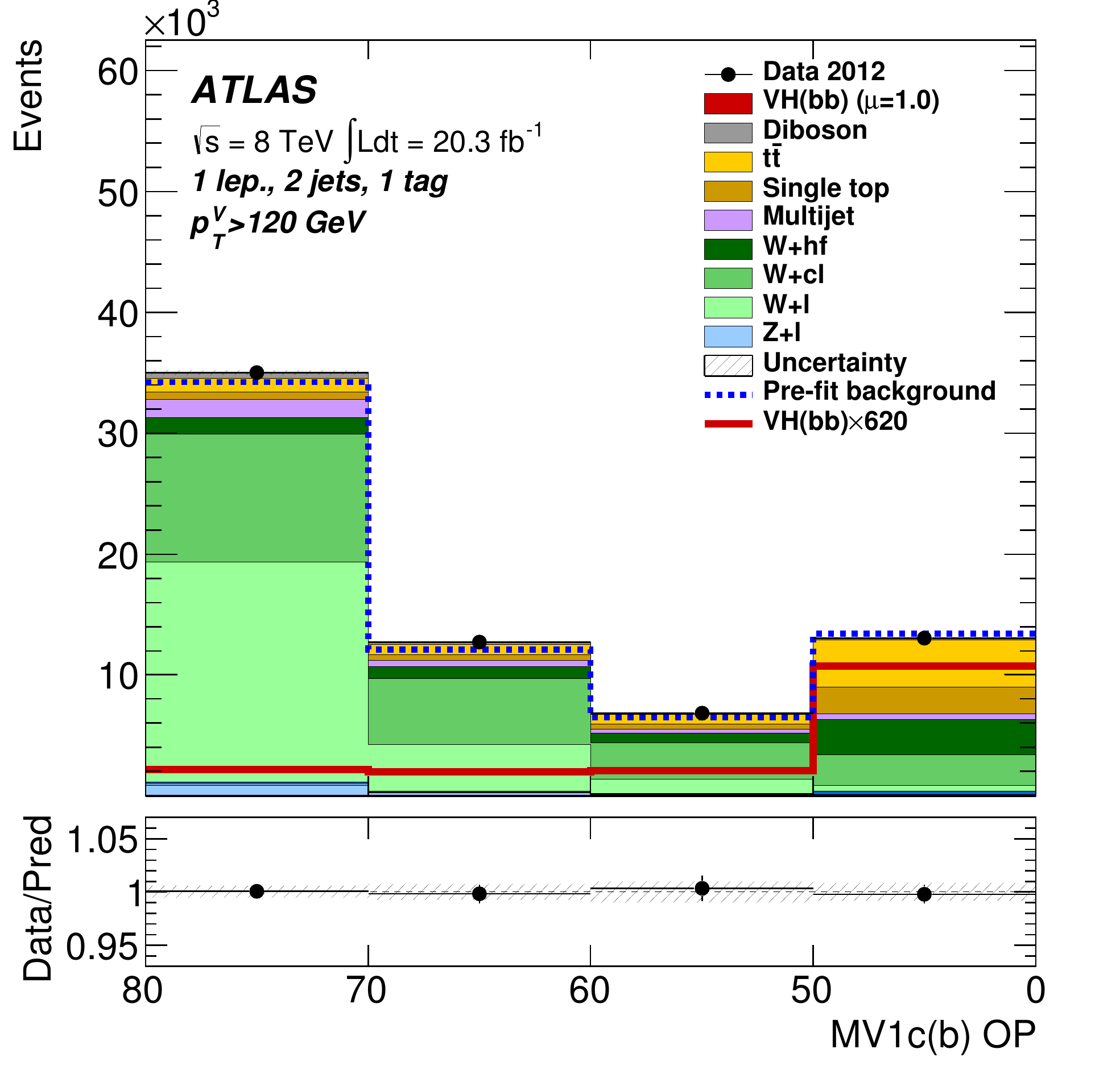}}
\hfill
\subfigure[]{\includegraphics[width=0.32\textwidth]{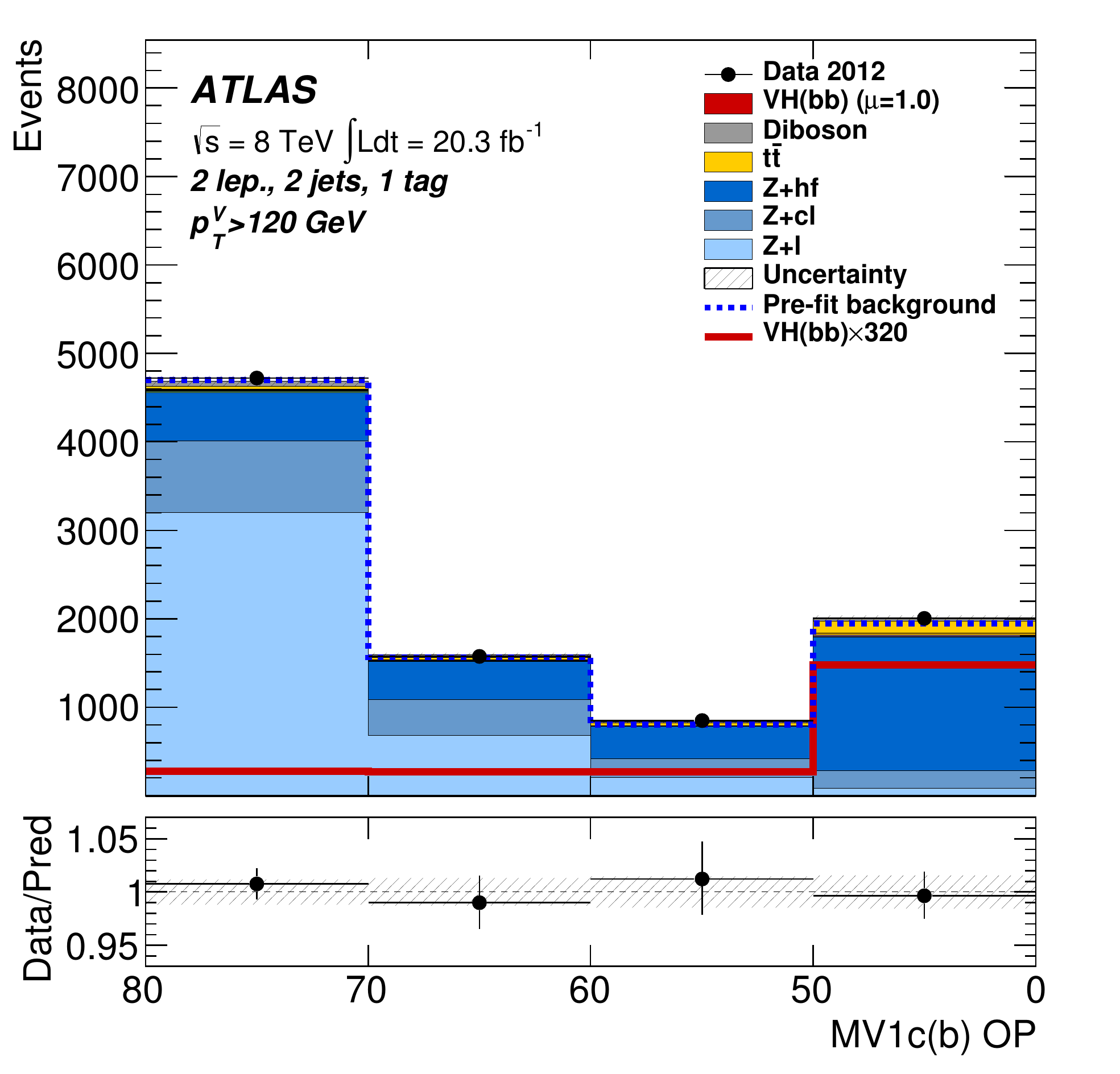}}
\end{center}
\caption{
Distribution of the output of the MV1c $b$-tagging algorithm observed in data
(points with error bars) and expected (histograms) for the 1-tag control regions of the MVA in the 2-jet
category with $\ptv >120$~GeV 
(a) in the 0-lepton channel, (b) in the 1-lepton channel, and (c) in the 2-lepton channel.
The left bin boundaries denote 
the operating points ({\it{MV1c(b) OP}}) as defined in section~\ref{sec:reco},
corresponding to 
$b$-tagging efficiencies of 80\%, 70\%, 60\%, 50\%,
i.e., the $b$-jet purity increases from left to right.      
The background contributions after the global fit of the MVA are shown 
as filled histograms.
The Higgs boson signal ($\mh = 125$~GeV) is shown as a filled histogram on top of 
the fitted backgrounds,  
as expected from the SM (indicated as $\mu=1.0$),
and, unstacked as an unfilled histogram, scaled by the factor indicated in the legend. 
The dashed histogram shows the total background as expected from the pre-fit 
MC simulation.
The size of the combined statistical and systematic uncertainty on the
sum of the signal and fitted background is indicated by the hatched band. The ratio
of the data to the sum of the signal and fitted background is shown in the lower panel.
\label{fig:MV1c_1tag}}
\end{sidewaysfigure}

\FloatBarrier

\section{Systematic uncertainties}
\label{sec:sys}
The systematic uncertainties discussed in this section are: those of
experimental origin; those related to the multijet background estimation;
and those associated with the modelling of the simulated backgrounds
and Higgs boson signal. 

\subsection{Experimental uncertainties}
\label{sec:sysexperimental}

All relevant experimental systematic uncertainties are considered, such
as those affecting the trigger selection, the object reconstruction and 
identification, and the object energy and momentum calibrations and resolutions.
The most relevant ones are discussed in the following.

For the \met\ trigger, an efficiency correction is derived from $W\to\mu\nu$+jets and $Z\to\mu^+\mu^-$+jets events.
This correction amounts to 4.5\% for events with an
\met\ of 100 GeV, the threshold required in the analysis, and is below 1\% for $\met > 120$~GeV.
The associated uncertainties arise from the statistical
uncertainties of this method and differences observed in the two event classes. 
They are very small (below 1\%) for the high \met\ (and thus high \ptv) intervals, 
and reach about 3\% for the low \met\ interval of the 0-lepton channel (100--120~GeV).    

For electrons and muons, uncertainties associated with the corrections 
for the trigger, reconstruction, identification and isolation efficiencies are taken into account. 
Uncertainties on energy and resolution corrections of the leptons are also considered.
The impact of these uncertainties is very small, typically less than 1\%.

Several sources contribute to the uncertainty of the jet energy scale (JES)~\cite{Aad:2014bia}
related e.g. to uncertainties from in situ calibration analyses, pile-up-dependent
corrections and the flavour composition of jets in different event classes. 
After being decomposed into uncorrelated components, these are treated as independent sources in the analysis.
The total relative systematic uncertainties on the JES range from about 3\% to 1\%
for central jets with a \pt\ of 20~GeV and 1~TeV, respectively. 
An additional specific uncertainty of about 1\%--2\% affects the energy
calibration of $b$-jets. Small uncertainties on the corrections applied to improve the 
dijet-mass resolution are also included.
Corrections and uncertainties are also considered for the jet energy resolution (JER)~\cite{Aad:2012ag},
with a separate contribution for $b$-jets.
The total relative systematic uncertainty on the JER ranges from about 10\% to 20\%, 
depending on the $\eta$ range, for jets with $\pt=20$~GeV to less than 5\% for jets 
with $\pt>200$~GeV.

The JES uncertainties are propagated to the \met, 
as are the much smaller uncertainties related to the energy and momentum
calibration of leptons. 
An uncertainty on the \met\ also comes from the uncertainties 
on the energy calibration (8\%) and resolution (2.5\%) of calorimeter energy clusters 
not associated with any reconstructed object~\cite{ATLAS-CONF-2013-082}. 
 
The $b$-tagging efficiencies for the different jet flavours are measured in both data 
and simulation using dedicated event samples~\cite{btagnote2014b,Burmeister:1559721}.
The $b$-tagging efficiencies for simulated jets are corrected 
within intervals between operating points by MC-to-data SFs, 
which depend on the jet kinematics. 
For $b$-jets, the precision is driven by an analysis 
of \ttb\ events in final states containing two leptons. 
The MC-to-data SFs are close to unity,
with uncertainties at the level of 2--3\% over most of the jet \pt\ range, 
reaching 5\% for $\pt=20$~GeV and $8$\% above 200~GeV.
The uncertainties, which depend on \pt\ and on the interval between operating points, 
are decomposed into uncorrelated components and the ten most significant
ones are kept in the analysis. It was checked that the neglected components
have a negligible impact. 
The uncertainties on $c$-jets
are decomposed into 15 components, 
and the uncertainties on light jets, to which the analysis is much less sensitive, 
are decomposed into ten components,
accounted for in \pt\ and $\eta$ ranges.
For $b$- and $c$-jets further uncertainties are added for the application of
the additional MC-to-MC SFs to obtain generator 
specific MC-to-data SFs as explained in section~\ref{sec:reco}. 
Half of the correction is used as systematic uncertainty.
As discussed in section~\ref{sec:reco}, a correction to $c$-jets in the $Vcc$ samples, for which parameterised tagging is used, 
is applied at low $\Delta R$ to the closest jet.
Half of this correction is assigned as a systematic uncertainty.

The uncertainty on the integrated luminosity is 2.8\%.
It is derived, following the same methodology as that described in ref.~\cite{Aad:2013ucp},
from a preliminary calibration of the luminosity scale derived from beam-separation scans
performed in November 2012.  
It is applied to the signal and backgrounds estimates that are taken from simulation. 

A 4\% uncertainty on the average number of interactions per bunch crossing is taken into account.

\subsection{Uncertainties on the multijet backgrounds}

In the 0-lepton channel, the robustness of the MJ background estimation 
is assessed by varying the ${\rm min}[\Delta \phi(\metvec,{\rm jet})]$
values defining the B and D regions of the ABCD method, and by replacing
the $b$-tagging fractions measured in region D by those measured in region B.
A systematic uncertainty of 100\% is assessed for this small ($\sim 1\%$) background,
uncorrelated between 2- and 3-jet, 1- and 2-$b$-tag categories.
The MJ background in the 2-lepton channel is also at the per-cent level, 
and an uncertainty of 100\% is assigned.

In the 1-lepton channel, normalisation uncertainties arise from the statistical uncertainties
of the multijet fits and from uncertainties on the non-MJ background subtractions performed to 
construct the MJ templates. Normalisation uncertainties are also assessed in the LL, MM and TT
categories to cover differences between multijet fits performed inclusively in the 2-tag
regions and in the individual categories. In the 2-jet 2-tag region of the electron sub-channel, 
the overall normalisation uncertainties amount to 11\%, 14\% and 22\% 
in the LL, MM and TT categories, respectively.
In the muon sub-channel, the corresponding uncertainties are about three times larger because 
of the smaller size of the MJ-enriched samples.
 
In the 1-lepton channel, shape uncertainties are assessed in the various regions 
by comparison of evaluations obtained using MJ-enriched samples defined by isolation requirements 
different from those applied in the nominal selections. In the electron sub-channel, 
an alternative template is constructed with a track-based isolation in the 12\% to 50\% interval,
and another alternative template with a calorimeter-based isolation in the 0\% to 4\% interval.  
In the muon sub-channel, the results obtained with the nominal MJ template are compared
with those obtained with tighter or looser isolation requirements, defined by track-based
isolation intervals of 7\%--9.5\% and 9.5\%--50\%, respectively.
Furthermore, half of the $\Delta R(\mathrm{jet}_1,\mathrm{jet}_2)$ and $\ptw$ reweightings 
mentioned in section~\ref{sec:multijet} for the electron sub-channel are taken as systematic uncertainties.

\subsection{Uncertainties on the modelling of the simulated backgrounds}

The physics-modelling systematic uncertainties evaluated focus on the quantities that are used
in the global fit, i.e., those affecting the jet multiplicities, the \ptv\ distributions, 
the flavour composition and the \mbb\ distributions. For the MVA, systematic 
uncertainties affecting the other variables used as inputs to the BDTs are also 
considered. Whenever possible, dedicated control regions are used to extract 
information directly from the data. This is the case for $Z$+jets and $W$+light jets. 
In other cases, uncertainties are assessed by comparison of MC predictions based
on a variety of generators with the nominal ones. 

Details of the assessment of systematic uncertainties are provided below 
in the context of the MVA. 
When systematic uncertainties are derived from a comparison
between generators, all relevant variables are considered independently.
The variable showing the largest discrepancy in some generator 
with respect to the nominal generator is assigned an uncertainty covering this
discrepancy, which is symmetrised.
If, once propagated to the BDT$_{VH}$ discriminant, this uncertainty is 
sufficient to cover all variations observed with the different generators, 
it is considered to be sufficient. If not, an uncertainty is considered 
in addition on the next most discrepant variable and the procedure is iterated
until all variations of the BDT$_{VH}$ discriminant are covered by the assigned 
uncertainties.

A given source of systematic uncertainty can affect different analysis
regions. Whether such an uncertainty should be treated as correlated
or not depends on whether constraints resulting from the global fit
should be propagated from one region to another. Details of the
procedures leading to such decisions are provided in section~\ref{sec:techfit}.

A summary of the systematic uncertainties affecting the modelling of the backgrounds 
can be found in table~\ref{tab:systematics}.

{\bf Top-quark-pair background:} 
As explained in section~\ref{sec:bkg}, the top-quark \pt\ distribution
is reweighted at generator level to bring it into agreement with measurement~\cite{Aad:2014zka}. 
A systematic uncertainty amounting to half of this correction is assigned, correlated 
across channels.
 
The predictions of the nominal \ttb\ generator ({\sc powheg}+{\sc pythia}) are compared, 
focussing on the 1-lepton channel selection, with those obtained using a variety of generators 
differing 
by the PDF choice ({\sc powheg}+{\sc pythia} with HERAPDF~\cite{Aaron:2009aa}),
by the parton showering and hadronisation scheme ({\sc powheg}+{\sc herwig}),
by the implementation of the NLO matrix element and the matching scheme 
({\sc mc@nlo}~\cite{Frixione:2002ik}+{\sc herwig}), 
by the amount of initial- and final-state radiation (ISR/FSR) using
{\sc AcerMC}+{\sc pythia}, 
or by the implementation of higher-order tree-level matrix elements 
({\sc alpgen}~\cite{alpgen}+{\sc pythia}).
It is found that, in general, the largest deviations are observed for {\sc alpgen}, 
which is therefore used to assess further systematic uncertainties as explained below. 

In the global fit, the normalisation of the \ttb\ background in the 2-jet category is left
floating freely, independently in each of the lepton channels.
An uncertainty of 20\% on the 3-to-2-jet ratio is estimated from the generator comparisons explained 
above.
In the global fit, this uncertainty is treated as correlated between the 0- and 1-lepton channels, and 
uncorrelated with the 2-lepton channel. 

The shape of the \mbb\ distribution is also studied with the same set of generators, 
leading to
correlated shape uncertainties for 2- and 3-jet events, 
and for $\ptv < 120$~GeV and $\ptv > 120$~GeV. The associated variation is larger in the higher \ptv\
interval: for 2-jet events, when it increases the 
distribution by 3\% for $\mbb = 50$~GeV, it decreases it 
by 1\% at 200~GeV; the effect is similar, but of opposite sign, for 3-jet events. 

The same procedure is used for the \ptv\ distribution, from which a 7.5\% 
uncertainty is assessed on the normalisation of the $\ptv > 120$~GeV interval.
Finally, the same approach calls for a shape systematic uncertainty on the \met\ 
distribution in the 1-lepton channel, different but correlated between $\ptv < 120$~GeV
and $\ptv > 120$~GeV. This uncertainty is not applied in the 0- and 2-lepton channels.

{\bf Single-top-quark background:}
The theoretical uncertainties on the cross sections of the three processes contributing
to single-top production are 4\%, 4\%, and 7\% for the $s$-channel, $t$-channel,
and $Wt$ production, respectively~\cite{Kidonakis:2012db}. 
 
The predictions of the nominal generators ({\sc powheg}+{\sc pythia} for the $s$-channel
and for $Wt$ production; {\sc acerMC}+{\sc pythia} for the $t$-channel) are compared, 
after the 1-lepton channel selection, with those obtained using a variety of generators. 
For the $s$-channel, the comparison is made with {\sc acerMC} and {\sc mc@nlo}; 
for $Wt$ production with {\sc acerMC}, {\sc powheg}+{\sc herwig}, and {\sc mc@nlo}; 
and for the $t$-channel with a{\sc mc@nlo}\footnote {Event generation with a{\sc mc@nlo} 
is based on the {\sc mc@nlo} formalism and the {\sc madgraph-5} framework~\cite{Alwall:2011uj,Alwall:2014hca}.}
\cite{madloop,Frederix:2011qg}+{\sc herwig}.
For all three processes, the impact of ISR/FSR is evaluated using {\sc acerMC}.
For $Wt$ production, there are interference effects with \ttb\ production, which 
need to be considered. 
Two methods are available for this:
the Diagram Removal (DR) and the Diagram Subtraction (DS) schemes~\cite{Frixione:2008yi}.
The former is used in the nominal generation, and the second for comparison.

Uncertainties on the acceptance for each of the three processes are taken as the largest
deviations observed, separately for $\ptv < 120$~GeV and $\ptv > 120$~GeV, and for 2- and 3-jet events. 
They can be as large as 52\% for 2-jet events in the $t$-channel at low \ptv,
of the order of 5\% for $Wt$ production (except for 3-jet events at high \ptv: 15\%),
and typically 20\% for the $s$-channel.

In addition to the acceptance uncertainties, the effects of the model
variations described above on variables input to the BDT are evaluated
and three shape systematic uncertainties are found to be needed in
$Wt$ production. 
The first uncertainty is on the shape of the \mbb\ distribution in the
high \ptv\ interval for 2-jet events where, when a shift from the nominal
model increases the rate by 20\% for $\mbb = 50$~GeV, it decreases it
by 40\% at 200~GeV. 
A second uncertainty is on the \mbb\ shape for 3-jet events, where the
corresponding shifts are 25\% and 20\%. Finally, a third uncertainty
is on the \pt\ distribution of the leading jet in the
low \ptv\ interval for 2-jet events.

{\bf $Z$+jets background:}
As explained in section~\ref{sec:bkg}, \dphi\ and \ptz\ reweightings are applied to the
$Zl$ and $Zc+Zb$ components, respectively. 
For the \dphi\ reweighting, a systematic uncertainty amounting to half of the correction 
is assigned to the $Zl$ component, while an uncertainty amounting to the full correction 
is assigned to the $Zc+Zb$ components. This is done separately for 2- and 3-jet events, 
and all these uncertainties are treated as uncorrelated. 
For the \ptz\ reweighting, uncorrelated systematic uncertainties of half the correction are 
assigned to the $Zl$ and $Zc+Zb$ components. The notation $Zc+Zb$ is meant to indicate
that a systematic uncertainty is treated as correlated between the $Zc$ and $Zb$ components.

The normalisation and the 3-to-2-jet ratio for the $Zl$ background are determined 
from data in the 0-tag region of the 2-lepton channel, both with an uncertainty of 5\%.
The normalisations of the $Zcl$ and $Zbb$ backgrounds are left free in the global fit.
The uncertainties on the 3-to-2-jet ratios for the $Zcl$ and $Z$+hf 
components are
assessed through a comparison of {\sc alpgen} with the nominal {\sc sherpa} generator 
in the 2-tag region of the 2-lepton channel; these are 26\% for $Zcl$ and 20\% for $Z$+hf. The same
procedure is used to estimate uncertainties on the flavour fractions within $Z$+hf events, yielding
12\% for each of $bl$/$bb$, $cc$/$bb$ and $bc$/$bb$, 
with $bl$/$bb$ uncorrelated between 2- and 3-jet samples.

The shape of the \mbb\ distribution is compared between data and simulation in the 2-tag
region of the 2-lepton channel, excluding the 100--150~GeV range, from which a shape 
uncertainty is derived that, when it increases the dijet-mass distribution by 3\% 
at 50~GeV, it decreases it by 5\% at 200~GeV. This uncertainty is applied
uncorrelated to the $Zl$ and $Zb+Zc$ components. The differences between
{\sc alpgen} and {\sc sherpa} are covered by this uncertainty. 

{\bf $W$+jets background:}
As explained in section~\ref{sec:bkg}, a \dphi\ reweighting is applied to the $Wl$ and $Wcl$
components. Uncorrelated systematic uncertainties amounting to half of the correction are assigned 
to these two components, for each of the 2- and 3-jet categories. For the $Wcc+Wb$ component, 
no reweighting is applied but a systematic uncertainty is assigned, equal to the full correction 
applied to the $Wl$ and $Wcl$ components, uncorrelated between 2- and 3-jet events. 

The normalisation and the 3-to-2-jet ratio for the $Wl$ background are taken directly from simulation,
both with a 10\% uncertainty. This is based on the agreement observed between data and prediction 
in the 0-tag sample. The 3-to-2-jet ratio for the $Wcl$ background is also assigned an uncertainty 
of 10\%. The normalisations of the $Wcl$ and $Wbb$ backgrounds are left free in the global fit. 

To assign further uncertainties on the $Wbb$ background, for which dedicated control regions 
are not available in the data, extensive comparisons are performed at generator level, with 
kinematic selections mimicking those applied after reconstruction. The predictions
of the {\sc sherpa} generator are compared to those of {\sc powheg}+{\sc pythia8}, of 
a{\sc mc@nlo}+{\sc herwig++}~\cite{Arnold:2012fq} and of {\sc alpgen}+{\sc herwig}. 
Comparisons are also made between samples generated with a{\sc mc@nlo} with renormalisation 
($\mu_{\mathrm{R}}$) and factorisation ($\mu_{\mathrm{F}}$) scales\footnote{The nominal scales are taken as 
$\mu_{\mathrm{R}}=\mu_{\mathrm{F}}=[m_W^2+p_{\mathrm T}(W)^2+m_b^2+(p_{\mathrm T}(b)^2+p_{\mathrm T}(\overline{b})^2)/2]^{1/2}$.} 
independently modified by factors of 2 or 0.5 and also with different PDF sets 
(CT10, MSTW2008NLO and NNPDF2.3~\cite{Ball:2012cx}). 
As a result, a 10\% uncertainty is assigned to the 3-to-2-jet ratio, taken as correlated between all 
$W$+hf processes. Shape uncertainties are also assessed for the \mbb\ and \ptw\ distributions.
When the former increases the dijet-mass distribution by 23\% at 50~GeV, it decreases it by 28\% at 200~GeV. 
It is taken as uncorrelated for $Wl$, $Wcl$, $Wbb+Wcc$ and $Wbl+Wbc$. 
For $Wbb+Wcc$, it is furthermore uncorrelated among \ptw\ intervals 
(with the three highest intervals correlated for the dijet-mass analysis).
When the latter shape uncertainty increases the \ptw\ 
distribution by 9\% for $\ptw = 50$~GeV, it decreases it by 23\% at 200~GeV. It is taken
as correlated for all $W$+hf processes, and uncorrelated between the 2- and 3-jet samples.

Predictions using the inclusive production of all flavours by {\sc sherpa} and 
{\sc alpgen} 
are compared after full reconstruction and event selection to assign uncertainties on the 
flavour fractions that take properly into account heavy-flavour production at both the matrix-element 
and parton-shower levels. 
(For {\sc alpgen}, the production of light flavours and heavy flavours are
performed separately at the matrix-element level; a dedicated procedure, based on the $\Delta R$
separation between $b$-partons, is used to remove the overlap between \bb\ pairs produced at 
the matrix-element and parton-shower levels.)
The following uncertainties are assigned in the $W$+hf 
samples: 35\% for $bl/bb$ and 12\% for each of $bc/bb$ and $cc/bb$. 
The uncertainty on $bl/bb$ is uncorrelated between \ptw\ intervals 
(with the three highest intervals correlated for the dijet-mass analysis).

{\bf Diboson background:}
The uncertainties on the cross sections for diboson production ($WW$, $WZ$ and $ZZ$) are assessed
at parton level using {\sc mcfm} at NLO in QCD. 
The sources of uncertainty considered are the renormalisation 
and factorisation scales and the choice of PDFs. The nominal scales are
dynamically set to half of the invariant mass of the diboson system and the nominal PDFs are 
the CT10 set. 

The scale uncertainties are evaluated by varying simultaneously $\mu_{\mathrm{R}}$ and $\mu_{\mathrm{F}}$ by 
factors of 2 or 0.5.
Since the analysis is performed in \ptv\ intervals and in exclusive 2- and 3-jet categories, 
the uncertainties are evaluated for each channel separately in those intervals and categories 
(2 and 3 final-state partons within the nominal selection acceptance) following the 
prescription of ref.~\cite{Stewart:2011cf}. This procedure leads, in each \ptv\ interval, to two 
uncorrelated uncertainties in the 2-jet category, one for 2+3 jets inclusively and one 
associated with the removal of 3-jet events, and to one in the 3-jet category anti-correlated
with the latter uncertainty in the 2-jet category. These
uncertainties are largest at high \ptv. For $\ptv > 200$~GeV, the two uncertainties affecting
the 2-jet category can be as large as 29\% and 22\% in the $WZ$ channel, roughly half this size
in the $ZZ$ channel and intermediate for $WW$; and the uncertainty affecting the 3-jet
category is about 17\% in all channels. 

The uncertainties due to the PDF choice are evaluated according to the PDF4LHC 
recommendation~\cite{Botje:2011sn}, i.e., using the envelope of predictions from
the CT10, MSTW2008NLO, and NNPDF2.3 PDF sets and their associated 
uncertainties. They range from 2\% to 4\%, with no \ptv\ dependence observed.

The shape of the reconstructed $Z\to\bb$ lineshape in $VZ$ production is affected by the parton-shower
and hadronisation model. A shape-only systematic uncertainty is assessed by comparing the 
lineshapes obtained with the nominal {\sc powheg}+{\sc pythia8} generator and with {\sc herwig}. 
The relative difference between the shapes is 20\% for a dijet mass around 125~GeV. 

\subsection{Uncertainties on the signal modelling}

The $\qqb\to WH$, $\qqb\to ZH$, and $gg\to ZH$ signal samples are normalised respectively to their inclusive cross sections as explained in section~\ref{sec:samples}.
The uncertainties on these cross sections~\cite{LHCHiggsCrossSectionWorkingGroup:2011ti} 
include those arising from the choice of scales $\mu_{\mathrm{R}}$ and $\mu_{\mathrm{F}}$ and of PDFs. 

The scale uncertainty is 1\% for $WH$ production. It is larger (3\%) for $ZH$ production, 
due to the contribution of the gluon-gluon initiated process.
Under the assumption that the scale uncertainties are similar (1\%) for $\qqb\to WH$
and $\qqb\to ZH$, a conservative uncertainty of 50\% is inferred for $gg\to ZH$.
The same procedure 
leads to PDF uncertainties of 2.4\% for $\qqb\to (W/Z)H$ and 17\% for $gg\to ZH$.
The relative uncertainty on the Higgs boson branching ratio to \bb\ 
is 3.3\% for $m_H=125$~GeV~\cite{Djouadi:1997yw}. 
The contribution of decays to final states other than \bb\ is verified to amount 
to less than 1\% after selection.

Acceptance uncertainties due to the choice of scales are determined from signal samples 
generated with {\sc powheg} interfaced to {\sc pythia8}, with $\mu_{\mathrm{R}}$ and $\mu_{\mathrm{F}}$ varied 
independently by factors of 2 or 0.5. The 
procedure advocated in ref.~\cite{Stewart:2011cf} is used, 
after kinematic selections applied at generator level, leading
to acceptance uncertainties of 3.0\%, 3.4\% and 1.5\% for $\qqb\to WH$, $\qqb\to ZH$ and 
$gg\to ZH$, respectively, for the 2- and 3-jet categories combined, and of 4.2\%, 3.6\% and 3.3\% 
for the 3-jet category. The latter
uncertainty is anti-correlated with an acceptance uncertainty associated with the
removal of 3-jet events from the 2+3-jet category to form the 2-jet category.
In addition, the \ptv\ spectrum is seen to be affected, and shape uncertainties are derived. 
For the $\qqb\to (W/Z)H$ samples, 
when they increase the distribution by 1\% for $\ptv = 50$~GeV, they decrease 
it by 3\% at 200~GeV. These variations are 2\% and 8\%, respectively, for the 
$gg\to ZH$ samples.

Acceptance uncertainties due to the PDF choice are determined in a similar way, following
the PDF4LHC prescription.
They range from 2\% in the 2-jet $gg\to ZH$
samples to 5\% in the 3-jet $\qqb\to ZH$ samples. There is no evidence of a need for
\ptv\ shape uncertainties related to the PDFs.

The applied uncertainties on the shape of the \ptv\ spectrum associated with the NLO electroweak 
corrections~\cite{Denner:2011rn} are typically at the level of 2\%, increasing with \ptv\ 
to reach 2.5\% in the highest \ptv\ interval. 

The effect of the underlying-event modelling is found to be negligible, using various {\sc pythia} 
tunes. The effect of the parton-shower modelling is examined by comparison of simulations
by {\sc powheg} interfaced with {\sc pythia8} and with {\sc herwig}. Acceptance
variations of 8\% are seen, except for 3-jet events in the $\ptv > 120$~GeV interval, 
where the variation is at the level of 13\%. These variations are taken as systematic uncertainties. 
  
A summary of the systematic uncertainties affecting the modelling 
of the Higgs boson signal is given in table~\ref{tab:systematics}.

\begin{table}[b!]
\begin{center}
\begin{tabular}{ l | c}
	\hline\hline
	\multicolumn{2}{c}{Signal}\\
	\hline
	Cross section (scale) & 1\% (\qqb), 50\% ($gg$) \\
	Cross section (PDF) & 2.4\% (\qqb), 17\% ($gg$) \\
	Branching ratio & 3.3 \% \\
	Acceptance (scale) & 1.5\%--3.3\% \\
	3-jet acceptance (scale) & 3.3\%--4.2\% \\
        \ptv\ shape (scale) & S \\
        Acceptance (PDF) & 2\%--5\% \\
	\ptv\ shape (NLO EW correction) & S \\
	Acceptance (parton shower) & 8\%--13\% \\
	\hline\hline
	\multicolumn{2}{c}{$Z$+jets}\\
	\hline
	$Zl$ normalisation, 3/2-jet ratio & 5\% \\
	$Zcl$ 3/2-jet ratio & 26\% \\
	$Z$+hf 3/2-jet ratio & 20\% \\
	$Z$+hf/$Zbb$ ratio & 12\% \\
	\dphi, \ptv, \mbb  & S \\
	\hline\hline
	\multicolumn{2}{c}{$W$+jets}\\
	\hline
	$Wl$ normalisation, 3/2-jet ratio & 10\% \\
	$Wcl$, $W$+hf 3/2-jet ratio & 10\% \\
	$Wbl$/$Wbb$ ratio & 35\% \\
	$Wbc$/$Wbb$, $Wcc$/$Wbb$ ratio & 12\% \\
	\dphi, \ptv, \mbb  & S \\
	\hline\hline
	\multicolumn{2}{c}{\ttb}\\
	\hline
	3/2-jet ratio & 20\% \\
	High/low-\ptv\ ratio & 7.5\% \\
	Top-quark \pt, \mbb, \met & S	\\
	\hline\hline
	\multicolumn{2}{c}{Single top}\\	
	\hline
	Cross section & 4\% ($s$-,$t$-channel), 7\% ($Wt$) \\
	Acceptance (generator) & 3\%--52\% \\
        \mbb, $\pt^{b_1}$ & S \\
	\hline\hline
	\multicolumn{2}{c}{Diboson}\\	
	\hline
        Cross section and acceptance (scale) & 3\%--29\% \\
	Cross section and acceptance (PDF) & 2\%--4\% \\
       	\mbb & S \\
	\hline\hline
	\multicolumn{2}{c}{Multijet}\\	
	\hline
	0-, 2-lepton channels normalisation & 100\% \\
	1-lepton channel normalisation & 2\%--60\% \\
	Template variations, reweighting & S \\
	\hline\hline
\end{tabular}
	\caption
	{Summary of the systematic uncertainties on the signal and background modelling. 
         An ``S'' symbol is used when only a shape uncertainty is assessed.
	\label{tab:systematics}}
	\end{center}
	\end{table}

\FloatBarrier

\section{Statistical procedure}
\label{sec:fit}

\subsection{General aspects}
A statistical fitting procedure 
based on the Roostats framework~\cite{Moneta:2010pm,Verkerke:2003ir} 
is used to extract the signal strength from the data.
The signal strength is a parameter, $\mu$, that multiplies the SM
Higgs boson production cross section times branching ratio into \bb.
A binned likelihood function is constructed as the product of
Poisson-probability terms over the bins of the input distributions
involving the numbers of data events and the expected
signal and background yields, taking into account the effects of the floating 
background normalisations and the systematic uncertainties.

The different regions entering the likelihood fit are summarised in table~\ref{tab:fitmodel}.
In the dijet-mass analysis, the inputs to the 
``global fit'' are
the \mbb\ distributions in the 
81 2-tag signal regions 
defined by three channels (0, 1 or 2 leptons), up to five \ptv\ intervals, 
two number-of-jet categories (2 or 3), and three $b$-tagging categories (LL, MM and TT). 
Here and in the rest of this section, 
\mbb\ distributions 
are to be understood as transformed distributions, as explained in section~\ref{sec:mva}.
In the MVA, the inputs are the BDT$_{VH}$ discriminants in the
24 2-tag signal regions 
defined by the three lepton channels, up to two \ptv\ intervals, the two number-of-jet categories,
and $b$-tagging categories. In the 1-lepton channel, the $b$-tagging categories are LL, MM and TT.
In the 0- and 2-lepton channels, they are the LL category and a combined MM and TT category 
(MM+TT).\footnote{While keeping distinct MM and TT categories in the 1-lepton channel improves 
the sensitivity, this is not observed for the 0- and 2-lepton channels. Keeping the LL category 
separated from the others improves the sensitivity in all lepton channels.}
These BDT$_{VH}$-discriminant distributions are 
supplemented by the three \mbb\ distributions in the 100--120~GeV \ptv\ interval 
of the 2-jet 2-tag categories (LL, MM, and TT) of the 0-lepton channel.
For the MVA, additional inputs are the $MV1c$ distributions of the $b$-tagged jet in the 
11 1-tag control regions of the MVA selection and in
the 100--120~GeV \ptv\ interval of the 2-jet 1-tag category of the 0-lepton channel.
In the dijet-mass analysis, the $MV1c$ distributions are combined in each of the 
$\ptv < 120$~GeV and $\ptv > 120$~GeV intervals, which also results in 11 1-tag control regions. 
Altogether, there are 584 \mbb\ and $MV1c$ bins in the 92 regions of the dijet-mass analysis, 
and 251 BDT$_{VH}$-discriminant and $MV1c$ bins in the 38 regions of the MVA, 
to be used in the global fits.

\begin{table}[tb!]
\begin{center}
\begin{tabular}{  l|l | c | c | c | c | c | c }
\hline\hline
   \multicolumn{2}{c|}{} & \multicolumn{3}{|c|}{Dijet-mass analysis} &  \multicolumn{3}{|c}{MVA} \\
\hline
 \multicolumn{2}{c|}{Channel} & 0-lepton & 1-lepton & 2-lepton & 0-lepton & 1-lepton & 2-lepton \\
\hline\hline
   \multicolumn{2}{c|}{1-tag}	 &\multicolumn{3}{c|}{$MV1c$} &\multicolumn{3}{c}{$MV1c$} \\
\hline
LL &  \multirow{3}{*}{2-tag} &  \multicolumn{3}{c|}{\mbb} &  BDT$^{(\ast)}$ & \multicolumn{2}{c}{BDT} \\
\cline{1-1}\cline{3-8}

MM &   			 &  \multicolumn{3}{c|}{\mbb} & \multirow{2}{*}{BDT$^{(\ast)}$} & BDT & \multirow{2}{*}{BDT} \\
\cline{1-1}\cline{3-5}\cline{7-7}
TT &  			 &  \multicolumn{3}{c|}{\mbb} &  & BDT & \\ 
\hline\hline
\end{tabular}
\caption[Regions used in profile likelihood]
{The distributions used in each region by the likelihood fit in the dijet-mass analysis 
and in the MVA applied to the 8~TeV data. Here, ``BDT'' stands for ``BDT$_{VH}$ discriminant''.
For each entry listed, there are additional divisions into \ptv\ intervals: five in the
dijet-mass analysis and two in the MVA, as shown in table~\ref{tab:selKinEvt}.
These distributions are input to the fit for the 2-jet and 3-jet categories separately,
except in the low \ptv\ interval (100--120~GeV) of the 0-lepton channel where only
the 2-jet category is used.
In the 0- and 2-lepton channels, the MM and TT 2-tag categories are combined in the MVA.
$(\ast)$ In the low \ptv\ interval of the 0-lepton channel, the MVA uses 
the \mbb\ distributions in the LL, MM and TT 2-tag categories 
as well as the $MV1c$ distribution in the 1-tag category. 
\label{tab:fitmodel}}
\end{center}
\end{table}

The impact of systematic uncertainties on the signal and background 
expectations is described by nuisance parameters (NPs), $\vec{\theta}$, which
are constrained by Gaussian or log-normal probability density functions, the latter being 
used for normalisation uncertainties to prevent
normalisation factors from becoming negative in the fit.
The expected
numbers of signal and background events in each bin are functions of $\vec{\theta}$.
The parameterisation of each NP is chosen such that the predicted signal and background
yields in each bin are log-normally distributed for a normally distributed $\theta$.
For each NP, the prior is added as a penalty term to the likelihood, 
$\mathcal{L} (\mu,\vec{\theta})$, which decreases it
as soon as $\theta$ is shifted away from its nominal value. 
The statistical uncertainties of background predictions from simulation 
are included through bin-by-bin nuisance parameters.

The test statistic $q_\mu$ is then constructed from 
the profile likelihood ratio 
$$q_\mu = - 2 \ln \Lambda_\mu \mathrm{~~with~~} \Lambda_\mu =
\mathcal{L} (\mu,\hat{\hat{\vec{\theta}}}_\mu)/\mathcal{L} (\hat{\mu},\hat{\vec{\theta}}),$$ 
where $\hat{\mu}$ and $\hat{\vec{\theta}}$ are the parameters that maximise the
likelihood with the constraint $0 \leq \hat{\mu} \leq \mu$, and
$\hat{\hat{\vec{\theta}}}_\mu$ are the nuisance parameter values that maximise the
likelihood for a given $\mu$.
This test statistic is used for exclusion intervals derived with the $CL_s$
method~\cite{Read:2002hq,Cowan:2010js}. 
To measure the compatibility of the background-only hypothesis with the observed
data, the test statistic used is 
$q_0 = -2\ln\Lambda_0$.
The results are presented in terms of: the 95\% confidence level (CL) 
upper limit on the signal strength; the probability $p_0$ of the background-only hypothesis;
and the best-fit signal-strength value $\hat\mu$ with its associated uncertainty $\sigma_\mu$.
The fitted $\hat\mu$ value is obtained  by maximising the likelihood function with respect to all parameters. 
The uncertainty $\sigma_\mu$ is obtained from the variation of $2 \ln \Lambda_\mu$ by one unit,
where $\Lambda_\mu$ is now defined without the constraint $0 \leq \hat{\mu} \leq \mu$. 
Expected results are obtained in the same way as the observed results by replacing the data
in each input bin by the expectation from simulation with all NPs set to their best-fit values,
as obtained from the fit to the data.\footnote{This type of pseudo-data
sample is referred to as an Asimov dataset in ref.~\cite{Cowan:2010js}.} 

While the analysis is optimised for a Higgs boson of mass 125~GeV, results are also extracted
for other masses. These are obtained without any change to the dijet-mass analysis,
except for the binning of the transformed \mbb\ distribution, which is reoptimised. For the
MVA, it is observed that the performance degrades for masses away from 125~GeV, for which
the BDTs are trained. This is largely due to the fact that \mbb\ is an input to the BDTs.
The MVA results for other masses are therefore obtained using BDTs
retrained for each of the masses tested at 5~GeV intervals between 100 and 150~GeV.
The details provided in the rest of this section refer to the analysis performed 
for a Higgs boson mass of 125~GeV.  

\subsection{Technical details}
\label{sec:techfit}

The data have sufficient statistical power to constrain the largest background-normali\-sation NPs, which are
left free to float in the fit. This applies to the \ttb, $Wbb$, $Wcl$, $Zbb$ and $Zcl$ processes.
The corresponding factors applied to the nominal background normalisations 
as resulting from the global fit of the MVA to the 8~TeV data, are shown in table~\ref{tab:resultscalefactorsMVA}.
As stated in section~\ref{sec:sys}, 
the \ttb\ background is normalised in the 2-jet category independently in each of the lepton channels.
The reason for uncorrelating the normalisations in the three lepton channels 
is that the regions of phase space probed
in the 2-jet category are very different between the three channels. In the 2-lepton channel, the \ttb\
background is almost entirely due to events in which both top quarks decay into $(W\to\ell\nu)b$ 
(fully leptonic decays) with all final-state objects detected (apart from the neutrinos). 
In the 1-lepton channel, it is in part due to fully leptonic decays with one of the leptons 
(often a $\tau$ lepton) undetected, and in part to cases where one of the top quarks decays as above
and the other into $(W\to q\overline{q}')b$ (semileptonic decays) with a missed light-quark jet. Finally,
in the 0-lepton channel, the main contributions are from fully leptonic decays with the two leptons 
undetected and from semileptonic decays with a missed lepton and a missed light-quark jet; here again,
the missed leptons are often $\tau$ leptons.
Futhermore, the \ptv\ range probed is different in the 0-lepton channel: 
$\ptv>100$~GeV in contrast to being inclusive in the 1- and 2-lepton channels. 

\begin{table}[tb!]
\begin{center}
\begin{tabular}{l|c}
\hline\hline
Process & Scale factor \\
\hline
\ttb\ 0-lepton  & $1.36 \pm 0.14$ \\ 
\ttb\ 1-lepton  & $1.12 \pm 0.09$ \\ 
\ttb\ 2-lepton  & $0.99 \pm 0.04$ \\ 
$Wbb$ &  $0.83 \pm 0.15$ \\ 
$Wcl$ &  $1.14 \pm 0.10$ \\
$Zbb$ &  $1.09 \pm 0.05$ \\ 
$Zcl$ &  $0.88 \pm 0.12$ \\
\hline\hline
\end{tabular}
\end{center}
\caption{Factors applied to 
the nominal normalisations of the \ttb, $Wbb$, $Wcl$, $Zbb$, and $Zcl$ backgrounds,
as obtained from the global MVA fit to the 8~TeV data. The
\ttb\ background is normalised in the 2-jet category independently in each of the lepton channels.
The errors include the statistical and systematic uncertainties. 
\label{tab:resultscalefactorsMVA}}
\end{table}

As described in detail in section~\ref{sec:sys}, a large number of sources 
of systematic uncertainty are considered. The number of nuisance parameters 
is even larger because care is taken to appropriately uncorrelate the impact of the same source 
of systematic uncertainty across background processes or across regions 
accessing very different parts of phase space. This avoids unduly propagating constraints. 
For instance, the \ttb\ background contributes quite differently in the 2-tag 3-jet regions 
of the 0- and 1-lepton channels on one side, and of the 2-lepton channel on the other. 
In the 0- and 1-lepton channels, it is likely that a jet from a $t\to b(W\to\qqb)$ decay is missed, 
while in the 2-lepton channel it is likely that an ISR or FSR jet is selected. 
This is the reason for not correlating,
between these two sets of lepton channels, the systematic uncertainty attached to the 3-to-2 jet 
ratio for the \ttb\ background. Another example is the $\Delta\phi$
reweighting in the $W$+jets processes, which is derived in the 0-tag sample and applied to the 
$Wcl$ and $Wl$ backgrounds. As explained in section~\ref{sec:bkg}, this reweighting is not 
applied to the $Wcc$ and $Wb$ backgrounds but, in the 
absence of further information, an uncertainty is assessed for the $\Delta\phi$ 
distributions of the $Wcc$ and $Wb$ backgrounds, uncorrelated with the uncertainty applied
to the $Wcl$ and $Wl$ backgrounds.
Altogether, the fit has to handle almost 170 NPs, with roughly half of those being of
experimental origin.

The fit uses templates constructed from the predicted yields for the signal and the 
various backgrounds in the bins of the input 
distribution in each region.
The systematic uncertainties are encoded in templates of variations relative to the nominal template for each up-and-down ($\pm 1\sigma$) variation.
The limited size of the MC samples for some simulated background processes in some regions can cause large local fluctuations in templates of systematic variations.
When the impact of a systematic variation translates into a reweighting of the nominal 
template, no statistical fluctuations are expected beyond those already present in the
nominal template. This is the case, for instance, for the $b$-tagging uncertainties. For those, no specific action is taken.
On the other hand, when a systematic variation may introduce changes in the events selected, 
as is the case for instance with the JES uncertainties, additional statistical fluctuations
may be introduced, which affect the templates of systematic variations.
In such cases, a smoothing procedure is applied to each systematic-variation template
in each region. Bins are merged based on the constraints that the statistical uncertainty 
in each bin should be less than 5\% and that the shapes of the systematic-variation templates 
remain physical: monotonous for a BDT$_{VH}$ discriminant, and with at most one local extremum for a dijet mass. 

Altogether, given the number of regions and NPs, the number of systematic-variation template pairs 
($+1\sigma$ and $-1\sigma$) is close to twenty thousand, 
which renders the fits highly time consuming.
To address this issue, systematic uncertainties that have a negligible impact on the final 
results are pruned away, region by region.
A normalisation (shape) uncertainty is dropped if the associated template variation is below 0.5\%
(below 0.5\% in all bins). Additional pruning criteria are applied to regions where 
the signal contribution is less than 2\% of the total background and where the systematic 
variations impact the total background prediction by less than 0.5\%.
Furthermore, shape uncertainties are dropped if the up- and down-varied shapes are more 
similar to each other than to the nominal shape.
This is only done
for those systematic uncertainties where opposite-sign variations are expected.
This procedure reduces the number of systematic-variation templates by a factor of two.

The behaviour of the global fit is evaluated by a number of checks, including
how much each NP is pulled away from its nominal value,
how much its uncertainty is reduced with respect to its nominal uncertainty, and which
correlations develop between initially uncorrelated systematic uncertainties.
To assess these effects, comparisons are made between the expectations from simulation
and the observations in the data.
When differences arise, their source is investigated, and this leads in a number of cases 
to uncorrelating further systematic uncertainties by means of additional NPs.
This is to prevent a constraint from being propagated from 
one kinematic region to another if this is not considered well motivated. 
The stability of the results is also tested
by performing fits for each lepton channel independently, which can also help 
to identify from which region each constraint originates.

It is particularly useful to understand 
which systematic uncertainties have the largest impact on the final results, and therefore should
be considered with greater care. For this purpose, a so-called ranking of the NPs is established. 
For each systematic uncertainty, the fit is performed again with the corresponding NP fixed 
to its fitted value, $\hat{\theta}$, shifted up or down by its fitted uncertainty,
with all the other parameters allowed to vary
so as to take properly into account the correlations between systematic uncertainties.
The magnitude of the shift in the fitted signal strength $\hat\mu$ is a measure of the observed 
impact of the considered NP.
The same procedure is repeated, using the nominal values of the NP and of its associated 
uncertainty to provide its expected impact. 
To reduce the computation time and therefore to enable more detailed fit studies, some of the NPs 
which have a negligible effect on the expected fitted uncertainty on $\hat\mu$ are dropped: those associated 
with the muon momentum scale and resolution and with the electron energy resolution; one of those
associated with the jet energy scale; and those associated with the quark--gluon composition of
the backgrounds, which turn out to be fully correlated with those associated with the difference in
energy response between quark and gluon jets.
The ranking of the systematic uncertainties obtained with the MVA 
applied to the 8~TeV data is shown in figure~\ref{fig:ranking} 
with the NPs ordered by decreasing post-fit impact on $\hat\mu$.
The five systematic uncertainties with the largest impact are, in
descending order, those:
on the dijet-mass shape for the $Wbb$ and $Wcc$ backgrounds for $\ptw > 120$~GeV; 
on the $Wbl/Wbb$ normalisation ratio for $\ptw > 120$~GeV;
on the $Wbb$ background normalisation;
on the \ptw\ shape in the 3-jet category for the $W$+hf background;
and on the signal acceptance due to the parton-shower modelling.

\begin{figure}[tb!]
\begin{center}
\includegraphics[width=0.68\textwidth]{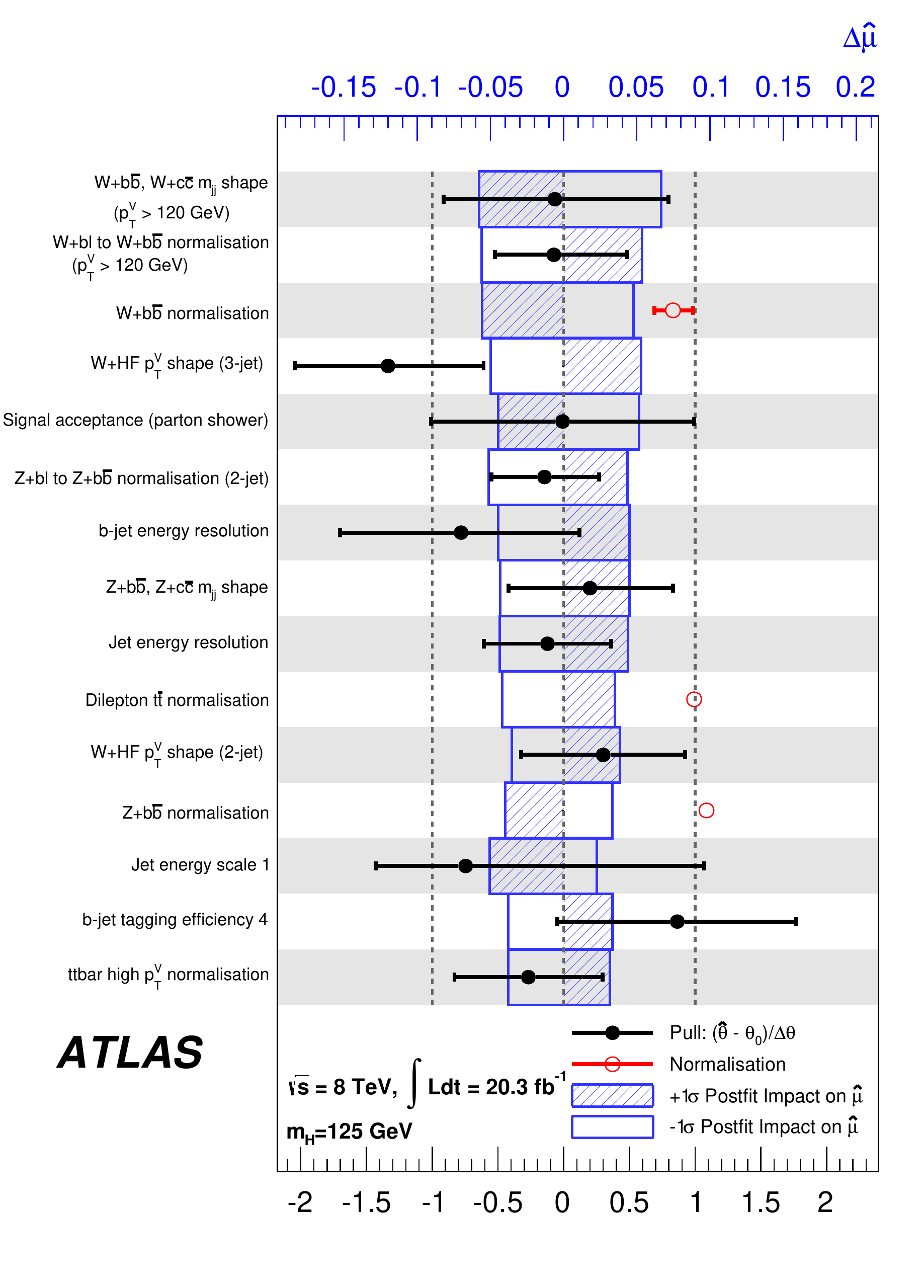}
\end{center}
\caption{Impact of systematic uncertainties on the fitted signal-strength parameter $\hat\mu$ for the MVA applied to the 8~TeV data.
         The systematic uncertainties are listed in decreasing order of their impact on $\hat\mu$ on the $y$-axis.
         The boxes show the variations of $\hat\mu$, referring to the top $x$-axis, when fixing 
         the corresponding individual nuisance parameter $\theta$ to its post-fit value $\hat\theta$ modified upwards or downwards
         by its post-fit uncertainty, and repeating the fit as explained in the text.
         The hatched and open areas correspond to the upwards and downwards variations, respectively. 
         The filled circles, referring to the bottom $x$-axis, show the deviations 
         of the fitted nuisance parameters $\hat\theta$ from their nominal values $\theta_0$, 
         expressed in terms of standard deviations with respect to their nominal uncertainties $\Delta\theta$. 
         The associated error bars show the post-fit uncertainties 
         of the nuisance parameters, relative to their nominal uncertainties. 
         The open circles with their error bars, also referring to the bottom $x$-axis, show the fitted
         values and uncertainties of the normalisation parameters that are freely floating in the fit. 
	 The normalisation parameters have a pre-fit value of one.
As explained in section~\ref{sec:sysexperimental}, the jet energy
scale and $b$-tagging uncertainties are decomposed into uncorrelated 
components; the labels 1 and 4 refer to such components.
         }
\label{fig:ranking}
\end{figure}

Since the same data sample is used for both the dijet-mass analysis and the MVA, the
consistency of the two final results, i.e., the two fitted signal strengths, 
is assessed using
the ``bootstrap'' method~\cite{bootstrap}.
A large number of event samples are randomly extracted from the simulated samples, 
with the signal strength $\mu$ set to unity, the SM value. Each of them is representative of the
integrated luminosity used for the data analysis in terms of expected yields as well as of 
associated Poisson fluctuations. Each of these event samples is subjected to both the 
dijet-mass analysis and the MVA, thus allowing the two fitted $\hat\mu$ values to be compared and their 
statistical correlation to be extracted. At the same time, the expected distributions of $\hat\mu$ and 
of its uncertainty are determined for both the dijet-mass analysis and the MVA. 

\subsection{Cross checks using diboson production}
\label{sec:vzfit}

Diboson production with a $Z$ boson decaying to a pair of $b$-quarks
and produced in association with either a $W$ or $Z$ boson has a signature
very similar to the one considered in this analysis,
but with a softer $p_{\mathrm T}^{bb}$ spectrum and with a  
\mbb\ distribution peaking at lower values. 
The cross section is about five times larger than for the SM Higgs boson
with a mass of 125~GeV.
Diboson production is therefore used as a validation of the analysis procedure. 
For the dijet-mass analysis, the binning of the transformed \mbb\ distribution
is reoptimised for the $Z$ boson mass.
For the MVA, the BDTs are retrained to discriminate the diboson signal from 
all backgrounds (including the Higgs boson). So-called 
``$VZ$ fits'' are performed, where the normalisation of the diboson contributions 
is allowed to vary with a multiplicative scale factor $\mu_{VZ}$ 
with respect to the SM expectation, 
except for the small contribution from $WW$ production,
which is treated as a background and constrained within its uncertainty. 
A SM Higgs boson with $m_H = 125$~GeV is included as a background, with
a production cross section at the SM value with an uncertainty of $50$\%.
Distributions of the BDT$_{VZ}$ discriminants of the MVA are shown in figure~\ref{fig:BDT_VZ}
for 2-tag signal regions with $\ptv > 120$~GeV in the 2-jet category 
of the 0-, 1- and 2-lepton channels.

\begin{sidewaysfigure}[p]
\begin{center}
\hfill
\subfigure[]{\includegraphics[width=0.32\textwidth]{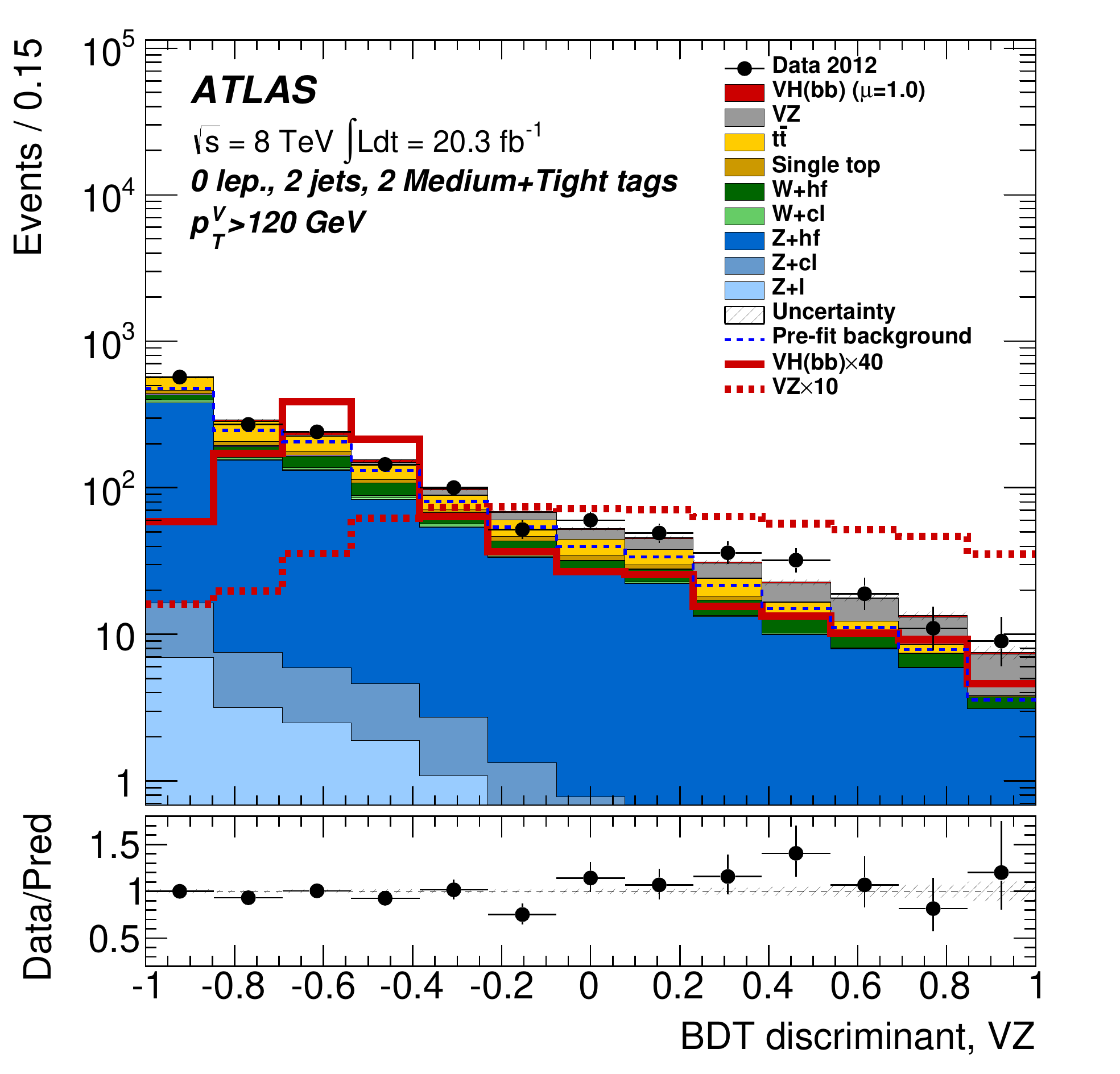}}
\hfill
\subfigure[]{\includegraphics[width=0.32\textwidth]{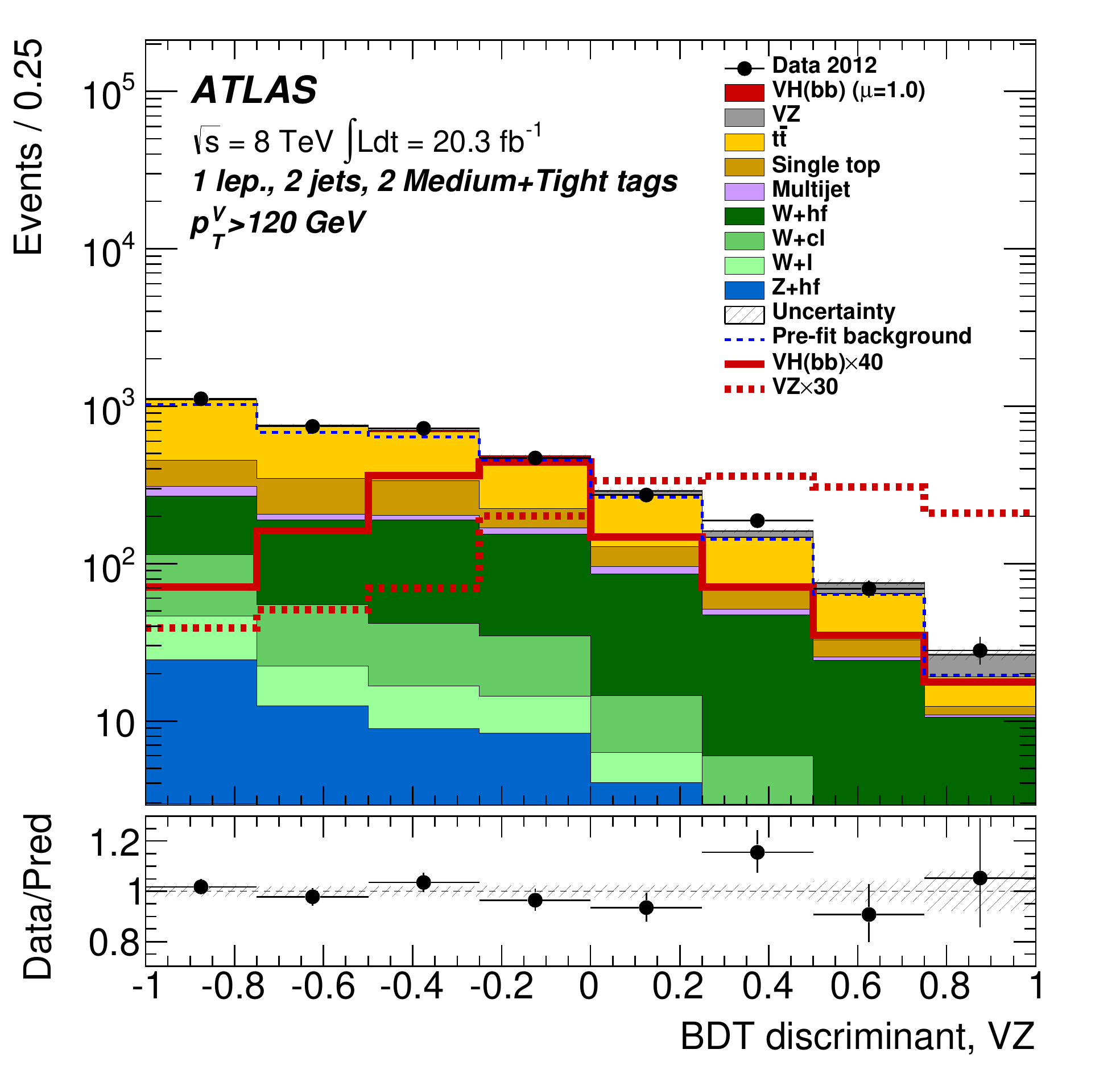}}
\hfill
\subfigure[]{\includegraphics[width=0.32\textwidth]{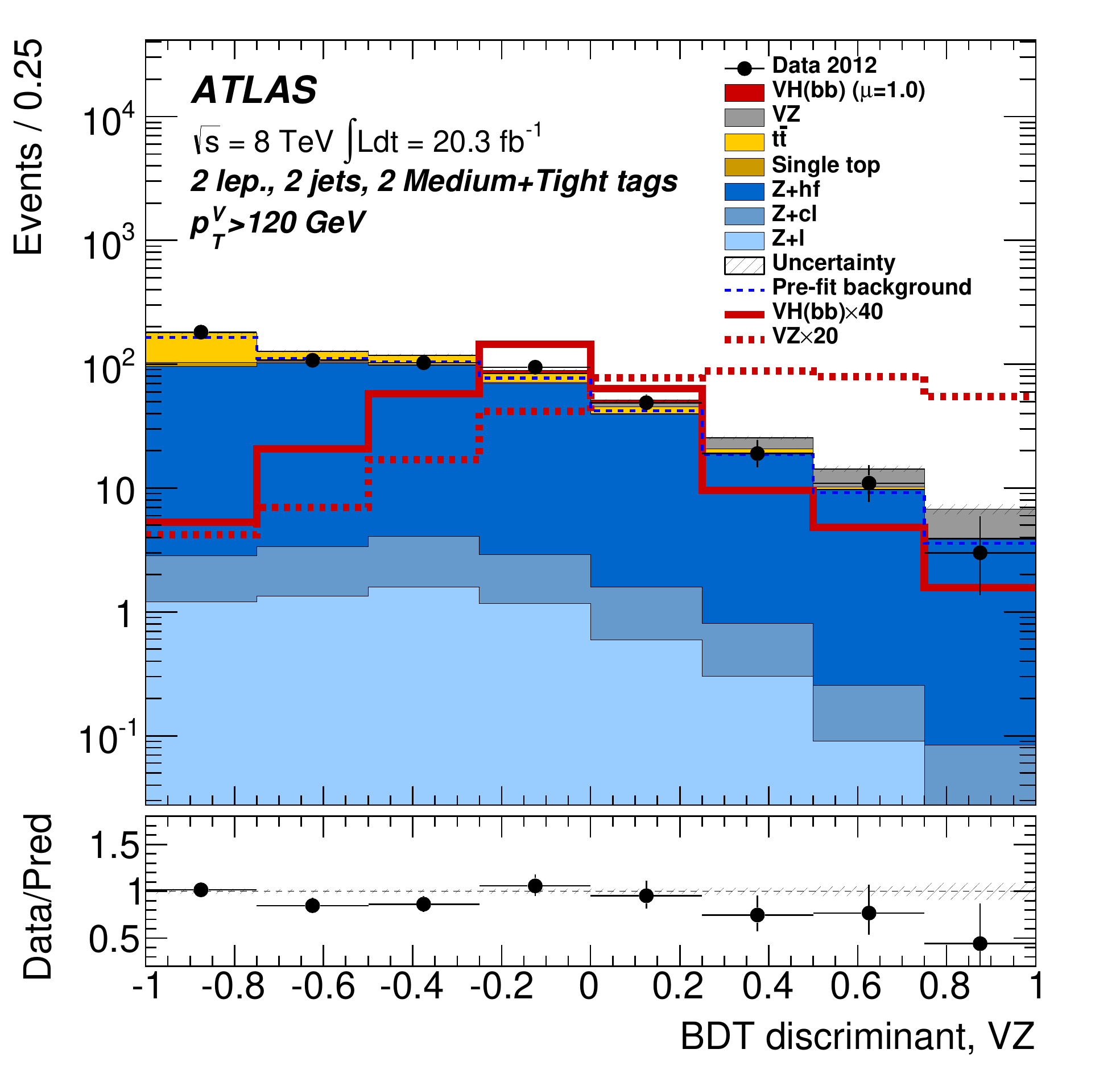}}
\hfill
\end{center}
\caption{
Distribution of the BDT$_{VZ}$ discriminant used in the $VZ$ fit 
observed in data (points with error bars) and expected (histograms) 
for the 2-jet 2-tag signal regions with $\ptv > 120$~GeV 
in the Medium (MM) and Tight (TT) categories combined of 
(a) the 0-lepton channel, 
(b) the 1-lepton channel,
(c) the 2-lepton channel.
The $VZ$ signal and background contributions are shown 
as filled histograms after the global $VZ$ fit of the MVA, except for
the Higgs boson contribution ($\mh = 125$~GeV). 
The latter is shown as expected from the SM (indicated as $\mu=1.0$)
both as a filled histogram on top of the sum of the fitted $VZ$ signal and other backgrounds, 
and unstacked as an unfilled histogram scaled by the factor indicated in the legend.
The expected $VZ$ contribution is shown in a similar way as an unfilled dotted histogram. 
The dashed histogram shows the sum of the total $VZ$ signal and background as expected from the pre-fit 
MC simulation.
The size of the combined statistical and systematic uncertainty on the
fitted $VZ$ signal and background is indicated by the hatched band. The ratio
of the data to the sum of the fitted $VZ$ signal and background is shown in the lower panel.
\label{fig:BDT_VZ}}
\end{sidewaysfigure}

As an additional check, fits are also performed with both the diboson and Higgs boson
signal-strength parameters $\mu_{VZ}$ and $\mu$ left freely floating, to study the
correlation between the two strength parameters. 
The fits in the dijet-mass analysis use the \mbb\ distributions with binning optimised
for a Higgs boson mass of 125 GeV. The fits in the MVA use BDTs trained for that same mass,
as well as the associated optimised binnings.

\FloatBarrier

\section{Analysis of the 7 TeV data}\label{sec:seven}

For the 7 TeV dataset, only a dijet-mass analysis is performed.
It is similar but not identical to the corresponding analysis for the 8~TeV data,
since some of the object reconstruction tools, such as the simultaneous use of multiple $b$-tagging
operating points, are not available for the 7 TeV data. 
In this section, the main differences between the two analyses are summarised.

\subsection{Object reconstruction}
The three categories of electrons are selected according to the loose, medium, and tight 
criteria defined in ref.~\cite{Aad:2014fxa}. The transverse energy threshold for loose 
electrons is set at 10~GeV, instead of 7~GeV. For tight electrons and muons, the calorimeter
isolation requirement is loosened from 4\% to 7\%. The procedure used to avoid double-counting
of reconstructed muon and jet objects removes muons separated by $\Delta R < 0.4$ from any jet, 
irrespective of the multiplicity of tracks associated with the jet. 
For jets, the global sequential calibration is not used
and the requirement on the fraction of track \pt\ carried by tracks 
originating from the primary vertex is raised from 50\% to 75\%. 
The $b$-tagging algorithm used is
MV1~\cite{ATLAS:2012aoa,btagnote2012b,btagnote2012c,btagnote2012l} 
instead of MV1c, with a single 
operating point to define $b$-tagged jets corresponding to an efficiency of 70\%.

\subsection{Event selection}
The selection criteria are those used in the dijet-mass analysis of the 8 TeV data, with the 
following differences.
With only one $b$-tagging operating point, a single 2-tag category is defined. 
In the 0-lepton channel, the 100--120~GeV \ptv\ interval is not used,
and the criterion for $\sum\pt^{\mathrm{jet}_i}$ is not applied.
In the 1-muon sub-channel, the \met\ trigger is used only in the 2-jet 2-tag category 
for $\ptw > 160$~GeV, 
and the events selected only by the \met\ trigger constitute distinct signal regions.
In the 1-lepton channel, $\mtw > 40$~GeV is required for $\ptw < 160$~GeV; there is no 
requirement on $H_{\mathrm{T}}$, but $\met > 25$~GeV is imposed for $\ptw < 200$~GeV. 
In the 2-lepton channel, no kinematic fit is performed.
Different lepton-flavour events are used to define a 2-tag
\ttb-dominated $e$--$\mu$ control region in the 2-lepton channel; 
the region is defined to be inclusive in jet multiplicity ($\geq 2$).

\subsection{Background composition and modelling}
The templates used to model the MJ background in the 1-lepton channel are obtained by
inversion of the track-based isolation criterion, and the normalisations are performed on the
\mtw\ and \met\ distributions in the electron and muon sub-channels, respectively. 

Corrections to
the simulation of the $V$+jet backgrounds are determined in the 1- and 2-lepton 0-tag samples
inclusively in \ptv, and applied as \dphi\ reweightings to the $W$+jet and $Z$+jet components 
in all channels. Selected dijet-mass distributions showing the background composition in 
various analysis regions are shown in figure~\ref{fig:seventevplots}. 

\begin{sidewaysfigure}[tb!]
\begin{center}
\subfigure[]{\includegraphics[width=0.32\textwidth]{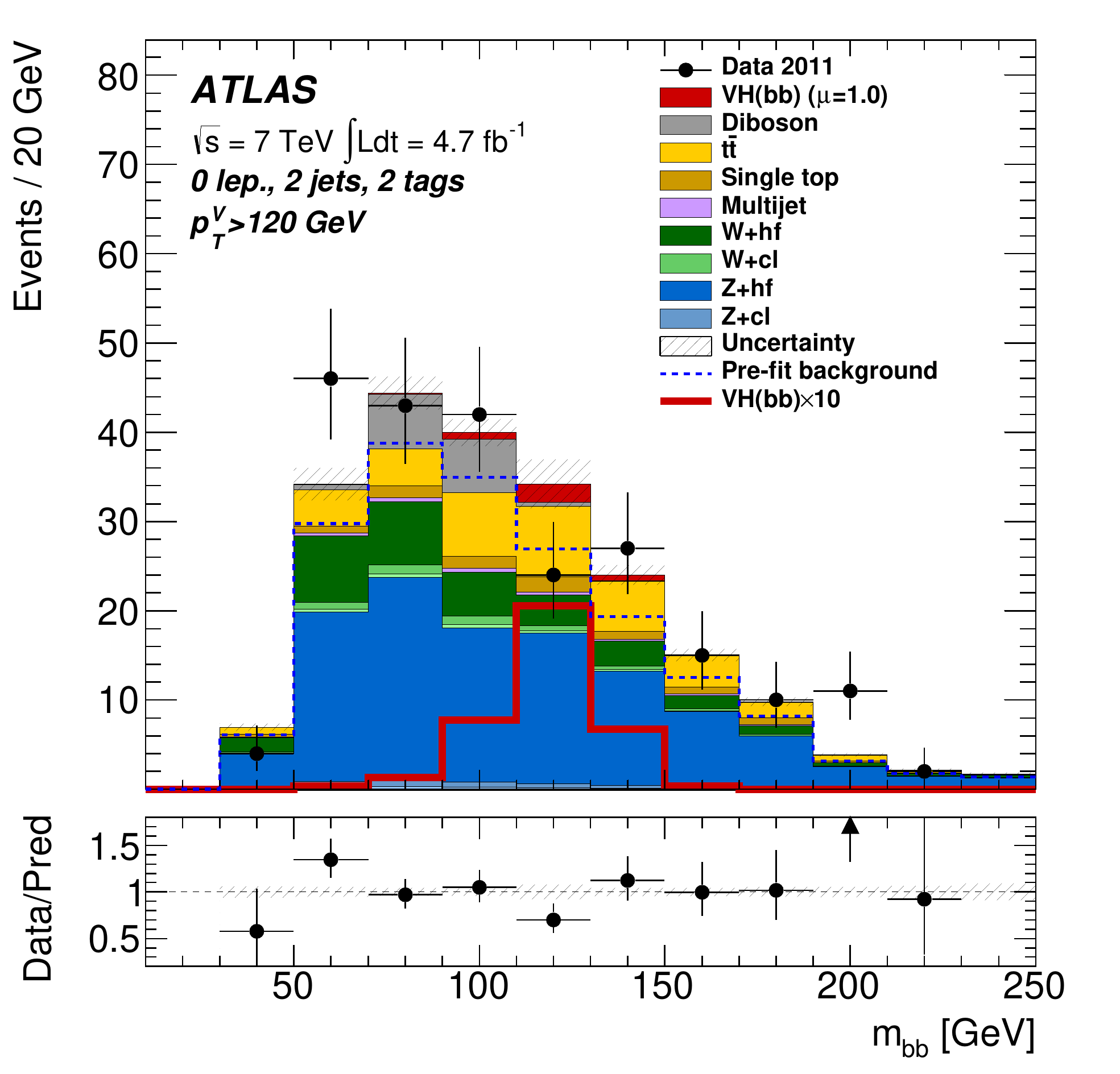}}
\hfill
\subfigure[]{\includegraphics[width=0.32\textwidth]{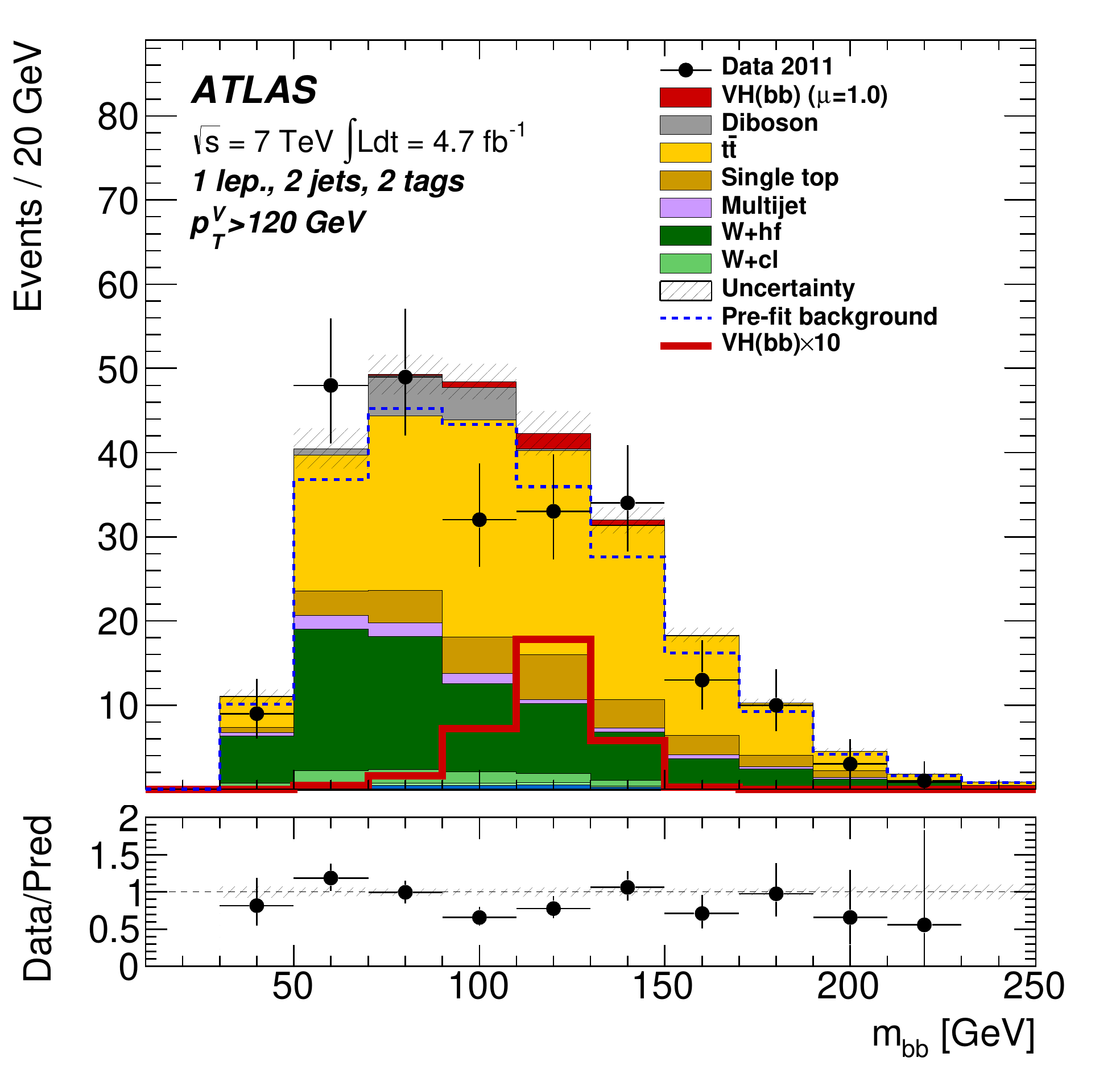}}
\hfill
\subfigure[]{\includegraphics[width=0.32\textwidth]{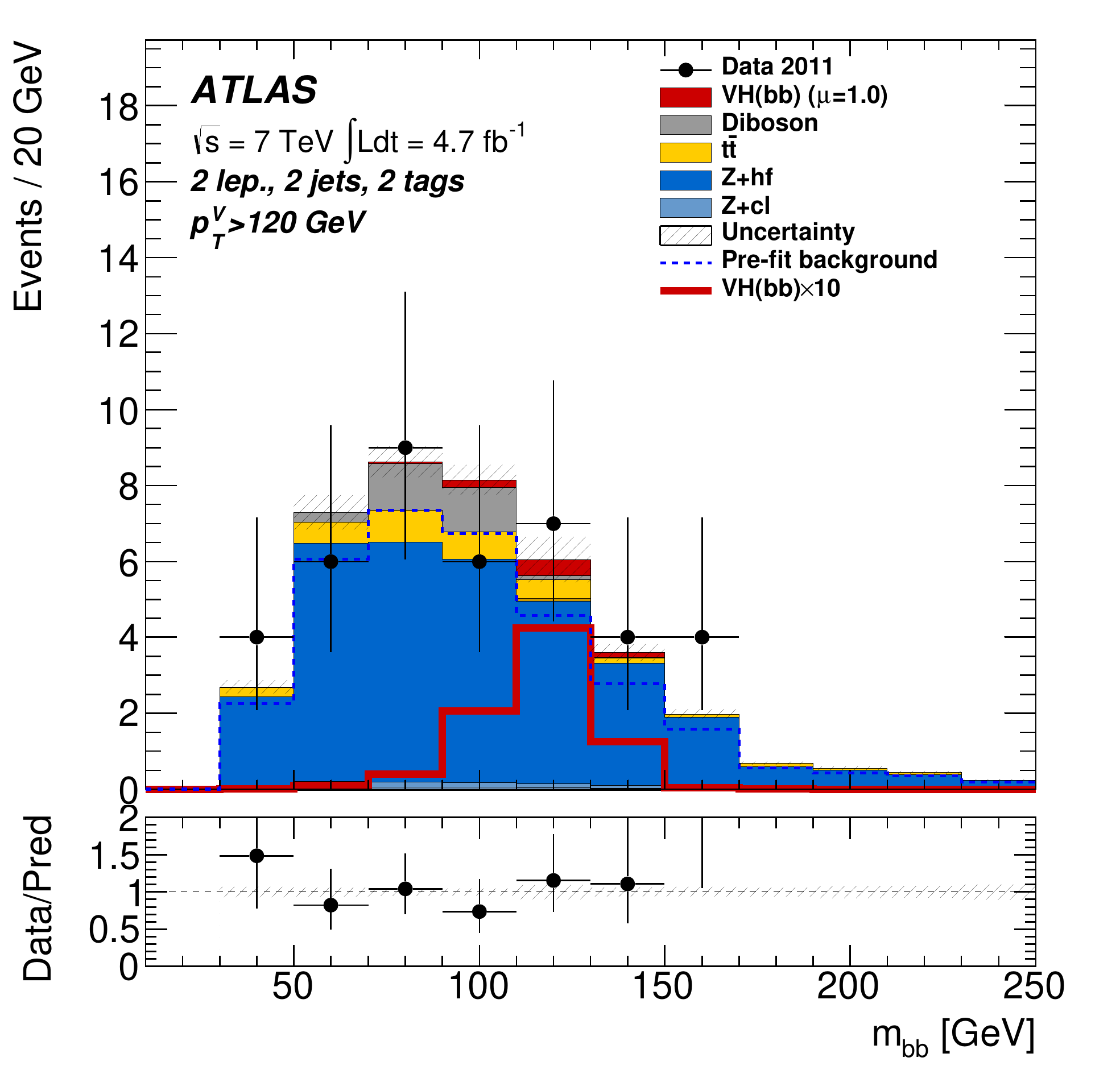}}
\end{center}
\caption{
The dijet-mass distributions observed in the 7~TeV data (points with error bars) and expected (histograms) 
for the 2-jet 2-tag signal regions with the $\ptv > 120$~GeV intervals combined: 
(a) 0-lepton channel, (b) 1-lepton channel, and (c) 2-lepton channel. 
The background contributions after the global fit of the dijet-mass analysis are shown 
as filled histograms.
The Higgs boson signal ($\mh = 125$~GeV) is shown as a filled histogram on top of 
the fitted backgrounds,
as expected from the SM (indicated as $\mu=1.0$),
and, unstacked as an unfilled histogram, scaled by the factor indicated in the legend. 
The dashed histogram shows the total background as expected from the pre-fit 
MC simulation. The entries in overflow are included in the last bin.
The size of the combined statistical and systematic uncertainty on the
sum of the signal and fitted background is indicated by the hatched band. The ratio
of the data to the sum of the signal and fitted background is shown in the lower panel.
\label{fig:seventevplots}}
\end{sidewaysfigure}

\subsection{Systematic uncertainties}
The differences with respect to the 8~TeV data analysis arise mainly from experimental systematic
uncertainties. Many of them are evaluated using independent data samples (7~TeV data vs. 8~TeV 
data), e.g., \met\ trigger efficiencies or JES. Others refer to different identification 
algorithms, e.g., electron identification or $b$-tagging. The uncertainty on the integrated luminosity is 1.8\% for the 2011 dataset~\cite{Aad:2013ucp}. 

The uncertainties affecting the signal and background simulation are estimated in a similar way 
as for the 8~TeV data, i.e., from comparisons between the baseline and alternative generators.
For $V$+jets, the $Vbc$ and $Vbb$ backgrounds are merged into a single component.
For dibosons, the baseline generator is {\sc herwig} instead of {\sc pythia8}; 
systematic uncertainties on the 3-to-2-jet ratios and on the \ptv\ distributions
are estimated at generator level for the different diboson processes by comparison with
{\sc mcfm} at NLO.
For the signal, the $gg\to ZH$ samples are generated with 
{\sc pythia8} instead of {\sc powheg}; for all processes, the
alternative generators used are {\sc pythia6} and {\sc herwig}.
  
Due to these differences, and because the phase space within which the systematic uncertainties are 
evaluated is more restricted than for the MVA applied to the 8~TeV data, all systematic 
uncertainties, except for the theoretical uncertainties on the signal, 
are treated as uncorrelated between the analyses of the 7~TeV and 8~TeV data
in the global fit to the combined dataset, in which the MVA is used for the 8~TeV data. 

\subsection{Statistical procedure}
The inputs to the likelihood fits are the \mbb\ distributions (not
transformed) in the 28 \ptv\ intervals of the 2-tag signal regions. 
Additional inputs are the event yields in the five \ptv\ intervals of
the 2-tag $e$--$\mu$ control region
and the 26 \ptv\ intervals of the 1-tag control regions. 
For the \ttb\ background, a single floating normalisation is determined
by the global fit, instead of one in each of the 0-, 1-, and 2-lepton channels.
In addition to the other floating normalisations mentioned for the
8~TeV data analysis, the MJ background normalisation 
is also left freely floating in all regions of the
1-lepton channel, except in the 2-tag 3-jet regions where the
statistical power of the data is not sufficient to provide a reliable
constraint. 
In these regions, an uncertainty of 30\% is assigned to the MJ background normalisation, 
using a method similar to what is done for the analysis of the 8~TeV data.

\FloatBarrier

\section{Results}
\label{sec:results}

As explained in section~\ref{sec:fit}, the results are obtained from maximum-likelihood 
fits to the data, where the inputs are the distributions of final discriminants in the 
2-tag signal regions and the $MV1c$ distributions of the $b$-tagged jet in the 1-tag 
control regions, with nuisance parameters either floating or constrained by priors. 
The final discriminants are the transformed \mbb\ for the dijet-mass analysis and the 
BDT$_{VH}$ discriminants for the MVA.
Results are extracted independently for the dijet-mass and multivariate analyses.
Since the MVA has better expected sensitivity to a Higgs boson signal, it is used for the
nominal results, while the dijet-mass analysis provides a cross-check 
(cf. section~\ref{sec:Xchecks}). 
For the 7 TeV data,
however, only a dijet-mass analysis is performed.
Unless otherwise specified, all results refer to a Higgs boson mass of 125~GeV.

In the following, the fitted signal-strength parameters are simply denoted $\mu$ and $\mu_{VZ}$,
rather than $\hat\mu$ and $\hat\mu_{VZ}$.

\subsection{Nominal results}

The nominal results are obtained from global fits using the MVA for the 8~TeV data 
and the dijet-mass analysis of the 7~TeV data.

\begin{figure}[Htp!]
\begin{center}
\includegraphics[width=0.62\textwidth]{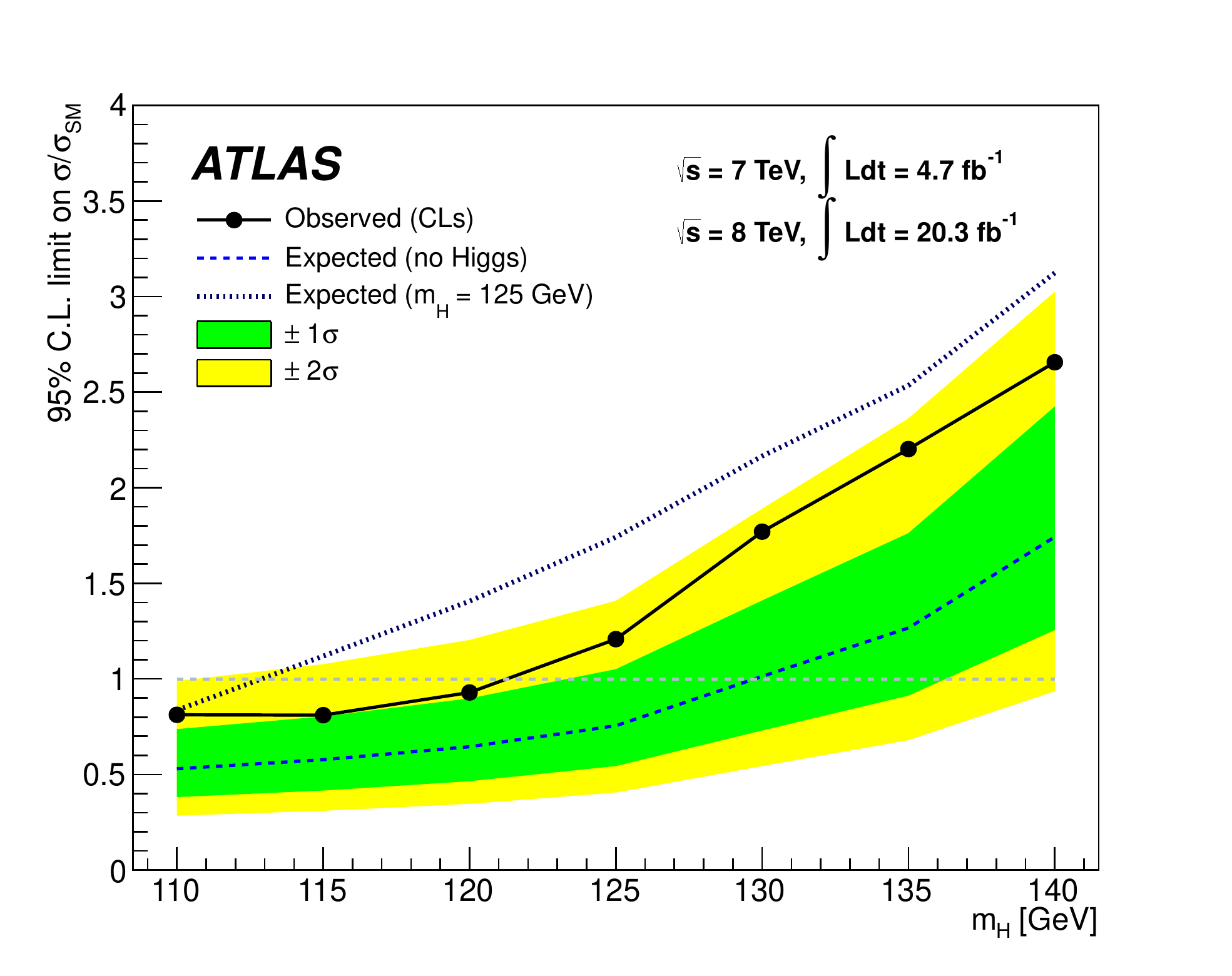}
\caption{Observed (solid) and expected 95\% CL cross-section 
upper limits, normalised to the SM Higgs boson production cross section, 
as a function of \mh\ for all channels and data-taking periods combined,
as obtained using the dijet-mass analysis for the 7 TeV dataset and
BDTs trained at each individual mass for the 8 TeV dataset.
The expected upper limit is given for the background-only hypothesis (dashed) 
and with the injection of a SM Higgs boson signal at a mass of 125~GeV (dotted).
The dark and light shaded bands represent the 1$\sigma$ and 2$\sigma$ ranges of the
expectation in the absence of a signal.
For all curves shown, the results obtained at the tested masses are linearly interpolated.
\protect\label{fig:limits}}

\includegraphics[width=0.62\textwidth]{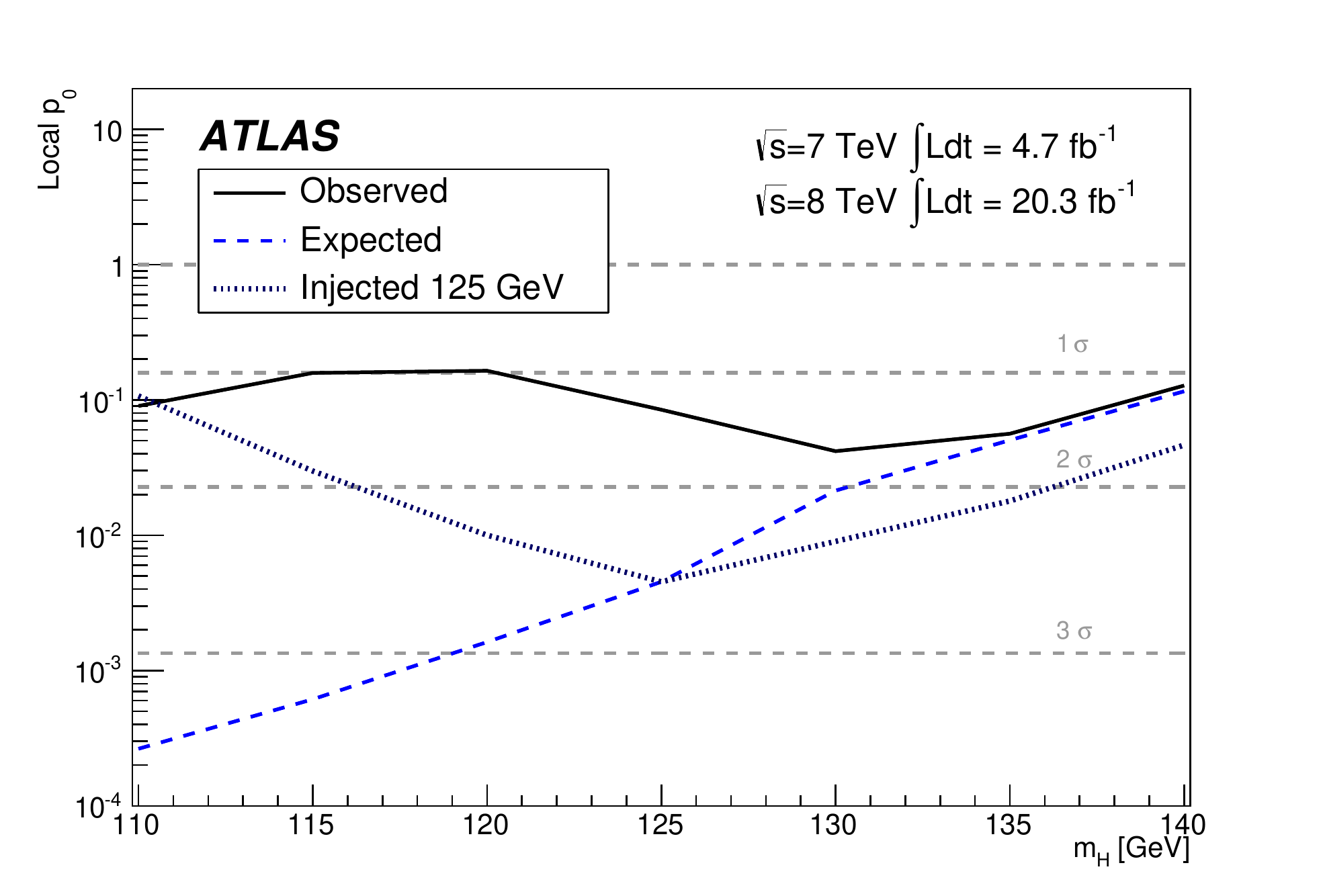}
\end{center}
\caption{Observed (solid) and expected $p_0$ values  
as a function of \mh\ for all channels and data-taking periods combined,
as obtained using the dijet-mass analysis for the 7 TeV dataset and
BDTs trained at each individual mass for the 8 TeV dataset.
The expected $p_0$ values are given for the background-only hypothesis 
in the presence of a SM Higgs boson: 
for the dashed curve the Higgs boson mass corresponds to each tested mass point in turn; 
for the dotted curve the Higgs boson mass is 125~GeV.
For all curves shown, the results obtained at the tested masses are linearly interpolated.
\protect\label{fig:pvalues}}
\end{figure}

Distributions of the BDT$_{VH}$ discriminant and of
$MV1c$, with background normalisations 
and nuisance parameters adjusted by the global fit to the 8 TeV 
data were already presented in section~\ref{sec:BDT}.
Dijet-mass distributions in the 7~TeV data analysis were shown in section~\ref{sec:seven}. 
Agreement between data and estimated background is observed within 
the uncertainties shown by the hatched bands. 

Figure~\ref{fig:limits} shows the 95\% CL upper
limits on the cross section times branching ratio for $pp\to(W/Z)(H\to\bb)$ 
in the Higgs boson mass range 
110--140 GeV.   
The observed limit for $\mh=125$~GeV is 1.2 times the SM value, 
to be compared to an expected limit, in the absence of signal, of 0.8.
For the 8~TeV (7~TeV) data only, the observed and expected limits are 1.4 (2.3) and 0.8 (3.2), 
respectively. 

The probability $p_0$ of obtaining from background alone a result at least as signal-like 
as the observation is 8\% for a tested Higgs boson mass of 125 GeV;
in the presence of a Higgs boson with that mass and the SM signal strength, 
the expected $p_0$ value is 0.5\%.
This corresponds to an excess observed with a significance of $1.4\sigma$, 
to be compared to an expectation of $2.6\sigma$. For the 8~TeV data alone, the observed and expected levels of significance are $1.7\sigma$ and $2.5\sigma$, respectively. For the 7~TeV data alone, the
expected significance is $0.7\sigma$ and there is a deficit rather
than an excess in the data, as can be seen in figure~\ref{fig:seventevplots}.
Figure~\ref{fig:pvalues} shows the $p_0$ values in the mass range 
110--140 GeV, as obtained for the 7~TeV and 8~TeV combined dataset. 

\begin{figure}[tb!]
\begin{center}
\includegraphics[width=0.7\textwidth]{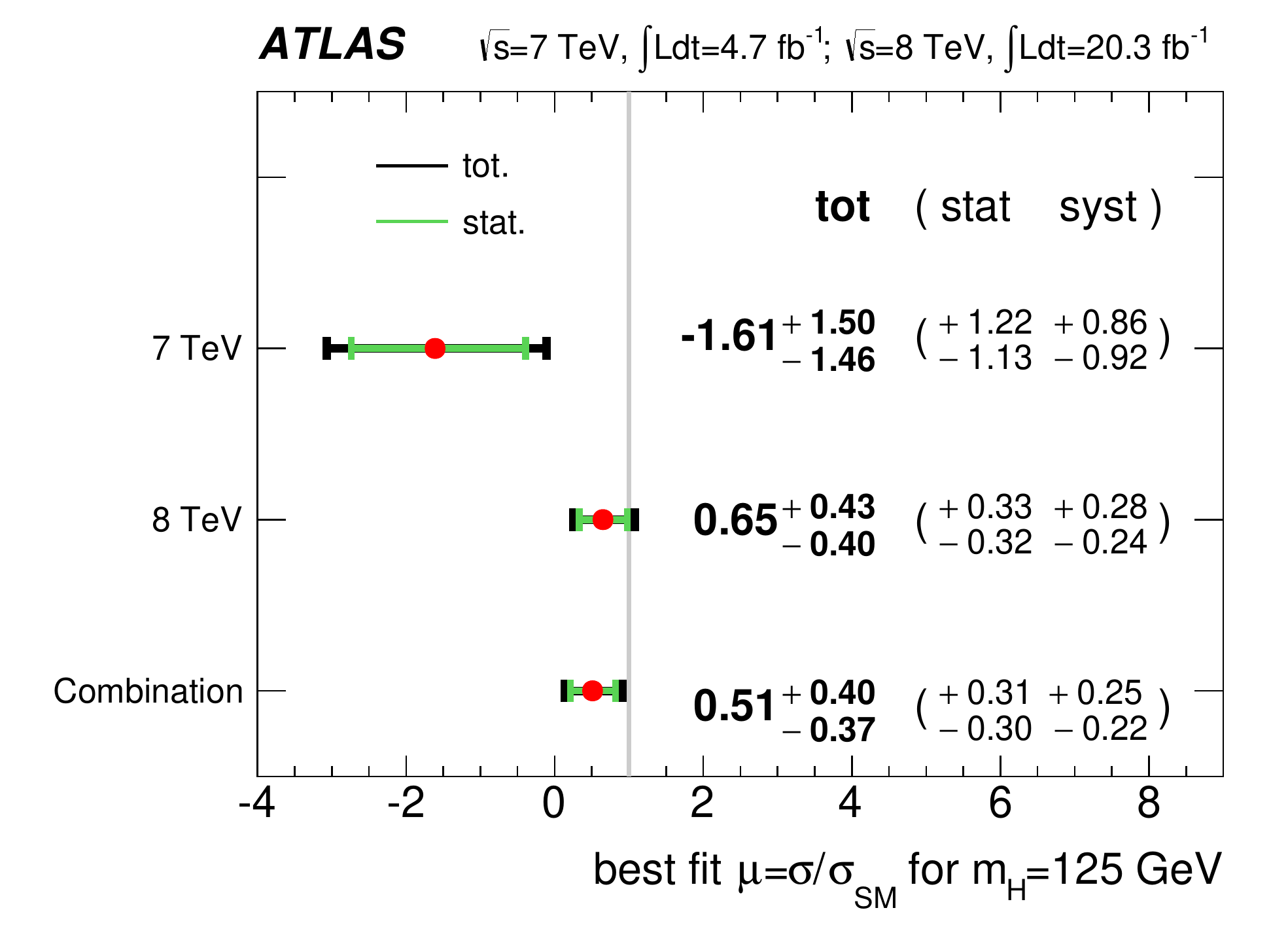}
\end{center}
\caption{The fitted values of the Higgs boson signal-strength parameter $\mu$ for $\mh=125$~GeV for the 7~TeV and 8~TeV datasets
and the combination of the 7~TeV and 8~TeV datasets.}
\label{fig:mu-higgs-a}
\end{figure}

The fitted $\mu$ values for $\mh = 125$~GeV are shown in 
figure~\ref{fig:mu-higgs-a} for the 7~TeV, 8~TeV and combined datasets.
With all lepton channels and data-taking periods combined, 
the fitted value of the signal-strength parameter 
is\footnote{The uncertainties of the normalisations of the floating backgrounds 
are included in the systematic
uncertainties; their contribution is 0.07.} 
$\mu = 0.51 \pm 0.31 \mathrm{(stat.)} \pm 0.24 \mathrm{(syst.)}$.
For the 8~TeV data, the fitted value of the signal-strength 
parameter is $\mu=0.65 \pm 0.32 \mathrm{(stat.)} \pm 0.26 \mathrm{(syst.)}$. 
For the 7~TeV data, it is $\mu= -1.6 \pm 1.2 \mathrm{(stat.)} \pm 0.9 \mathrm{(syst.)}$.

For a Higgs boson with a mass of 125.36~GeV, as measured by ATLAS~\cite{Aad:2014aba}, 
the signal-strength parameter is 
$\mu = 0.52 \pm 0.32 \mathrm{(stat.)} \pm 0.24 \mathrm{(syst.)}$.

Fits are also performed where the signal strengths are floated independently for
(i) the $WH$ and $ZH$ production processes, or (ii) the three lepton channels. 
The results of these fits are shown in figures~\ref{fig:mu-higgs-c} and~\ref{fig:mu-higgs-b} respectively.
The consistency of the fitted signal strengths in the $WH$ and $ZH$ processes
is at the level of 20\%. For the lepton channels, the consistency
between the three fitted signal strengths is at the level of 72\%
for the 7~TeV data, and of 8\% for the 8~TeV data. 
The low values of the fitted signal strengths for the $ZH$ process and 
in the 0-lepton channel are associated with the data deficit observed in 
the most sensitive bins of the BDT$_{VH}$ discriminant in the 0-lepton channel, 
shown in figure~\ref{fig:BDT_2tag_0lep}(a).

\begin{figure}[tbhp!]
\begin{center}
\includegraphics[width=0.7\textwidth]{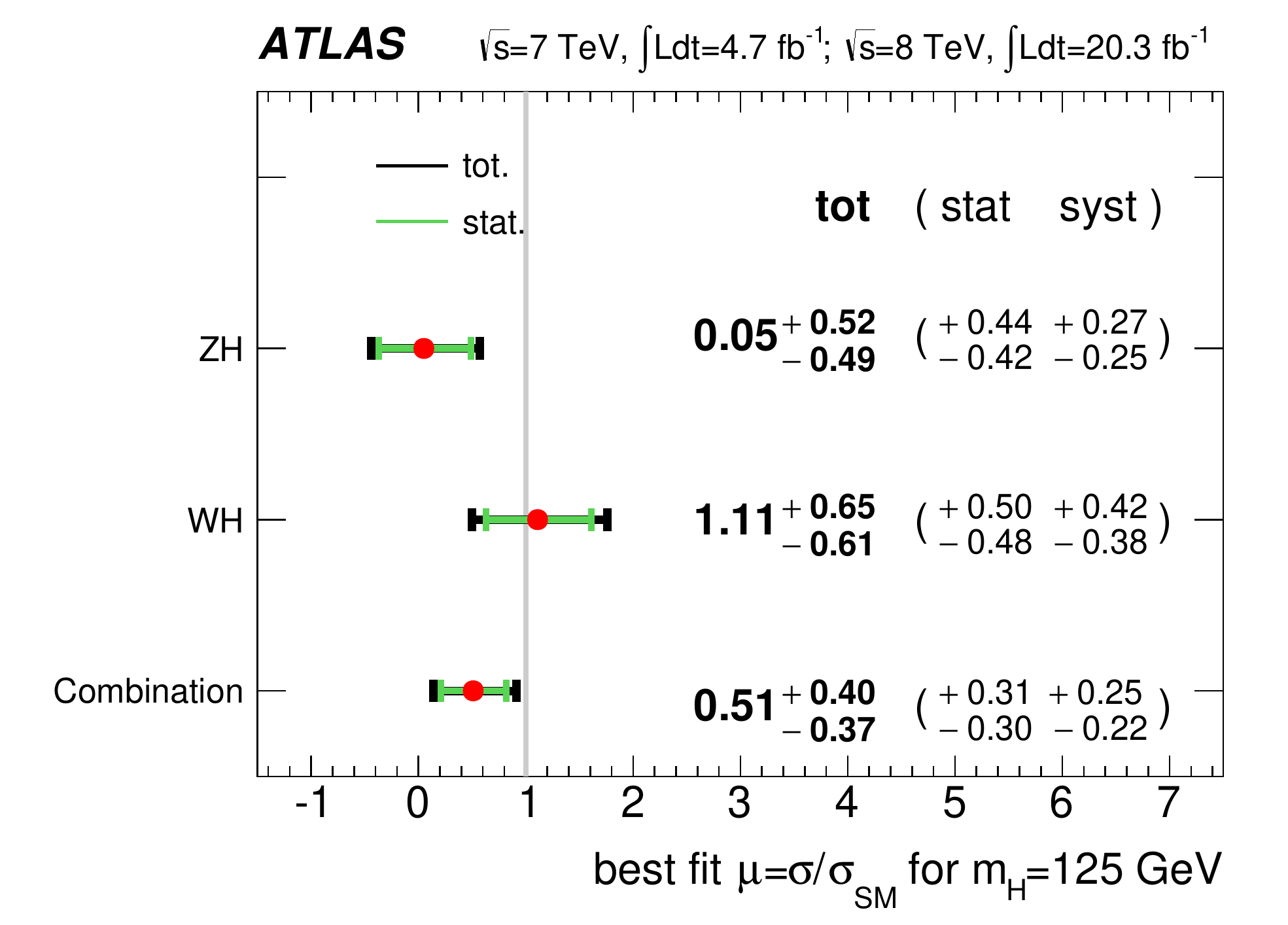}
\end{center}
\caption{The fitted values of the Higgs boson signal-strength parameter $\mu$ for $\mh=125$~GeV for the $WH$ and $ZH$ processes
and the combination of the $WH$ and $ZH$ processes, with the 7 and 8~TeV datasets combined. The individual $\mu$ values for the $(W/Z)H$ processes
are obtained from a simultaneous fit with the signal strength for each of the $WH$ and $ZH$ processes floating independently.\label{fig:mu-higgs-c}}
\begin{center}
\includegraphics[width=0.7\textwidth]{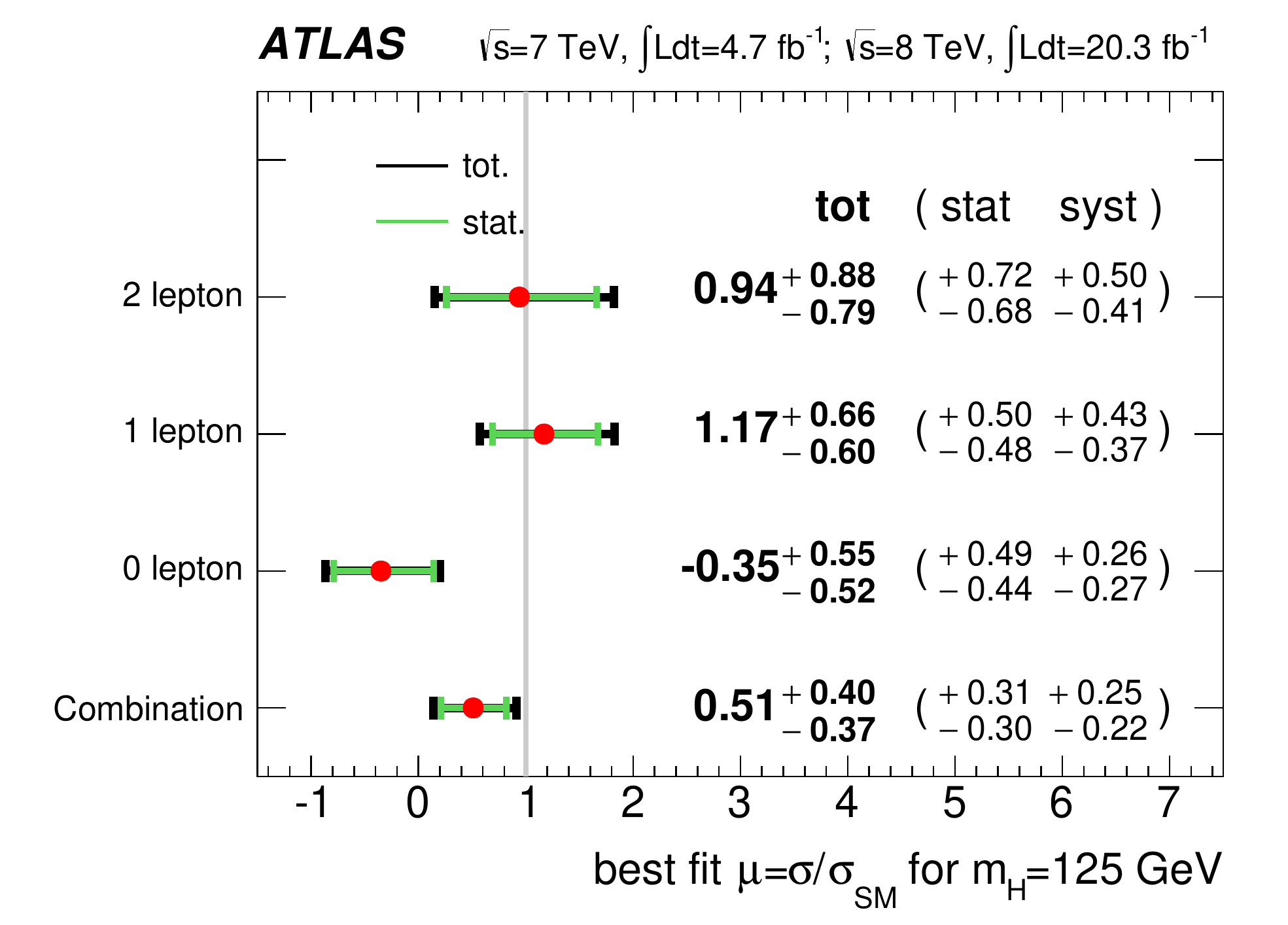}
\end{center}
\caption{The fitted values of the Higgs boson signal-strength parameter $\mu$ for $\mh=125$~GeV for the 0-, 1- and 2-lepton channels
and the combination of the three channels, with the 7 and 8~TeV datasets combined.The individual $\mu$ values for the lepton channels
are obtained from a simultaneous fit with the signal strength for each of the
lepton channels floating independently.\label{fig:mu-higgs-b}}
\end{figure}

\clearpage

Figure~\ref{fig:logstob} shows the data, background and signal yields, where the 
final-discriminant bins in all signal regions are combined into bins 
of $\log(S/B)$, separately for the 7 and 8~TeV datasets.
Here, $S$ is the expected signal yield and $B$ is the fitted background yield.
Details of the fitted values of the signal and of the various background
components are provided in table~\ref{tab:details}. 

\begin{figure}[tHb!]
\begin{center}
\hfill
\subfigure[]{\includegraphics[width=0.49\textwidth]{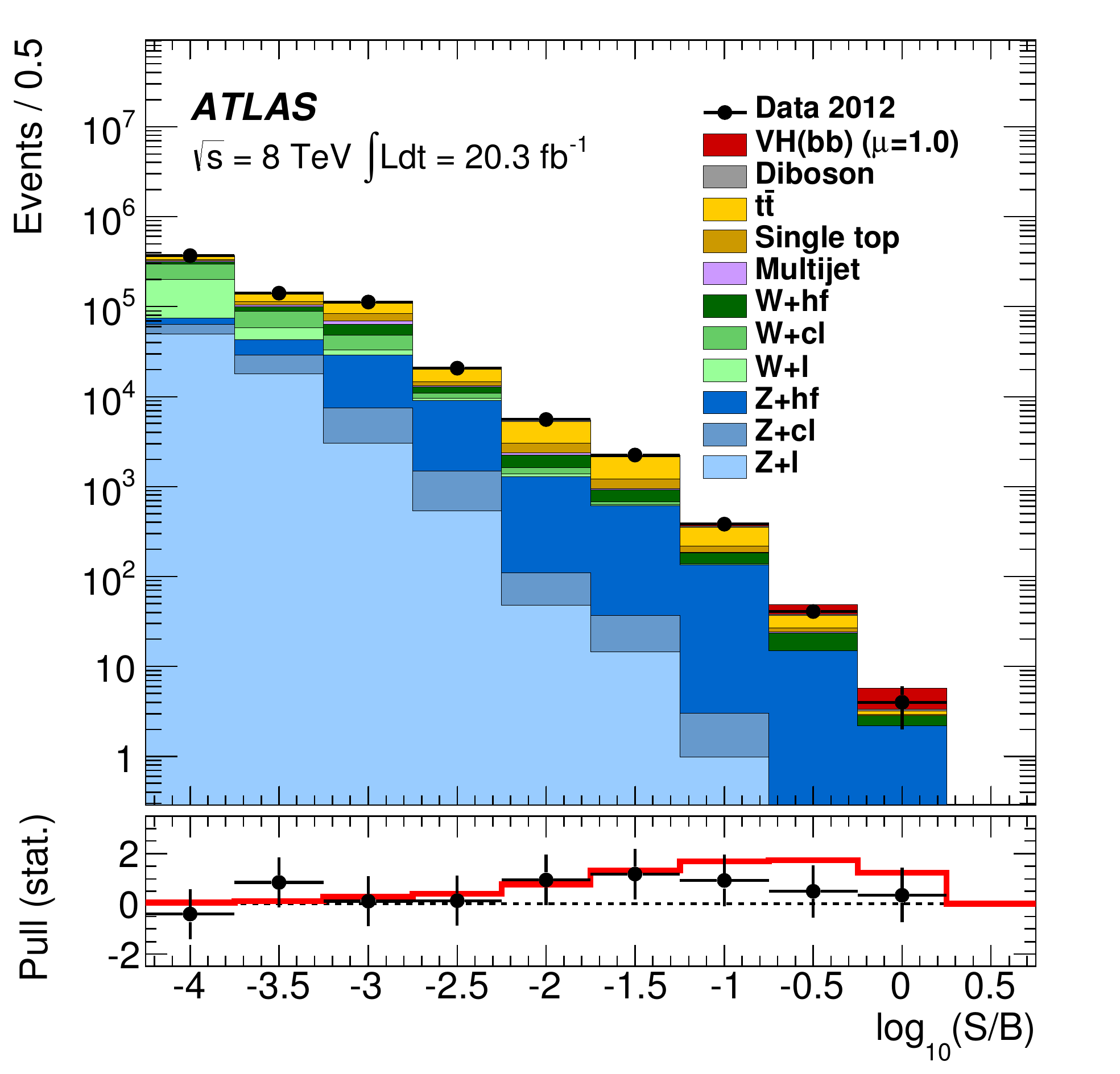}}
\hfill
\subfigure[]{\includegraphics[width=0.49\textwidth]{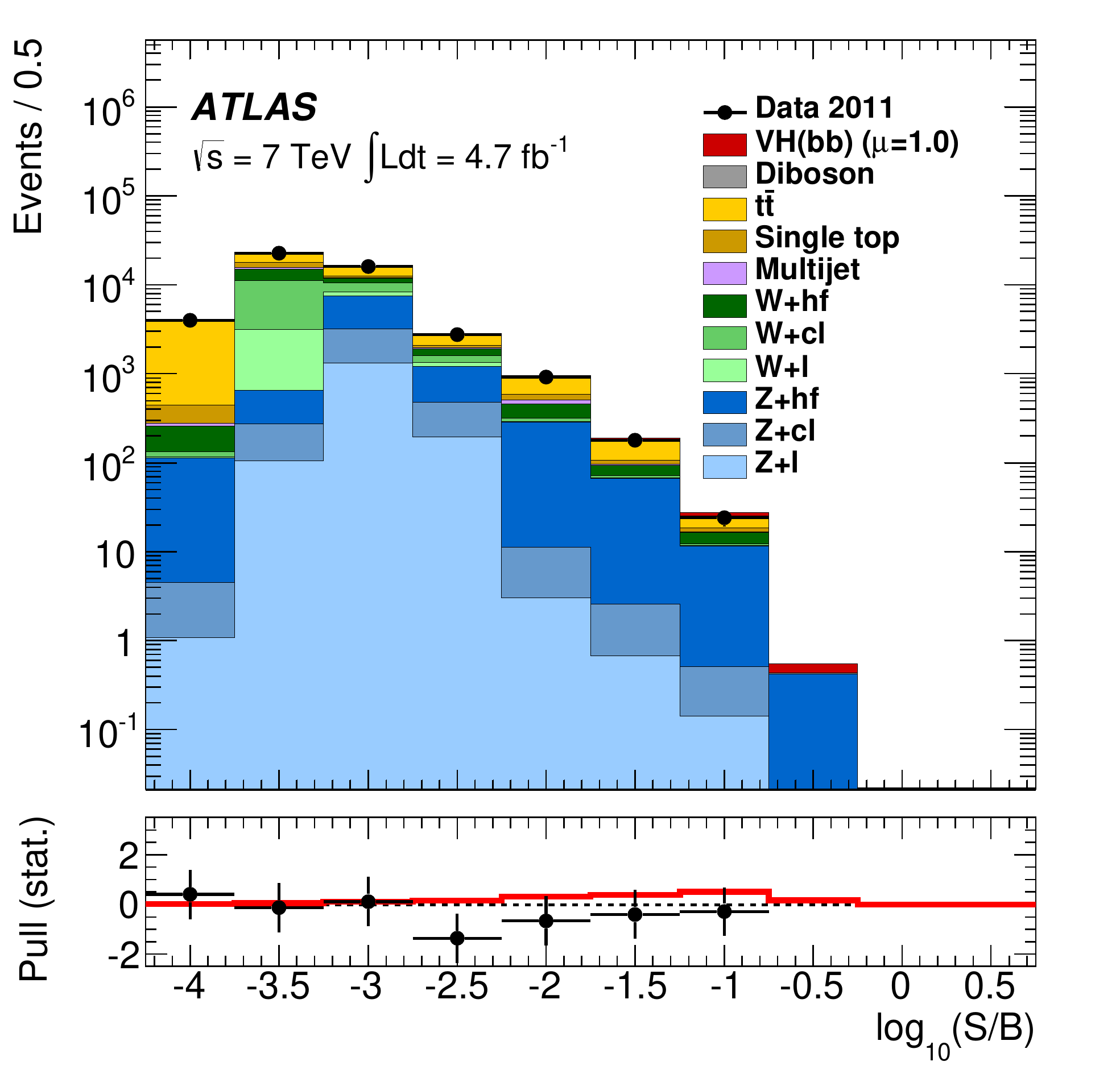}}	
\hfill
\end{center}

\caption{Event yields as a function of $\log(S/B)$ for data, background
and Higgs boson signal with $\mh=125$~GeV for the (a) 8~TeV data and (b) 7~TeV data. 
Final-discriminant bins in all signal regions 
are combined into bins of $\log(S/B)$.
The signal $S$ and background $B$ yields are expected and fitted, respectively.
The Higgs boson signal contribution is shown as expected for the SM cross section 
(indicated as $\mu=1.0$). The pull of the data with respect to the
background-only prediction is also shown with statistical
uncertainties only. 
The full line indicates the pull of the prediction for signal ($\mu=1.0$) and background
with respect to the background-only prediction.
}
	\label{fig:logstob}
\end{figure}

\begin{table}[hbpt!]
\begin{center}
		\begin{tabular}{l|ccccccccc}
		\hline\hline
		Process & Bin 1& Bin 2 & Bin 3 & Bin 4 & Bin 5 & Bin 6 & Bin 7 & Bin 8 & Bin 9 \\
		\hline
		Data & 368550& 141166 & 111865 & 20740 & 5538 & 2245 & 382 & 41 & 4 \\
		Signal & 29 & 43 & 96 & 57 & 58 & 62 & 32 & 10.7 & 2.3 \\
		Background & 368802 & 140846 & 111831 &  20722 & 5467 & 2189 & 364 & 37.9 & 3.4 \\
		$S/B$ &$8\times 10^{-5}$& 0.0003 & 0.0009 & 0.003 & 0.01 & 0.03 & 0.09 & 0.3 & 0.7 \\
		\hline
		$W$+hf & 14584 & 10626 & 15297 & 1948 & 618 & 250 & 45 & 8.2 & 0.7 \\
		$Wcl$ & 96282 & 30184 & 15227 & 1286 & 239 & 47 & 4.2 & 0.2 & 0.005 \\
		$Wl$ & 125676 & 14961 & 3722 & 588 & 107 & 16 & 1.3 & 0.03 & 0.001 \\
		$Z$+hf & 10758 & 14167 & 21684 & 7458 & 1178 & 577 & 130 & 14.8 & 2.2 \\
		$Zcl$ & 13876 & 11048 & 4419 & 941 & 61 & 22 & 2.1 & 0.1 & 0.008 \\
		$Zl$ & 49750 & 18061 & 3044 & 537 & 48 & 15 & 1 & 0.05 & 0.004 \\
		\ttb\ & 30539 & 24824 & 26729 & 5595 & 2238 & 922 & 137 & 10 & 0.3 \\
		Single top & 10356 & 9492 & 14279 & 1494 & 688 & 252 & 31 & 2.7 & 0.1 \\
		Diboson & 4378 & 1831 & 1247 &  474 & 186 & 62 & 9.7 & 1 & 0.2 \\
		Multijet & 12603 & 5650 & 6184 & 400 & 103 & 26 & 3 & 0.9 & 0 \\
		\hline\hline
		\end{tabular}
\end{center}
\caption{The numbers of expected signal and fitted background events and the observed 
numbers of events after MVA selection in the bins of figure~\protect\ref{fig:logstob}(a). 
These numbers are for both the 1-tag and 2-tag events in the 8~TeV dataset, corresponding 
to an integrated luminosity of 20.3~\ifb.
\label{tab:details}}
\end{table}

\subsection{Cross-check with the dijet-mass analysis}
\label{sec:Xchecks}

The distributions of \mbb\ in the dijet-mass analysis, 
with background normalisations
and nuisance parameters adjusted by the global fit to the 8 TeV 
data were already presented in section~\ref{sec:dijet}. 
Agreement between data and estimated background is observed within 
the uncertainties shown by the hatched bands. 

In the dijet-mass analysis, a $\mu$ value of $1.23 \pm 0.44 \mathrm{(stat.)} \pm  0.41 \mathrm{(syst.)}  $ is obtained for the 8~TeV dataset.
The consistency of the results of the three lepton channels is at the level of 8\%. 
Using the ``bootstrap'' method mentioned in section~\ref{sec:techfit},  
the results for the 8~TeV data with the dijet-mass analysis and with the MVA 
are expected to be 67\% correlated, and 
the observed results are found to be statistically consistent at the
level of 8\%.
The observed significance in the dijet-mass analysis is $2.2\sigma$.
The expected significance is $1.9\sigma$, 
to be compared to $2.5\sigma$ for the MVA,
which is the reason for choosing the MVA for the nominal results.

Figure~\ref{fig:diboson} shows the \mbb\ distribution in
data after subtraction of all backgrounds except for 
diboson production for the 7 and 8 TeV data, 
as obtained with the dijet-mass analysis.
In this figure, the contributions of all 2-tag signal regions in all channels 
are summed weighted by their respective ratios of expected Higgs boson signal 
to fitted background. 
The $VZ$ contribution is clearly seen, located at the expected $Z$ mass.
The Higgs boson signal contribution is shown as expected for the SM cross section.

\begin{figure}[tb!]
\begin{center}
\hfill
\subfigure[]{\includegraphics[width=0.49\textwidth]{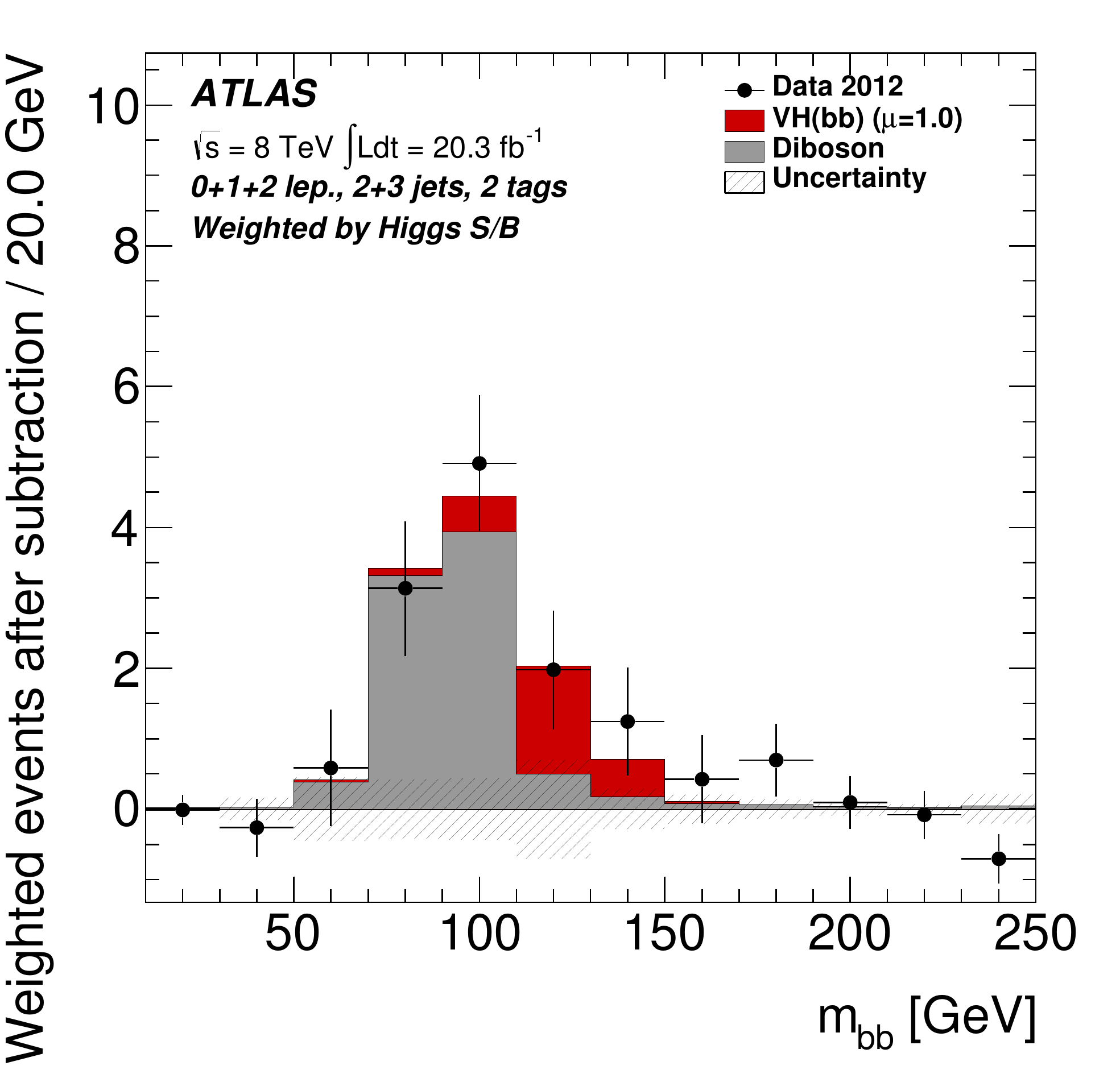}}
\hfill
\subfigure[]{\includegraphics[width=0.49\textwidth]{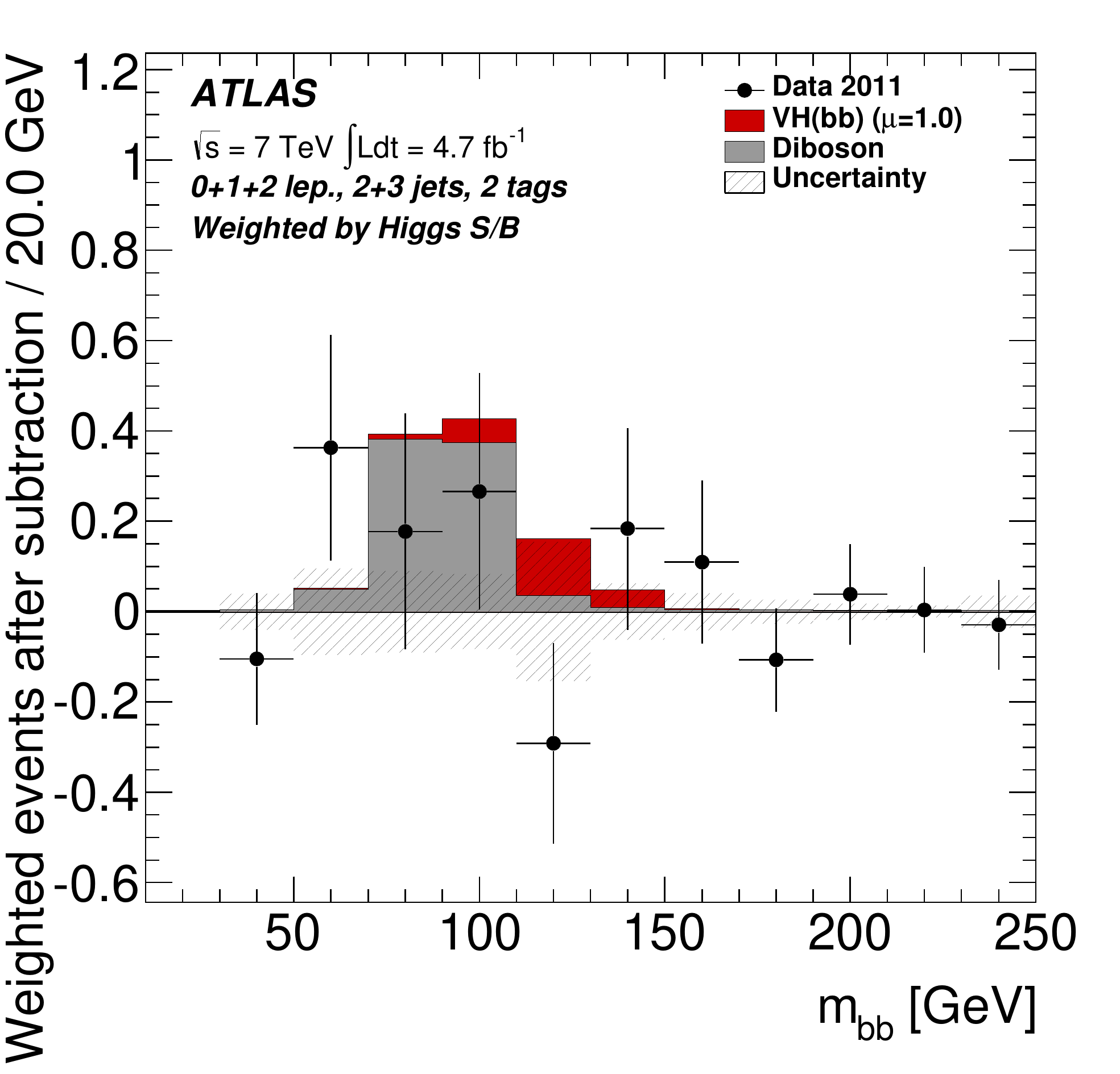}}
\hfill
\end{center}
\caption{The distribution of \mbb\ in data after subtraction of all backgrounds
except for the diboson processes, 
as obtained with the dijet-mass analysis for the (a) 8 TeV and (b) 7 TeV data. 
The contributions from all lepton channels, \ptv\ intervals, 
number-of-jets and 2-tag $b$-tagging categories are summed weighted by 
their respective values of the ratio of expected Higgs boson signal to fitted background.
The contribution of the associated $WH$ and $ZH$ production of a SM Higgs boson with $\mh=125$~GeV
is shown as expected for the SM cross section (indicated as $\mu=1.0$).
The size of the combined statistical and systematic uncertainty
on the fitted background is indicated by the hatched band.
}
	\label{fig:diboson}
\end{figure}

\smallskip

\subsection{Cross-check with the diboson analysis}
To validate the analysis procedures, $VZ$ fits are performed,
the technical details of which were discussed in section~\ref{sec:vzfit}.

The measured signal strength for the 8~TeV dataset with the MVA 
is $\mu_{VZ} = 0.77 \pm 0.10 \mathrm{(stat.)} \pm 0.15 \mathrm{(syst.)}$.
This result is consistent with the observations already made on figure~\ref{fig:diboson}.  
The signal strengths obtained for the three lepton channels are consistent at the 85\% level.
In the dijet-mass analysis at 8~TeV, a $\mu_{VZ}$ value of 
$0.79 \pm 0.11 \mathrm{(stat.)}  \pm 0.16 \mathrm{(syst.)}$ is obtained.
The correlation of the systematic uncertainties on $\mu_{VZ}$ and $\mu$
is 35\% in the MVA and 67\% in the dijet-mass analysis.

Fits are performed with the same final discriminants as used to obtain the 
results for the Higgs boson based on the 8~TeV dataset, 
but with both the $VZ$ and Higgs boson
signal-strength parameters $\mu_{VZ}$ and $\mu$ left freely floating.
The result for the Higgs boson signal strength is unchanged from the
nominal result, and the statistical correlation between the two signal-strength
parameters is found to be $-3$\% in the MVA and 9\% in the dijet-mass analysis. 
The main reason for these low correlations is the different shape of
the \ptv\ distributions for $VZ$ and for the Higgs boson signal, the \ptv\
variable being used by both the MVA and the dijet-mass analysis.
The yield tables in the appendix show that the ratio of the
diboson contribution to that of the Higgs boson is indeed smaller in
the higher \ptv\ interval than in the lower one. The additional variables input to
the BDT provide further separation in the MVA, leading to a very small
diboson contribution in the most significant bins of the
BDT$_{VH}$ discriminant, as can be seen in table~\ref{tab:details}. 

A value of $\mu_{VZ} = 0.50 \pm 0.30 \mathrm{(stat.)} \pm 0.38 \mathrm{(syst.)}$ 
is obtained for the 7~TeV dataset.
The signal strength obtained for the combined 7 and 8~TeV dataset is
$0.74 \pm 0.09 \mathrm{(stat.)} \pm 0.14 \mathrm{(syst.)}$
The $VZ$ signal is observed with a significance of 4.9$\sigma$, to be compared to
an expected significance of 6.3$\sigma$. 
The fitted $\mu_{VZ}$ values are shown in 
figure~\ref{fig:mu-diboson} for the 7~TeV, 8~TeV and combined  
datasets, and for the three lepton channels separately for the combined dataset,
all with the MVA used for the 8~TeV data.
A measurement of $VZ$ production in $pp$ collisions at
$\sqrt{s}=8$~TeV in final states with $b$-tagged jets
was recently reported by the CMS Collaboration~\cite{Chatrchyan:2014aqa}.

\begin{figure}[tb!]
\begin{center}
\hfill
\subfigure[]{\includegraphics[width=0.49\textwidth]{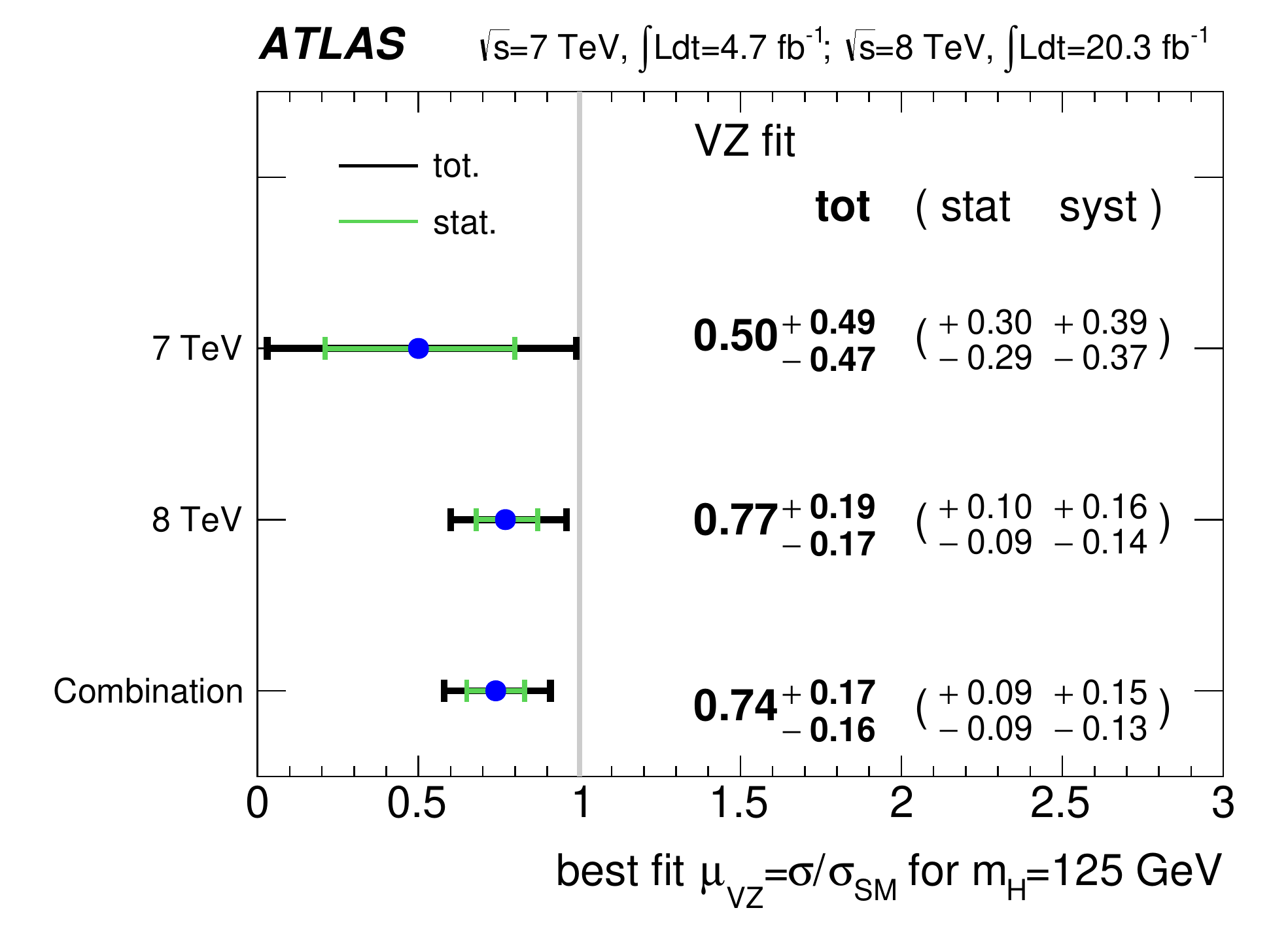}}
\hfill
\subfigure[]{\includegraphics[width=0.49\textwidth]{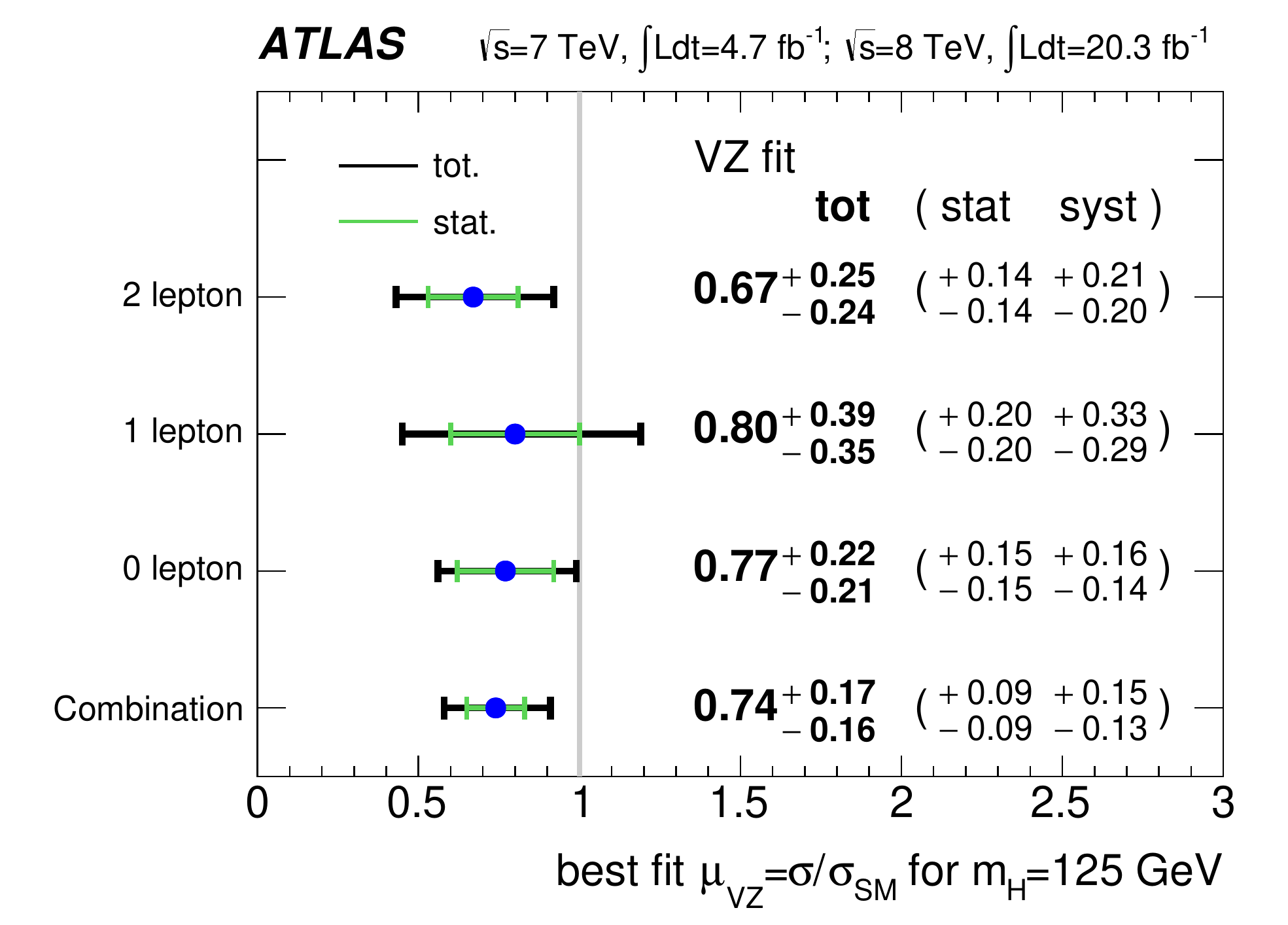}}
\hfill
\end{center}
\caption{The fitted values of the diboson signal strength $\mu_{VZ}$ for 
(a) the 7~TeV, 8~TeV and combined datasets, and 
(b) for the three lepton channels separately and combined, for the combined dataset.
The MVA is used for the 8~TeV data.
The individual $\mu_{VZ}$ values for the lepton channels are obtained from a 
simultaneous fit with the signal strength for each floating independently.}
\label{fig:mu-diboson}
\end{figure}

\FloatBarrier

\section{Summary}\label{sec:summary}

A search for the Standard Model Higgs boson produced in 
association with a $W$ or $Z$ boson and decaying into \bb\ has been presented.
The $(W/Z)$ decay channels considered are \wln, \zll\ and \znn.
The dataset corresponds to 
integrated luminosities of 4.7~\ifb\ and 
20.3~\ifb\ from $pp$ collisions at 7~TeV and 8~TeV, respectively, 
recorded by the ATLAS experiment during Run~1 of the LHC. 

The analysis is carried out 
in event categories based on the numbers of leptons, jets, and jets 
tagged as originating from $b$-quark fragmentation, and on the transverse 
momentum of the vector-boson candidate. 
A multivariate analysis provides the nominal results.
An alternative analysis using invariant-mass distributions
of the Higgs boson candidates leads to consistent results.

For a Higgs boson mass of 125.36~GeV,
the observed (expected) deviation from the background-only hypothesis corresponds
to a significance of 1.4 (2.6) standard deviations and the ratio of 
the measured signal yield
to the Standard Model expectation is found to be 
$\mu = 0.52 \pm 0.32 \mathrm{(stat.)} \pm 0.24 \mathrm{(syst.)}$.
The analysis procedure is validated by a measurement of the yield 
of $(W/Z)Z$ production with $Z\to\bb$, 
from which the ratio of the observed signal yield to the Standard Model 
expectation is found to be $0.74 \pm 0.09 \mathrm{(stat.)} \pm 0.14 \mathrm{(syst.)}$.

\clearpage

\acknowledgments
%

\section{Acknowledgements}

We thank CERN for the very successful operation of the LHC, as well as the
support staff from our institutions without whom ATLAS could not be
operated efficiently.

We acknowledge the support of ANPCyT, Argentina; YerPhI, Armenia; ARC,
Australia; BMWFW and FWF, Austria; ANAS, Azerbaijan; SSTC, Belarus; CNPq and FAPESP,
Brazil; NSERC, NRC and CFI, Canada; CERN; CONICYT, Chile; CAS, MOST and NSFC,
China; COLCIENCIAS, Colombia; MSMT CR, MPO CR and VSC CR, Czech Republic;
DNRF, DNSRC and Lundbeck Foundation, Denmark; EPLANET, ERC and NSRF, European Union;
IN2P3-CNRS, CEA-DSM/IRFU, France; GNSF, Georgia; BMBF, DFG, HGF, MPG and AvH
Foundation, Germany; GSRT and NSRF, Greece; ISF, MINERVA, GIF, I-CORE and Benoziyo Center,
Israel; INFN, Italy; MEXT and JSPS, Japan; CNRST, Morocco; FOM and NWO,
Netherlands; BRF and RCN, Norway; MNiSW and NCN, Poland; GRICES and FCT, Portugal; MNE/IFA, Romania; MES of Russia and ROSATOM, Russian Federation; JINR; MSTD,
Serbia; MSSR, Slovakia; ARRS and MIZ\v{S}, Slovenia; DST/NRF, South Africa;
MINECO, Spain; SRC and Wallenberg Foundation, Sweden; SER, SNSF and Cantons of
Bern and Geneva, Switzerland; NSC, Taiwan; TAEK, Turkey; STFC, the Royal
Society and Leverhulme Trust, United Kingdom; DOE and NSF, United States of
America.

The crucial computing support from all WLCG partners is acknowledged
gratefully, in particular from CERN and the ATLAS Tier-1 facilities at
TRIUMF (Canada), NDGF (Denmark, Norway, Sweden), CC-IN2P3 (France),
KIT/GridKA (Germany), INFN-CNAF (Italy), NL-T1 (Netherlands), PIC (Spain),
ASGC (Taiwan), RAL (UK) and BNL (USA) and in the Tier-2 facilities
worldwide.


\bibliographystyle{JHEP}
\bibliography{bibliography}

 
\clearpage
\appendix
\section{Tables of event yields}\label{app:additionalplots}

The event yields in each category for the multivariate analysis are shown in 
tables~\ref{tab:0lepyield}--\ref{tab:2lepyield}.

\begin{landscape} 

 \begin{table}[htbp] 
\small
 \centering 
 \centerline{ 
 \begin{tabular}{l|ccccccc} 
\hline \hline
Sample 	&	 \multicolumn{2}{c}{1-tag} 	&	 \multicolumn{2}{c}{LL-tag} 	&	 MM-tag 	&	 MM$+$TT-tag 	&	 TT-tag\\
  \ptv     	&	 $100-120$~GeV 	&	 $> 120$~GeV 	&	 $100-120$~GeV 	&	 $ > 120$~GeV 	&	 $100-120$~GeV 	&	 $> 120$~GeV 	&	 $100-120$~GeV\\\hline
    	&	 \multicolumn{7}{c}{2-jet} \\ 
 \hline
$VH$ &7.9 &23 &2.2 &6.9 &3.5 &23 &4.0\\
$WZ,ZZ,WW$ &235 &635 &18 &49  &14.7 &81 &13.3\\
$t\bar{t}$ &840 &1520 &114 &183 &129 &332 &116\\
Single top &531 &704 &40 &56 &32.6 &66 &22.8\\
$Wl$ &5470 &7100 &159 &206  &14.4 &16 &0.2\\
$Wcl$ &2230 &3710 &106 &159 &23.4 &27 &1.1\\
$W+$hf &762 &1520 &54 &124 &33.6 &128 &21.8\\
$Zl$ &3890 &10750 &96 &272 &6.1 &17 &0.1\\
$Zcl$ &1590 &3990 &59 &162 &9.4 &24 &0.3\\
$Z+$hf &2550 &6510 &225 &607 &186 &876 &151\\
\hline
Total &18340 $\pm$ 150 &36890 $\pm$ 200 &886 $\pm$ 17 &1841 $\pm$ 25 &458.5 $\pm$ 9.8 &1599 $\pm$ 28 &333.2 $\pm$ 8.9\\
Data &18343 &36903 &887 &1860  &477  &1592  &306 \\
\hline
    	&	 \multicolumn{7}{c}{3-jet} \\ 
 \hline
$VH$ & --  &8 & --  &2 & --  &7 & -- \\
$WZ,ZZ,WW$  & --  &260 & --  &17 & --  &20 & -- \\
$t\bar{t}$ & --  &1670 & --  &186 & --  &315 & -- \\
Single top & --  &318 & --  &25 & --  &30 & -- \\
$Wl$ & --  &2280 & --  &59 & --  &4.3 & -- \\
$Wcl$ & --  &1240 & --  &53 & --  &8.9 & -- \\
$W+$hf & --  &750 & --  &60 & --  &62 & -- \\
$Zl$ & --  &3190 & --  &79 & --  &4.5 & -- \\
$Zcl$ & --  &1620 & --  &65 & --  &9.8 & -- \\
$Z+$hf & --  &1890 & --  &170 & --  &259 & -- \\
\hline
Total & --  &13310 $\pm$ 100 & --  &718 $\pm$ 12 & --  &719 $\pm$ 17 & -- \\
Data & --  &13344 & --  &657 & --  &710 & -- \\
\hline
\hline \hline 
\end{tabular} 
 } 
\caption{The expected signal and fitted background yields for each category of the 0-lepton channel after the full selection of the multivariate analysis. 
The \mbb\ distribution is used in the 100 $< \ptV < 120$~GeV interval. 
The background yields are normalised by the results of the global likelihood fit. 
All systematic uncertainties are included in the indicated uncertainties.\label{tab:0lepyield}} 
\end{table} 
 \end{landscape}

 \begin{landscape} 

  \begin{table}[htbp] 
			 \small
  \centering 
  \centerline{ 
  \begin{tabular}{l|cccccccc} 
 \hline \hline
 Sample 	&	 \multicolumn{2}{c}{1-tag} & \multicolumn{2}{c}{LL-tag} & \multicolumn{2}{c}{MM-tag} & \multicolumn{2}{c}{TT-tag}\\
   \ptv     & $< 120$~GeV & $> 120$ & $< 120$~GeV & $> 120$~GeV & $< 120$~GeV & $> 120$~GeV & $< 120$~GeV & $> 120$~GeV\\\hline
     & \multicolumn{8}{c}{2-jet} \\ 
  \hline
$VH$ &39 &28 &11 &9.3 &17 &14 &19 &16\\
$WZ,ZZ,WW$  &1950 &927 &103.6 &62 &64 &36 &52 &29.5\\
$t\bar{t}$ &11380 &6641 &1954 &1051 &2426 &1080 &2290 &890\\
Single top &13680 &3730 &1150 &398 &975 &307 &739 &219\\
$Wl$ &65980 &23702 &1603 &697 &124 &48 &3.1 &0.9\\
$Wcl$ &71930 &21650 &2630 &966 &465 &153 &23 &6.5\\
$W+$hf &16030 &6112 &1470 &648 &954 &402 &506 &227\\
$Zl$ &3940 &1223 &101 &37 &7.4 &2.8 &0.1 &0.0\\
$Zcl$ &1350 &333 &53 &18 &10 &3.1 &0.5 &0.2\\
$Z+$hf &2080 &475 &161 &45 &126 &30 &85 &24\\
MJ ($e$) & --  &2618 & --  &162 & --  &89 & --  &40\\
MJ ($\mu$) &10230 &164 &721 &16.0 &329 &4.8 &178 &1.3\\
\hline
Total &198540 $\pm$ 500 &67600 $\pm$ 290 &9953 $\pm$ 91 &4106 $\pm$ 50 &5492 $\pm$ 66 &2161 $\pm$ 33 &3889 $\pm$ 55 &1448 $\pm$ 27\\
Data &198544  &67603 &9941 &4072  &5499  &2199  &3923 &1405 \\
 \hline
     	&	 \multicolumn{8}{c}{3-jet} \\ 
  \hline
VH &15 &14 &3.2 &3.8 &4.8 &5.8 &5.4 &6.5\\
WZ,ZZ,WW  &1100 &689 &50 &39.6 &22.6 &18 &14 &14\\
$t\bar{t}$ &18660 &10490 &3240 &1622 &4119 &1670 &4181 &1388\\
Single top &7390 &2815 &66 &318 &619 &261 &503 &188\\
$W+$l &24980 &11320 &588 &322 &42 &20 &1.1 &0.4\\
$W+$cl &25900 &10080 &952 &454 &164 &72 &7.7 &3.2\\
$W+$hf &6530 &4740 &576 &490 &353 &297 &187 &168\\
$Z+$l &1780 &572 &43 &18.1 &2.8 &1.4 &0.0 &0.0\\
$Z+$cl &690 &193 &27 &9.8 &4.5 &1.6 &0.2 &0.1\\
$Z+$hf &1024 &272 &77 &25.9 &54 &18.8 &40 &14\\
MJ ($e$) & --  &1290 & --  &68.6 & --  &36 & --  &15\\
MJ ($\mu$) &5300 &91 &227 &4.9 &117 &3.2 &58 &0.8\\
\hline
Total &93350 $\pm$ 320 &42570 $\pm$ 200 &6447 $\pm$ 57 &3376 $\pm$ 43 &5501 $\pm$ 50 &2405 $\pm$ 33 &4995 $\pm$ 55 &1796 $\pm$ 30\\
Data &93359 &42557 &6336 &3472 &5551 &2356 &4977 &1838 \\
 \hline \hline 
 \end{tabular} 
  } 
 \caption
 {The expected signal and fitted background yields for each category
   of the 1-lepton channel after the full selection of the multivariate analysis.
The background yields are normalised by the results of the global likelihood fit. 
All systematic uncertainties are included in the indicated uncertainties.\label{tab:1lepyield}
 } 
 \end{table} 
  \end{landscape}

\begin{landscape} 
 \begin{table}[htbp]
	 \small 
 \centering 
 \centerline{ 
 \begin{tabular}{l|cccccc} 
\hline \hline
Sample 	&	 \multicolumn{2}{c}{1-tag} 	&	 \multicolumn{2}{c}{LL-tag} 	&	 \multicolumn{2}{c}{MM$+$TT-tag}\\
   \ptv    	&	 $< 120$~GeV 	&	 $ > 120$~GeV 	&	 $< 120$~GeV 	&	 $> 120$~GeV 	&	 $< 120$~GeV 	&	 $> 120$~GeV\\\hline
    	&	 \multicolumn{6}{c}{2-jet} \\ 
 \hline
VH &20 &7.4 &5.5 &2.0 &19 &7.0\\
WZ,ZZ,WW  &800 &135 &73 &12.2 &129 &22\\
$t\bar{t}$ &3490 &213 &1023 &49 &3180 &137\\
Single top &385 &44 &40 &5.8 &96 &12\\
$W+$l &245 &12 &3.7 &0.4 &0.3 &0.0\\
$W+$cl &166 &18 &5.8 &0.6 &1.0 &0.1\\
$W+$hf &90 &13 &7.5 &0.7 &4.9 &0.6\\
$Z+$l &29410 &4180 &648 &101 &41 &6.2\\
$Z+$cl &12130 &1622 &421 &64 &71 &9.7\\
$Z+$hf &21090 &2853 &1916 &287.7 &2858 &414\\
\hline
Total &69360 $\pm$ 280 &9123 $\pm$ 86 &4271 $\pm$ 50 &524.6 $\pm$ 7.9 &6473 $\pm$ 76 &607 $\pm$ 13\\
Data &69313 &9150 &4313 &517 &6501 &570 \\
\hline
    	&	 \multicolumn{6}{c}{3-jet} \\ 
 \hline
VH &7.4 &3.9 &1.5 &0.9 &4.9 &2.8\\
WZ,ZZ,WW  &347 &100 &22 &6.9 &29 &8.7\\
$t\bar{t}$ &2170 &182 &486 &29 &1476 &74\\
Single top &144 &22 &18 &3.0 &41 &6.5\\
$W+$l &76 &7.1 &1.4 &0.2 &0.1 &0.0\\
$W+$cl &57 &8.9 &2.3 &0.4 &0.4 &0.1\\
$W+$hf &40 &6.4 &3.1 &0.8 &0.8 &0.4\\
$Z+$l &8870 &1913 &190 &46 &12 &2.7\\
$Z+$cl &4650 &949 &164 &38 &25 &5.7\\
$Z+$hf &5790 &1223 &497 &121 &766 &180\\
\hline
Total &22720 $\pm$ 150 &4430 $\pm$ 56 &1403 $\pm$ 20 &245.2 $\pm$ 5.2 &2365 $\pm$ 43 &280.9 $\pm$ 8.4\\
Data &22662 &4436 &1428 &253 &2394 &283 \\
\hline \hline 
\end{tabular} 
 } 
\caption
{The expected signal and fitted background yields for each category of the 2-lepton channel after the full selection of the multivariate analysis.
The background yields are normalised by the results of the global likelihood fit. 
All systematic uncertainties are included in the indicated uncertainties.\label{tab:2lepyield}
} 
\end{table} 
 \end{landscape}


\clearpage 
\begin{flushleft}
{\Large The ATLAS Collaboration}

\bigskip

G.~Aad$^{\rm 85}$,
B.~Abbott$^{\rm 113}$,
J.~Abdallah$^{\rm 152}$,
S.~Abdel~Khalek$^{\rm 117}$,
O.~Abdinov$^{\rm 11}$,
R.~Aben$^{\rm 107}$,
B.~Abi$^{\rm 114}$,
M.~Abolins$^{\rm 90}$,
O.S.~AbouZeid$^{\rm 159}$,
H.~Abramowicz$^{\rm 154}$
% ...
\bigskip
\\
% ...
$^{11}$ Institute of Physics, Azerbaijan Academy of Sciences, Baku, Azerbaijan\\
% ...
$^{85}$ CPPM, Aix-Marseille Universit{\'e} and CNRS/IN2P3, Marseille, France\\
% ...
$^{90}$ Department of Physics and Astronomy, Michigan State University, East Lansing MI, United States of America\\
% ...
$^{107}$ Nikhef National Institute for Subatomic Physics and University of Amsterdam, Amsterdam, Netherlands\\
% ...
$^{113}$ Homer L. Dodge Department of Physics and Astronomy, University of Oklahoma, Norman OK, United States of America\\
$^{114}$ Department of Physics, Oklahoma State University, Stillwater OK, United States of America\\
% ...
$^{117}$ LAL, Universit{\'e} Paris-Sud and CNRS/IN2P3, Orsay, France\\
% ...
$^{152}$ Institute of Physics, Academia Sinica, Taipei, Taiwan\\
% ...
$^{154}$ Raymond and Beverly Sackler School of Physics and Astronomy, Tel Aviv University, Tel Aviv, Israel\\
% ...
$^{159}$ Department of Physics, University of Toronto, Toronto ON, Canada\\
% ...
\end{flushleft}

\end{document}